\documentclass[10pt]{amsart}
\usepackage{hyperref}
\usepackage{float}

\usepackage{amscd,amssymb,amsmath,latexsym,bm}
\usepackage[mathcal,mathscr]{euscript}
\usepackage{lipsum}
\usepackage{amsfonts}
\usepackage{graphicx}
\usepackage{epstopdf}
\usepackage{algorithmic}
\usepackage{hyperref}
\usepackage{xcolor}
\usepackage{upgreek}
\usepackage{enumerate}
\usepackage{ulem}
\usepackage{gensymb}			%DJA added this for degrees symbol
\usepackage[utf8]{inputenc}
\usepackage[small,nohug]{diagrams}
\diagramstyle[labelstyle=\scriptstyle]
 
\newtheorem{theorem}{Theorem}[section]

\newtheorem{corollary}[theorem]{Corollary}
\newtheorem{proposition}[theorem]{Proposition}
\newtheorem{remark}[theorem]{Remark}
\newtheorem{definition}[theorem]{Definition}
\newtheorem{example}[theorem]{Example}

\newtheorem{assumption}[theorem]{Assumption}

\numberwithin{equation}{section}
\numberwithin{figure}{section}

\newcommand{\BM}{{\mathbb B}}
\newcommand{\CM}{{\mathbb C}}
\newcommand{\NM}{{\mathbb N}}

\newcommand{\RM}{{\mathbb R}}

\newcommand{\TM}{{\mathbb T}}
\newcommand{\ZM}{{\mathbb Z}}

\newcommand{\KM}{{\mathbb K}}

\newcommand{\Aa}{{\mathcal A}}

\newcommand{\Pp}{{\mathcal P}}

\newcommand{\BB}{{\bf B}}

\newcommand{\Dd}{{\mathcal D}}

\newcommand{\Gg}{{\mathcal G}}

\newcommand{\Tt}{{\mathcal T}}

\newcommand{\Mm}{{\mathcal M}}
\newcommand{\Cc}{{\mathcal C}}

\newcommand{\Kk}{{\mathcal K}}
\newcommand{\Hh}{{\mathcal H}}

\newcommand{\one}{{\bf 1}}

\begin{document}

\title[The $K$-Theoretic Bulk-Boundary Principle]{The $K$-Theoretic Bulk-Boundary Principle for Dynamically Patterned Resonators}

\author{Emil Prodan, Yitzchak Shmalo}
\address{Department of Mathematical Sciences, Yeshiva University, USA}

\thanks{Financial support from the W. M. Keck Foundation is greatly acknowledged.}

\date{\today}

\begin{abstract} Starting from a dynamical system $(\Omega,G)$, with $G$ a generic topological group, we devise algorithms that generate families of patterns in the Euclidean space, which densely embed $G$ and on which $G$ acts continuously by rigid shifts. We refer to such patterns as being dynamically generated. For $G=\ZM^d$, we adopt Bellissard's $C^\ast$-algebraic formalism to analyze the dynamics of coupled resonators arranged in dynamically generated point patterns. We then use the standard connecting maps of $K$-theory to derive precise conditions that assure the existence of topological boundary modes when a sample is halved. We supply four examples for which the calculations can be carried explicitly. The predictions are supported by many numerical experiments.
\\
\end{abstract}


\maketitle

\setcounter{tocdepth}{2}

{\scriptsize \tableofcontents}

\section{Introduction}

The bulk-boundary correspondence principle for the Integer Quantum Hall Effect (IQHE) was formulated by Hatsugai in 1993 \cite{Hat1}. It provided a topological explanation for the emergence of robust quantum modes along edges cut to IQHE systems. In the simplest terms, the phenomenon occurs because both the bulk and the edge modes carry topological invariants, whose numerical values coincide. Hence, absence of edge spectrum implies trivial edge topological invariant, which, at its turn, implies trivial bulk topological invariant. In other words, a non-trivial bulk topological invariant warrants the emergence of edge spectrum. In the $K$-theoretic framework, the general argument remains the same but the topological invariants are no longer numerical and instead are represented by the classes of certain $K$-groups. In this framework, the bulk-boundary correspondence principle was formulated by Kellendonk, Richter and Schulz-Baldes at the beginning of 2000's \cite{SKR,KRS,KS,Kel1} (see also \cite{KR} for an overview). It enabled extensions of the work by Hatsugai to the case of disordered crystals and irrational magnetic fluxes per unit cell and, for this reason, it is often referred to as the strong version of the bulk-boundary principle. The concepts and the strategy introduced in these works are quite general and have been followed up in \cite{PS} and \cite{BKR} to formulate the bulk-boundary correspondence principle for all classes of topological insulators and super-conductors in arbitrary dimensions. 

\vspace{0.2cm}

The works mentioned above dealt strictly with disordered crystals. Having a configuration space which is contractible to a point, the disordered crystals can be considered as the simplest class among the aperiodic systems. For example, from the $K$-theoretic point of view, they are mere perturbations of the ideal periodic crystals. The present work is part of a new and extremely vigorous effort from the topological meta-materials community
\cite{KLR2012,VZK2013,MBB,KRZ,Pro,HPW,KZ,TGB,TDG,VZL,
LBF,DLA,BRS,BLL,AS17,KellendonkArxiv2017,BP2017,
MNH,LSPZB2018,ZHGW2018,MaoPRL2018},  focused on going beyond the periodic table of disordered topological insulators \cite{SRFL2008,Kit2009,RSFL2010}. In \cite{ApigoArxiv2018}, the first author and his collaborators have pushed this effort one step forward, opening a program on exploring the bulk-boundary principle in aperiodic systems with the stated goal of generating topological boundary spectrum solely by smart patterning.  

\vspace{0.2cm}

In the present work, we introduce a broad class of patterns, which we call dynamically generated. Specifically, starting from a  topological dynamical system $(\Omega,G,\tau)$, we propose a class of algorithms to generate patterns in the Euclidean space, such that all points or a dense set of points of the pattern can be canonically labeled by the elements of the group $G$ and this group acts by rigid shifts on the family of dynamically generated patterns. Explicit examples show that extremely complex patterns can be generated by our proposed algorithms. Yet, the dynamics and the bulk-boundary principle over these patterns are tractable, because the algebras of observables are crossed products. While our general construction is for generic topological groups, the bulk of our analysis is focused on point patterns dynamically generated by $\ZM^d$. The algebras of the physical observables for these patterns are crossed products by $\ZM^d$ and, as such, the $K$-theoretic bulk-boundary principle can be easily adapted to the present context. Using these readily available tools, we show that, indeed, topological boundary spectrum can emerge solely from the patterning, without any tuning of the Hamiltonians. 

\vspace{0.2cm}

Key to all these is the concept of discrete hull of a pattern, defined as the topological subspace $\Xi$ traced by the rigid translates of the given pattern, inside the space of patterns. For dynamically generated patterns by $\ZM^d$, the discrete hull comes equipped with an action of this group and, as we shall see, whenever the topology of $\Xi$ is nontrivial, novel topological phases emerge, which are not accounted by the periodic table. These phases were called virtual topological phases in \cite{Pro}, and the methods introduced here can be seen as algorithms for producing such virtual topological phases.

\vspace{0.2cm}

The paper is organized as follows. In chapter~\ref{Ch:PattRez}, we introduce the idea of patterned resonators, which should give the experimentalists a quick guide on how to implement and observe our predictions. In fact, the latter has already been done in \cite{ApigoArxiv2018,QianArxiv2018}. We also use in this chapter simple physical arguments to show how the algebraic formalism emerges and what kind of results one can expect. In chapter~\ref{Ch:DGP}, we introduce the dynamically generated patterns and carefully study their main properties. In chapter~\ref{Ch:DGUPP}, we restrict ourselves to uniform point patterns generated by $\ZM^d$, whose properties are further studied. This chapter also introduces four particular patterns, which will serve as working examples throughout our exposition. Chapter~\ref{Ch-BulkAlgebra} reviews Bellissards $C^\ast$-algebraic formalism and computes the $C^\ast$-algebra for our working examples, showing that they are all isomorphic with non-commutative tori of various dimensions. Chapter~\ref{Ch:KTh} introduces the $K$-groups and explains the gap-labeling procedure. The latter generates all topological invariants that can be associated with a model. We show how to extract these labels from a simple computation of the integrated density of states. Chapter~\ref{Ch:BBCorrespondence} formulates the bulk-boundary correspondence for dynamically patterned resonators, which is put to the test on our working examples. Chapter~\ref{Ch:SingDS} presents two singular limits of the dynamically generated patterns, with drastically different outcomes regarding the boundary spectra.

\section{Patterned resonators}
\label{Ch:PattRez}

In this chapter we formalize the concept of a resonator and introduce the idea of patterned resonators. Leaving the technical details aside, we use simple physical arguments to show how the algebraic formalism championed by Jean Bellissard \cite{Bel1} emerges in this physical context. The chapter concludes with a heads-up on the follow-up chapters and how they relate to the physical systems introduced here. 

\subsection{Definitions, examples, dynamics}

In an attempt to cover a broad range of systems with one theory, we introduce an abstract definition of a resonator. Regardless of the nature of the resonators, which can be quantum, photonic, plasmonic, magnonic, acoustic or mechanical, the individual or the coupled resonators are formally all identical, if one focuses entirely on the dynamics of the discrete degrees of freedom.

\begin{figure}[t]
\center
\includegraphics[width=0.9\textwidth]{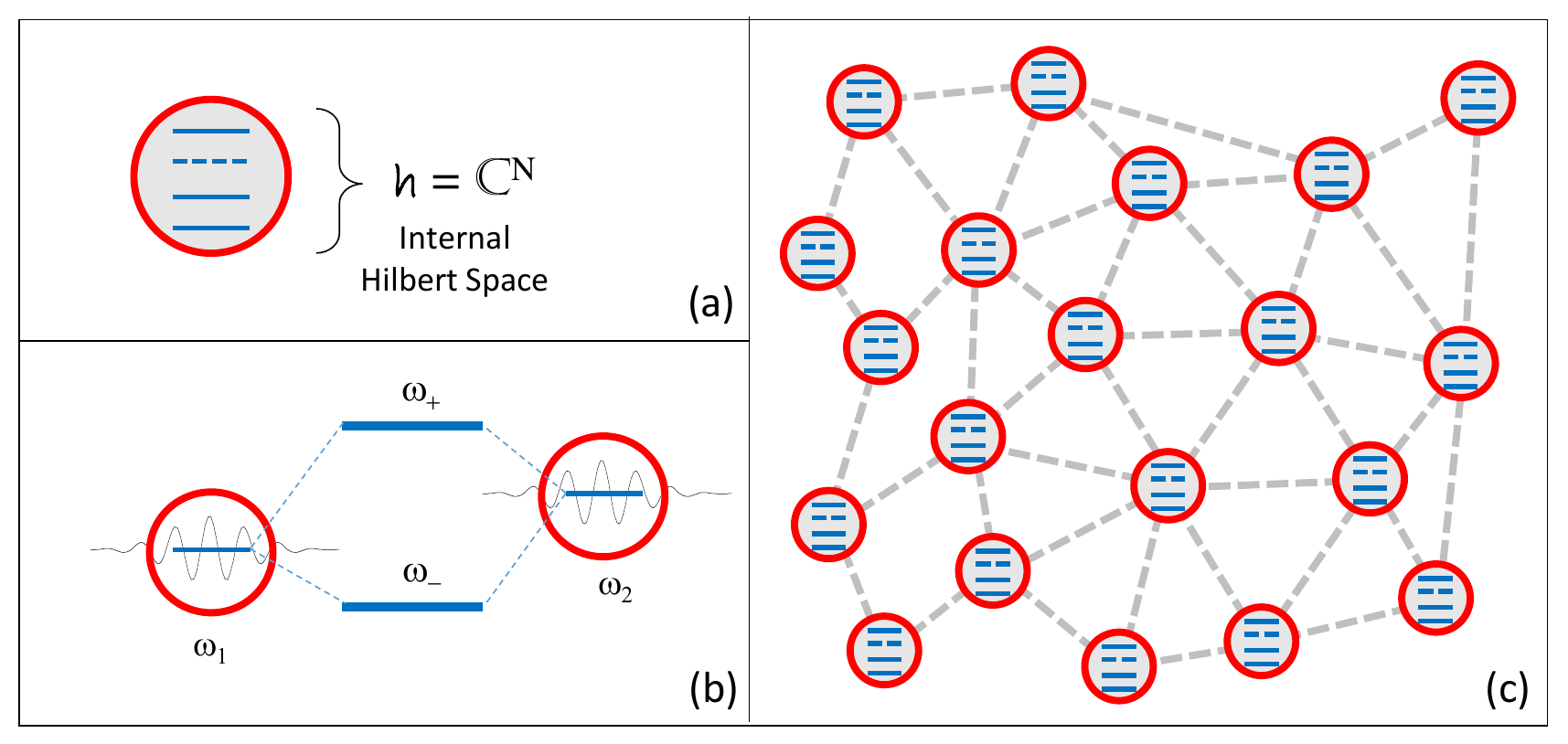}\\
  \caption{\small {\bf Patterned Resonators.} (a) Representation of a single resonator as a confined collection of $N$ resonant modes; (b) Coupling between two resonators results in a hybridization of the modes and a shift of the eigen-frequencies; (c) Identical resonators are placed in a point pattern. Coupling between resonators leads to collective resonant modes, which are the object of our investigation.
}
 \label{Fig-CoupledResonators}
\end{figure}

\begin{definition} In our view, a resonator is a $0$-dimensional physical system, {\it i.e.} a system confined to a small region of the physical space, whose physical observables and dynamics can be described by linear operators over a finite dimensional Hilbert space that, of course, can be chosen to be $\CM^N$. The number $N$ will be referred to as the number of internal degrees of freedom and $\CM^N$ as the internal Hilbert space. A resonator will be represented schematically as in panel (a) of Fig.~\ref{Fig-CoupledResonators}.
\end{definition}

\begin{example}{\rm A confined quantum mechanical system with a finite number of quantum states is the prototype of the resonator. Quite often and for a good reason, the atoms or the molecules in an extended condensed matter system are treated using such simplified representation. 
}$\Diamond$
\end{example}

\begin{example}{\rm A mechanical harmonic oscillator with $N$-degrees of freedom also fits our definition of a resonator. Indeed, if $(q_j,p_j)$, $j=1,\ldots,N$ are the generalized coordinates and the canonical momenta, then by passing to the complex coordinates:
\begin{equation}
(q_j,p_j) \rightarrow \xi_j = \frac{1}{\sqrt{2}}(q_j+\imath p_j), \quad j=1,\ldots,N,
\end{equation}
Hamilton's equations take the form:
\begin{equation}
\imath \frac{{\rm d}\xi_j}{{\rm d} t} =\frac{\partial H}{\partial \xi_j^\ast}, \quad j=1,\ldots,N.
\end{equation}
A harmonic oscillator is defined by a quadratic Hamiltonian of the form:
\begin{equation}
H(\xi_1,\xi_1^\ast,\ldots,\xi_N,\xi_N^\ast) = \sum_{i,j=1}^N h_{ij}\, \xi_i^\ast \xi_j, \quad h_{ij}^\ast = h_{ji},
\end{equation}
for which the Hamilton's equations reduce to:
\begin{equation}
\imath \frac{{\rm d}\psi}{{\rm d} t} = h \psi, \quad \psi=\begin{pmatrix} \xi_1 \\ \ldots \\ \xi_N \end{pmatrix} \in \CM^N.
\end{equation}
Above, $h$ is the $N \times N$ matrix with the entries $h_{ij}$. The eigenvectors of $h$ define the normal oscillating modes and the corresponding eigenvalues give the pulsations of the modes. Examples and analysis of patterned mechanical resonators can be found in \cite{ApigoArxiv2018,QianArxiv2018}.
}$\Diamond$
\end{example}

\begin{example}{\rm The dynamical Maxwell equations without sources can be cast in the form of a linear Schroedinger equation \cite{DL}. Then the discrete electromagnetic resonant modes inside a cavity with reflecting walls provide additional examples of resonators, provided the higher frequency modes can be neglected. 
}$\Diamond$
\end{example}

When two resonators are brought close to each other, the modes couple due to their overlap or/and because the force-fields extend beyond the confinement of the resonators. The signature of such coupling is the hybridization of the modes accompanied by shifts of the resonant frequencies, as schematically depicted in panel (b) of Fig.~\ref{Fig-CoupledResonators}. In the regime of weak coupling, the internal spaces are preserved and the dynamics of two coupled resonators takes place inside the Hilbert space:
\begin{equation}
\CM^{N}\oplus \CM^{N} \simeq \CM^N \otimes \CM^2,
\end{equation}
and is generated by Hamiltonians of the type:
\begin{equation}
 H = h_{11} \otimes |1\rangle \langle 1|  + h_{22} \otimes |2 \rangle \langle 2| + h_{12} \otimes |1 \rangle \langle 2| + h_{21} \otimes |2 \rangle \langle 1|.
 \end{equation} 
In the weak coupling regime, $h_{nn}$, $n=1,2$, coincide with the generators of the internal dynamics and $h_{12}=h_{21}^\ast$ can be mapped entirely from the shifts of the eigen-frequencies (see {\it e.g.} \cite{ApigoArxiv2018}). More generally, when a large number of identical resonators are placed in a certain point pattern $\Pp$, as in panel (c) of Fig.~\ref{Fig-CoupledResonators}, the dynamics of the collective modes takes place inside the Hilbert space:
\begin{equation}
\Hh = \CM^N \otimes \ell^2(\Pp),
\end{equation}
and is generated by Hamiltonians of the type:
\begin{equation}\label{Eq:Ham0}
H(\Pp) = \sum_{\bm p,\bm p'\in \Pp} h_{\bm p,\bm p'}(\Pp) \otimes |\bm p \rangle \langle \bm p' |, \quad h_{\bm p,\bm p'} \in \Mm_N(\CM).
\end{equation}
For now, one should think of a point pattern as a discrete subset of the physical space, without points of accumulations. The precise definition can be found in \ref{Def-PointPattern}. Throughout, $\Mm_N(\CM)$ will denote the algebra of $N \times N$ matrices with complex entries. 

\vspace{0.2cm}

Without any assumption on the hopping matrices $h_{\bm p,\bm p'}$, \eqref{Eq:Ham0} can generate any single Hamiltonian from $\BB(\Hh)$, the algebra of bounded operators over $\Hh$. However, the physical realities impose several important constraints.
\begin{itemize}

\item As one perhaps noticed, we used a notation that suggests that the hopping matrices $h_{\bm p,\bm p'}(\Pp)$ depend not just on the points $\bm p$ and $\bm p'$ but on the entire pattern. This is clearly the case because $\Pp$ contains all the geometric information.

\item While the above observation may seem trivial, it brings a subtle yet essential perspective: Once the type of the resonators was chosen, the hopping matrices are determined entirely by $\Pp$. As such, $h_{\bm p,\bm p'}$ are genuine functions defined over the space of point patterns. To put it in plain words, they are mapped experimentally once and then these functions are simply evaluated on the pattern at hand. Of course, the space of patterns is huge but see below.

\item The hopping matrices $h_{\bm p,\bm p'}(\Pp)$ must depend continuously on the pattern and that they become less significant as the distance between $\bm p$ and $\bm p'$ increases.
\end{itemize}
We will refer to the Hamiltonians which display the above characteristics as physical Hamiltonians.

\begin{example}{\rm If the coupling between single-state resonators ({\it i.e.} $N=1$) is due to the overlap of the exponentially decaying tails of the modes (assumed isotropic), then the Hamiltonian takes the form:
\begin{equation}
 H(\Pp) = \sum_{\bm p,\bm p'\in \Pp} e^{-\beta|\bm p-\bm p'|} \, |\bm p \rangle \langle \bm p' |,
\end{equation}
in some adjusted energy units.
}$\Diamond$
\end{example}

\begin{remark}{\rm If we adjust the length unit such that $\beta=1$ in the above example, then the hopping coefficients become less than $10^{-3}$ if $|\bm p -\bm p'| >7$ and, in many instances, they can be neglected entirely beyond this limit. If this is indeed the case, then these type of Hamiltonians are said to be of finite hopping range and simply called finite-range Hamiltonians.
}$\Diamond$
\end{remark}

\subsection{Point labeling problem.}

Assume now that a researcher is examining a micro-pattern of resonators under a microscope. As the patterns may drift during the observations, he or she must correlate the physics of the shifted patterns and make sure that the observations, measurements and conclusions are consistent. We assume that the pattern is much larger then the field of view of the microscope, so that the researcher cannot use the edges of the samples as references. Still, one sensible thing the researcher can always do is to move the slide on which the pattern rests until one resonator is exactly at the center of his field of view. The coordinates of this center is assumed to be $\bm 0$ in a cartesian coordinate system tide to the microscope. Next, the researcher might try to label the resonators, but this is not always an easy task. One goal of our work is to introduce a large class of patterns whose points can be canonically labeled by the elements of a discrete group $G$. This labeling is such that a natural action of the group $G$ emerges on the space of patterns. To see why such patterns are special, let us exemplify using the case $G=\ZM^d$. First, two examples.

\begin{example}\label{Ex-1DLabeling}{\rm Any 1-dimensional point pattern can be labeled by $\ZM$, by ordering the coordinates of the points.
}$\Diamond$
\end{example}

\begin{example}\label{Ex-PerturbedLattice}{\rm Patterns that are perturbations of the $\ZM^d$ lattice can be automatically labeled by $\ZM^d$. By perturbation we mean that there is one and only one point in each ball centered at $\bm n \in \ZM^d$ and of radius smaller 1/2.
}$\Diamond$
\end{example}

Let us now assume that the researcher has at his/her disposal a canonical way of labeling {\it all} the points of the shifted patterns by elements $\bm n$ of $\ZM^d$, such that:
\begin{itemize}
\item The resonator at the center of the field of view is always labeled by $\bm 0$.
\item The labelings of two patterns $\Pp=\{ \bm p_{\bm n}\}_{\bm n \in \ZM^d}$ and $\Pp'=\{ \bm p'_{\bm n}\}_{\bm n \in \ZM^d}$, connected by a shift, $\Pp' = \Pp - \bm p_{\bm k}$, are such that:
\begin{equation}\label{Eq-LabelRule}
\bm p'_{\bm n}=\bm p_{\bm n + \bm k}-\bm p_{\bm k}, \quad \forall \; \bm n \in \ZM^d.
\end{equation}
Note that this does not contradict the requirement $\bm p'_{\bm 0}=\bm 0$.
\item The labels reflect the spatial distribution of the points, in the sense that any finite-range Hamiltonian on $\CM^N \otimes \ell^2(\Pp)$:
\begin{equation}
H(\Pp) = \sum_{\bm p,\bm p'\in \Pp}^{\bm p - \bm p' \in \Kk} h_{\bm p,\bm p'}(\Pp) \otimes |\bm p \rangle \langle \bm p' |,
\end{equation}
with $\Kk$ a compact neighborhood of the origin, can be translated to a finite-range Hamiltonian on $\CM^N \otimes \ell^2(\ZM^d)$, using the labels alone:
\begin{equation}\label{Eq-LatticeHamiltonian0}
H(\Pp) = \sum_{\bm n,\bm m\in \ZM^d}^{\bm n - \bm m \in \Kk'} h_{\bm n,\bm m}(\Pp) \otimes |\bm n \rangle \langle \bm m |,
\end{equation}
for some compact neighborhood $\Kk'$ of the origin.
\end{itemize}
As we shall see next, in the above conditions, any physical Hamiltonian over the pattern $\Pp$ takes a very particular form.

\subsection{Canonical form of the Hamiltonians} The canonical labeling by $\ZM^d$ enables the researcher to represent the dynamics of the couple resonators on the fixed Hilbert space $\CM^N \otimes \ell^2(\ZM^d)$, as we have already seen in \eqref{Eq-LatticeHamiltonian0}. Now, Galilean invariance enforces:
\begin{equation}
h_{\bm p_{\bm n}, \bm p_{\bm m}}(\Pp) = h_{\bm p_{\bm n} - \bm p_{\bm k},\bm p_{\bm m} - \bm p_{\bm k}}(\Pp - \bm p_{\bm k}),
\end{equation}
which, together with the rule \eqref{Eq-LabelRule}, give:
\begin{equation}
h_{\bm p_{\bm n}, \bm p_{\bm m}}(\Pp) = h_{\bm p'_{\bm n - \bm k}, \bm p'_{\bm m - \bm k}}(\Pp - \bm p_{\bm k}). 
\end{equation}
When translated on $\CM^N \otimes \ell^2(\ZM^d)$, this reads:
\begin{equation}
h_{\bm n, \bm m}(\Pp) = h_{\bm n - \bm k, \bm m - \bm k}(\Pp - \bm p_{\bm k}), 
\end{equation}
or:
\begin{equation}
H(\Pp - \bm p_k) = S_{\bm k}^\ast H(\Pp) S_{\bm k}, \quad S_{\bm k}|\bm n \rangle = | \bm n + \bm k \rangle,
\end{equation}
which is often referred to as the covariance property. This can be used to re-write the Hamiltonian \eqref{Eq-LatticeHamiltonian0} in the following form:
\begin{equation}\label{Eq-LatticeHamiltonian}
H(\Pp) = \sum_{\bm q \in \Kk'} \sum_{\bm n \in \ZM^d} h_{\bm q}(\Pp - \bm p_{\bm n}) \otimes |\bm n \rangle \langle \bm n |S_{\bm q},
\end{equation}
where $h_{\bm q}$ stands for $h_{\bm 0,-\bm q}$. This is the canonical form we sought.

\vspace{0.2cm}

Let us point out a few remarkable facts about \eqref{Eq-LatticeHamiltonian}:
\begin{itemize} 

\item In order to reproduce any physical Hamiltonian over the pattern $\Pp$, we only need to evaluate the hopping matrices on the sub-manifold generated by the translations of $\Pp$. This sub-manifold $\Xi$ of the space of patterns is called the discrete hull of the pattern. Tremendous understanding is gain when $\Xi$ can be computed explicitly.

\item There exists a natural action of $\ZM^d$ by homeomorphisms on $\Xi$, provided by the rigid shift: $\tau_{\bm n} \Pp = \Pp - \bm p_{\bm n}$, $\bm n \in \ZM^d$. Hence, $(\Xi,\ZM^d,\tau)$ becomes a topological dynamical system, which will latter be computed explicitly for several patterns.

\item The only operators appearing in \eqref{Eq-LatticeHamiltonian} are diagonal operators of the form $\sum_{\bm n \in \ZM^d} f(\Pp - \bm p_{\bm n}) \otimes |\bm n \rangle \langle \bm n |$ and the shift operators $S_{\bm q}$, where $f \in C_N(\Xi)$, the algebra of continuous functions over $\Xi$ with values in $\Mm_N(\CM)$. As such, all physical Hamiltonians over $\Pp$ belong to the algebra generated by these few operators. It will be later computed explicitly for several patterns.
\end{itemize}

\subsection{A look ahead}

It was Jean Bellissard, in his pioneering work \cite{Bel1}, who realized that the algebra mentioned above relates to crossed product algebras (in our case, by $\ZM^d$). These algebras are extremely well studied in the mathematical literature \cite{Wil}, and that existing knowledge can be used to gain unprecedented insight into the spectral properties of the Hamiltonians \cite{Bel95} (see section~\ref{Sec-GapLabeling} on gap labeling), as well as into their topological classification. The $K$-theoretic bulk-boundary principle was also formulated \cite{KRS,PS,BKR} for crossed product algebras by $\ZM^d$, hence the whole machinery can be adopted to the present context, as it was already done in the literature \cite{ProdanPRB2015,KellendonkArxiv2017, ApigoArxiv2018}.

\vspace{0.2cm}

As we already mentioned, the labeling problem is highly non-trivial and, in many cases, it does not accept a solution. As such, the reader may ask if there are any interesting examples apart from \ref{Ex-1DLabeling} and \ref{Ex-PerturbedLattice}. In chapter~\ref{Ch:DGP}, we introduce the concept of dynamically generated patterns, which are constructed from a pre-defined dynamical system $(\Omega,G)$, with $G$ a topological group. This patterns came with a canonical $G$-labeling and $G$-action. Furthermore, if $(\Omega,G)$ is minimal, then it coincide with the discrete hull of the pattern. Put differently, chapter~\ref{Ch:DGP} introduces an algorithmic way to produce patterns with pre-defined discrete hulls. Using concrete examples, we will try to convince the reader of the variety of point patterns that can be generated in this way.

\section{Dynamically generated patterns}\label{Ch:DGP}

In this chapter, we briefly review the space of patterns, after which we introduce the concept of dynamically generated patterns. More precisely, starting from a topological dynamical system $(\Omega,G)$, we put forward an algorithm which generates patterns in the Euclidean space with a dense subset of points canonically labeled by the elements of $G$ and with a canonical $G$-action. The case when $G=\ZM^d$ is investigated in details.

\subsection{The space of patterns}

\begin{figure}[H]
\center
\includegraphics[width=\textwidth]{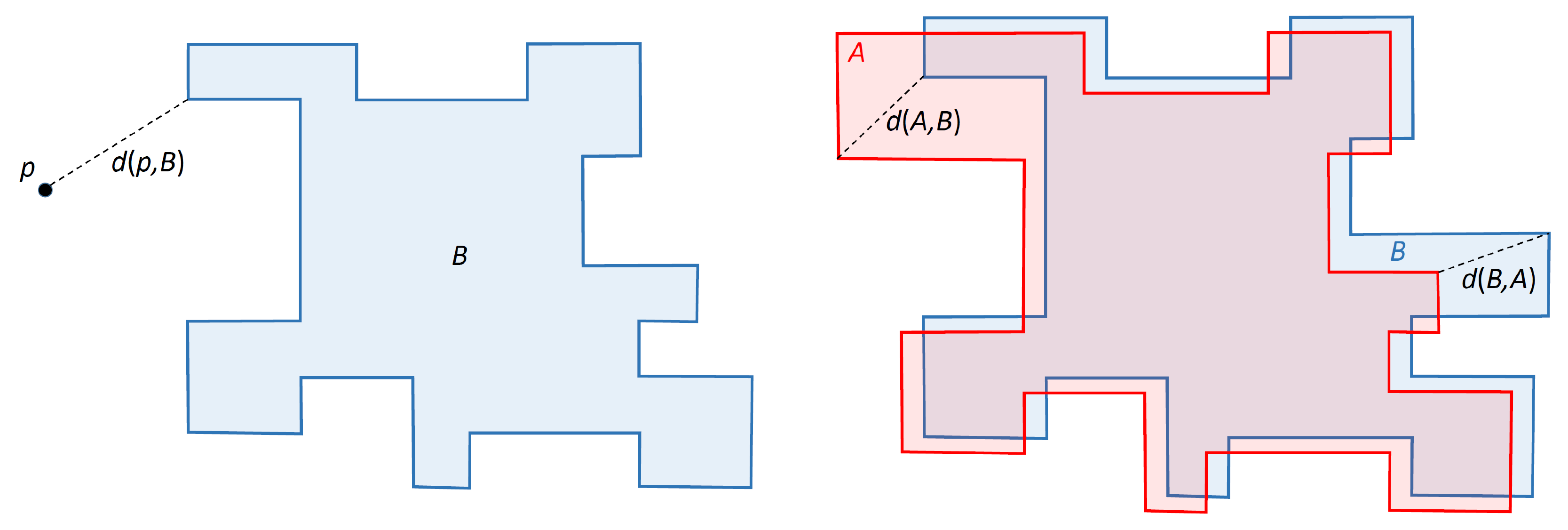}\\
  \caption{\small {\bf Housdorff metric}. (left) Illustration of the distance ${\rm d}(\bm p,B)$ for a point $\bm p\in \RM^2$ and a compact set $B \subset \RM^2$. (right) Illustration of $\tilde {\rm d}(A,B)$ as well as $\tilde {\rm d}(B,A)$. The Hausdorff distance between $A$ and $B$ will be the maximum between these two numbers.
  }
 \label{Fig-HausdorffMetric}
\end{figure}

Given a topological space $X$, we will denote its family of closed subsets by $\Cc(X)$, and its family of compact subsets by $\Kk(X)$. If $X$ is a complete metric space, then we denote by ${\rm d}_H$ the Hausdorff metric on $\Kk(X)$. We recall:

\begin{definition} The next definitions can be found in any standard textbook ({\it e.g.} \cite{Bar}):
	\begin{itemize}
		\item   The distance from a point $\bm p \in X$ to a compact subset $B \subseteq X$ is defined as:
\begin{equation}
{\rm  d}(\bm p,B)=\min\{ {\rm d}(\bm p,\bm b):  \bm b\in B\}.
\end{equation}
\item   The ``distance'' (see \ref{Re-Distance1}) from a compact subset $A$ to another compact subset $B$ is given by:
\begin{equation}
\tilde {\rm d}(A,B)=\max\{ {\rm d}(\bm a,B):  \bm a\in A\}.
\end{equation} 
\item Lastly, the Haudorff distance between the points of $\Kk(X)$ is given by:
\begin{equation}
{\rm d}_H(A,B) = \max\{\tilde {\rm d}(A,B),\tilde {\rm d}(B,A)\}, \quad A,B \in \Kk(X).
\end{equation} 
\end{itemize}
For the reader's convenience, we illustrate these concepts in Fig.~\ref{Fig-HausdorffMetric}.
\end{definition}

\begin{remark}\label{Re-Distance1}{\rm 
	$\tilde {\rm d}(A,B)$ is not a distance, given that $\tilde {\rm d}(A,B) \not= \tilde {\rm d}(B,A)$ and $\tilde {\rm d}(A,B)$ can be zero while $A \not= B$. The Hausdorff metric, however, satisfies all the axioms of a metric, hence $(\Kk(X),{\rm d}_H)$ becomes a metric space.$\Diamond$
	}
\end{remark}

We will exclusively deal with patterns in the $\alpha$-dimensional Euclidean space $\RM^\alpha$ ($\alpha \in \NM_+$). Many interesting patterns are not compact, hence they cannot be measured and compared using directly the Hausdorff metric. It is then useful to think of patterns as belonging to the larger space $\Cc(\RM^\alpha)$ of closed subsets of $\RM^\alpha$. This space can be topologized using several equivalent metrics, in particular the one defined explicitly below.

\begin{definition}[\cite{LS},\cite{FHK}~p.\,16] \label{Def-PatternMetric}  Let $B(\bm x,r)$ denote the closed ball centered at $\bm x \in \RM^\alpha$ and of radius $r$. Given a closed subset $\Lambda \subset \RM^\alpha$, define 
\begin{equation}
\Lambda[r] = \big (\Lambda \cap B(\bm 0,r)\big ) \cup \partial B(\bm 0,r),
\end{equation} 
and let ${\rm d}_H$ be the Hausdorff metric on $\Kk(\RM^\alpha)$. Then:
\begin{equation}\label{Eq-PatternMetric}
D(\Lambda,\Lambda')=\inf \big \{ 1/(1+r) \, | \, {\rm d}_H (\Lambda[r],\Lambda'[r])<1/r \big \}
\end{equation}
defines a metric on $\Cc(\RM^\alpha)$. 
\end{definition}

\begin{definition} We call the metric space $\big ( \Cc(\RM^\alpha),D \big )$ the space of patterns in $\RM^\alpha$.
\end{definition}

\begin{remark}{\rm The space of patterns is bounded, compact and complete. Furthermore, there is a continuous action of $\RM^\alpha$ by translations, that is, a homomorphism $T$ between topological groups: 
$$
T : \RM^\alpha \rightarrow {\rm Homeo}\big ( \Cc(\RM^\alpha),D \big ), \quad T_{\bm x}(\Lambda) = \Lambda + {\bm x}.
$$ 
These special properties are essential for the analysis of point patterns. $\Diamond$
}
\end{remark}

\begin{remark}{\rm  Throughout, if $X$ is a topological space, then ${\rm Homeo}(X)$ is equipped with the compact-open topology. We recall that, if $X$ is compact and Hausdorff, then ${\rm Homeo}(X)$ is a topological group. Same is true for noncompact, locally compact, and locally connected spaces.}
$\Diamond$
\end{remark}

\subsection{Dynamically generated patterns}

Let $G$ be a topological group and $\Omega$ a compact  and Hausdorff topological space. In this section, $G$ can be non-abelian and non-discrete. The starting point of our construction is a homomorphism between topological groups:
\begin{equation}\label{Eq:OrigAction}
W: G \rightarrow {\rm Homeo}(\dot \RM^\alpha \times \Omega),
\end{equation}
where the $\dot \RM^\alpha$ is the one-point compactification of the Euclidean space. Specifically, $W$ is a continuous map and has the property that $W_{gg'}= W_g \circ W_{g'}$. We will denote by $\pi_1$ and $\pi_2$ the standard projections from $\RM^\alpha \times \Omega$ to $\RM^\alpha$ and $\Omega$, respectively. The first working assumption on $W$ is:

\begin{assumption}\label{As:1} The maps $\pi_2 \circ W_g$ are all constant of the first argument.
\end{assumption}

\begin{proposition} Let $\bm p_0 \in \RM^\alpha$ arbitrary but fixed. Then:
\begin{equation}
\tau: G \rightarrow {\rm Homeo}(\Omega), \quad \tau_g(\omega) = (\pi_2 \circ W_g)(\bm p_0,\omega)
\end{equation}
defines a continuous action of $G$ on $\Omega$. It is independent of $\bm p_0$.
\end{proposition}

\proof First, we need to show that the map is well defined. Since $\pi_2$ is continuous, $\tau_g(\omega)$ is continuous of both $g$ and $\omega$. Furthermore:
\begin{equation}
(\tau_{g'} \circ \tau_g)(\omega) = (\pi_2\circ W_{g'})(\bm p_0,\tau_g\omega) = (\pi_2\circ W_{g'}) \big (W_g(\bm p_0,\omega) \big ),
\end{equation}
where in the second equality we used that $\pi_2 \circ W_{g'}$ is constant on the first argument. Then:
\begin{equation}
(\tau_{g'} \circ \tau_g)(\omega) = (\pi_2\circ W_{g'}\circ W_g)(\bm p_0,\omega) = (\pi_2\circ W_{g'g})(\bm p_0,\omega) = \tau_{g'g}(\omega).
\end{equation} 
In particular, $\tau_{g}\circ \tau_{g^{-1}}=\tau_{g^{-1}}\circ \tau_g ={\rm id}$, hence $\tau_g$'s are invertible with continuous inverse. The map $\tau$ takes indeed values in ${\rm Homeo}(\Omega)$ and we have already seen that $\tau$ respects the group operations. The map is also continuous.\qed

\begin{remark}{\rm  The topological dynamical system $(\Omega,G,\tau)$ is at the heart of our construction. It generates the dynamics that provides the ticks by which the points are laid down, while $W$ acts like a bridge, transferring this dynamics to the physical space. The precise mechanism is give below.}$\Diamond$
\end{remark}

\begin{definition} Let $W$ be as above. Then the $(\Omega,G,\tau)_W$-generated pattern seeded at $(\bm x,\omega) \in \RM^\alpha \times \Omega$ is defined as the closed sub-set  of $\RM^\alpha$:
\begin{equation}\label{Eq:DynPatterns1}
\Pp(\bm x,\omega) = \overline{\big \{ \RM^\alpha \ni \bm p_g(\bm x,\omega) = (\pi_1\circ W_g)(\bm x,\omega), \ g \in G \big \}}.
\end{equation}
\end{definition}

\begin{remark}{\rm Note that, for now, the points $\bm p_g(\bm x,\omega)$ are not required to be distinct as $g$ is varied. For example, note that the starting group action $W$ is only a homomorphism but this is, of course, not the only reason why the points might not be distinct.}
$\Diamond$
\end{remark}

\begin{example}\label{Ex:1} Let $\Lambda : G \rightarrow {\rm Homeo}(\dot \RM^\alpha)$ be a homomorphism of topological groups and $F : \Omega \rightarrow {\rm Homeo}(\dot \RM^\alpha)$ be a continuous map. Then:
\begin{equation}
W_g(\bm p,\omega) = \big ((F_{\tau_g \omega} \circ \Lambda_g \circ F_\omega^{-1})(\bm p),\tau_g \omega \big )
\end{equation}
is a group action on $\dot \RM^\alpha \times \Omega$ which obviously satisfies \ref{As:1}. As we shall see, many interesting patterns can be generated this way.
\end{example}

There are several properties one can be looking for in a pattern. The first one is the possibility of labeling its points by the elements of a topological group. Note that \eqref{Eq:DynPatterns1} can be viewed as an explicit way of indexing a dense subset of (possibly repeating) points by the group $G$. The second property is the existence of a group action on the family of patterns. The following is a preparatory statement in this direction.

\begin{proposition} The following identity holds:
\begin{equation}\label{Eq:Id1}
\Pp(\bm x, \omega) = \Pp(\bm p_a(\bm x,\omega),\tau_a \omega ), \quad \forall a \in G.
\end{equation}
\end{proposition}

\proof 
\begin{align}
\Pp(\bm x,\omega) & = \overline{\big \{ (\pi_1\circ W_g)(\bm x,\omega), \ g \in G \big \}} \\ \nonumber
& = \overline{\big \{ (\pi_1\circ W_{ga})(\bm x,\omega), \ g \in G \big \}} \\ \nonumber
& = \overline{\big \{ (\pi_1\circ W_{g})\big (W_a(\bm x,\omega)\big), \ g \in G \big \}} \\ \nonumber
& = \overline{\big \{ (\pi_1\circ W_{g})(\bm p_a(\bm x,\omega),\tau_a \omega), \ g \in G \big \}},
\end{align}
and the affirmation follows.\qed

\vspace{0.2cm}

Let us emphasize that, without additional assumptions, $\Pp(\bm x, \omega)$ depends in an essential way on the position $\bm x$ where the observer was initially located. This is physically sound in the presence of background fields. But if we want to be consistent with the Galilean invariance, then the dependence of $\bm x$ must be a simple rigid translation. For this, we need an additional assumption on $W$:

\begin{assumption}\label{As:2} The original group action \eqref{Eq:OrigAction} is such that:
\begin{equation}\label{Eq:As2}
(T_{\bm y} \times {\rm id}) \circ W_g = W_g \circ (T_{\bm y} \times {\rm id}), \quad \forall \, (\bm y, g) \in \RM^\alpha \times G.
\end{equation}
\end{assumption}

\begin{remark}\label{Re:Constraint1}{\rm The constraint in \eqref{Eq:As2} seems rather rigid. It forces $\pi_1 \circ W_g$ to take values in the group of translations, specifically:
\begin{equation}
(\pi_1 \circ W_g)(\bm x,\omega) = \bm x + (\pi_1 \circ W_g)(\bm 0,\omega).
\end{equation} 
As one can see, the maps are fully determined by the evaluation at the origin of $\RM^\alpha$. Note, however, that $\pi_1\circ W_g$ do not need to commute with each other.}
$\Diamond$
\end{remark}

\begin{remark}{\rm The assumption \label{As:2} can be relaxed to some degree, as for example:
\begin{equation}\label{Eq:As22}
(A \times {\rm id}) \circ W_g = W_g \circ (A \times {\rm id}), 
\end{equation}
with $A$'s belonging to a subgroup of ${\rm Homeo}(\dot\RM^\alpha)$. They can be, for example, general affine transformations. We will not explore these possibilities in the present work but we still want to point to the reader that all the statements below can be easily reformulated if 
 assumption \eqref{Eq:As22} is considered instead of \eqref{Eq:As2}.}
$\Diamond$
\end{remark}

\begin{proposition} With assumption $\ref{As:2}$:
\begin{equation}\label{Eq:PatternShift}
\Pp(\bm x',\omega) = \Pp(\bm x,\omega) + \bm x' - \bm x , \quad \forall \, \bm x,\bm x' \in \RM^\alpha, \ \omega \in \Omega.
\end{equation}
\end{proposition}

\proof From the very definition:
\begin{align}
\Pp(\bm x + \bm y,\omega) & = \overline{\big \{ (\pi_1\circ W_g)(T_{\bm y}\bm x,\omega), \ g \in G \big \}} \\ \nonumber
& = \overline{\big \{ (T_{\bm y} \circ\pi_1\circ W_{g})(\bm x,\omega), \ g \in G \big \}} \\ \nonumber
& = \overline{ T_{\bm y} \big (\big \{ (\pi_1\circ W_{g})(\bm x,\omega), \ g \in G \big \} \big ) } \\ \nonumber
& = T_{\bm y} \Big (\overline{\big \{ (\pi_1\circ W_{g})(\bm x,\omega), \ g \in G \big \}} \Big ),
\end{align}
and the affirmation follows if we take $\bm y = \bm x' -\bm x$.\qed

\vspace{0.2cm}

We will work from now on exclusively with the family of patterns $\Pp(\bm 0,\omega)$ seeded at the origin and we will let $\omega$ take all the values in $\Omega$. Henceforth, we will simplify the notation to $\Pp(\omega)$. We warn the reader that, even with the above simplifying assumptions, the range of patterns that can be generated with the proposed algorithms is very large. In particular, the following assumption cannot be taken for granted: 

\begin{assumption}\label{As:3} The map:
\begin{equation}\label{Eq:MainMap}
\Omega \ni \omega \rightarrow \Pp(\omega) \in \Cc(\RM^\alpha)
\end{equation}
is injective and continuous. We will denote this map by $\Pp$.
\end{assumption}

\begin{remark}{\rm As we shall see in the next section, if $\Pp(\omega)$ are all uniform point-patterns, then \ref{As:3} is automatically satisfied.}
$\Diamond$
\end{remark}

We denote the image of $\Omega$ through the map $\Pp$ by $\Xi$. The continuous image of a compact set is again a compact set, hence $\Xi$ is a compact subset of $\Cc(\RM^\alpha)$. In fact:

\begin{proposition}\label{Pro:TopEquiv} Topologically, $\Omega \simeq \Xi$.
\end{proposition}

\proof $\Pp$ maps continuously from a compact space to a Hausdorff space, hence it is closed, by closed map lemma. As such, when restricted to its image $\Xi$, the map $\Pp$ \eqref{Eq:MainMap} is a continuous closed bijection. Then the map $\Pp$ has a continuous inverse, hence it is a homeomorphism. \qed

\vspace{0.2cm}

Given the above, $\Omega$ and $\Xi$ are identical as topological sets. The next statement shows that in fact the whole dynamical system $(\Omega, G,\tau)$ can be ported over $\Xi$. Extremely important is the fact that the action of $G$ is implemented by rigid shifts of the patterns, as it is shown next.

\begin{proposition}\label{Pro:DynamicalConjg} Let $\xi \in \Xi$. Then necessarily $\xi=\Pp(\omega)$, with $\omega=\Pp^{-1}(\xi)$, and inside $\xi$ we can canonically identify a dense family of points $\{\bm p_g(\omega)\}_{g \in G}$, not necessarily distinct, such that $\bm p_e$ seats at the origin of $\RM^\alpha$, where $e$ is the identity element of $G$. With these notations, the following identity holds:
\begin{equation}\label{Eq:Id2}
\xi - \bm p_g\big(\Pp^{-1}(\xi) \big ) = (\Pp \circ \tau_g \circ \Pp^{-1})(\xi), \quad \forall \, (g,\omega) \in G \times \Omega.
\end{equation}
In particular, the space $\Xi$ is invariant with respect to such shifts, which define a topological dynamical system $(\Xi,G,\chi)$, conjugate to $(\Omega,G,\tau)$.
\end{proposition}

\proof With $\omega = \Pp^{-1}(\xi)$, the right hand side of \eqref{Eq:Id2} becomes $\Pp(\tau_g \omega)$. Expanding the notation and using \eqref{Eq:PatternShift}:
\begin{equation}
\Pp(\bm 0, \tau_g \omega) + \bm p_g(\bm 0,\omega) = \Pp(\bm p_g(\bm 0,\omega),\tau_g \omega)
\end{equation}
From here, the identity \eqref{Eq:Id2} follows from \eqref{Eq:Id1}.\qed

\begin{remark}{\rm It is instructive to look at two consecutive group actions:
\begin{equation}\label{Eq:Use1}
\Pp(\tau_{g'g}\omega) = \Pp(\tau_{g}\omega)- \bm p_{g'}(\tau_g\omega)=\Pp(\omega) - \bm p_g(\omega) - \bm p_{g'}(\tau_g \omega).
\end{equation}
This exercise helps to understand that, against the abelian nature of the translations, the group action may not be abelian. Indeed, switching $g$ and $g'$ above leads to a different outcome. The simple explanation is that after a shift, the points of the pattern are relabeled. 
}
$\Diamond$
\end{remark}

\subsection{Patterns labeled by $\ZM^d$}

We restrict now to the case $G=\ZM^2$ and provide a practical construction that covers all the cases that can steam from the general algorithm proposed in the previous section. Throughout we will use $\bm n = (n_1,\ldots,n_d)$ to specify the elements of $\ZM^d$ and will denote by $\bm e_i$ the standard generators, hence $\bm n = n_1 \bm e_1 \ldots n_d \bm e_d$. We let $\Gg_d = \{\pm \bm e_1, \ldots, \pm \bm e_d\}$ be the set of these  generators, taken with a sign (hence $2d$ in total). To get a sense of how the dynamical algorithms might look, let us start by adapting example \ref{Ex:1} to the present particular setting.

\begin{example}\label{Ex:X2}{\rm We take $G=\ZM^d$, $\Lambda_{\bm n}(\bm x) = \bm x + \bm n$, $\bm n \in \ZM^2$, and $F_\omega(\bm p) = \bm p + \Gamma(\omega)$, with $\Gamma : \Omega \rightarrow \RM^d$ a continuous map. Then the algorithm proposed in \ref{Ex:1} leads to:
\begin{equation}
\bm p_{\bm n}(\omega) = \bm n + \Gamma(\tau_{\bm n} \omega) - \Gamma(\omega).
\end{equation}
}
$\Diamond$
\end{example}

\begin{proposition}\label{Pro-Algorithm} When $G = \ZM^d$, any group action $W$ with the properties listed in the previous chapter is generated by a set of continuous maps  $\Gamma_{\bm e}: \Omega \rightarrow \RM^\alpha$, $\bm e \in \Gg_d$,
obeying the consistency relations:
\begin{equation}\label{Eq:Consistency}
\Gamma_{\bm e'} + \Gamma_{\bm e}\circ \tau_{\bm e'} = \Gamma_{\bm e} + \Gamma_{\bm e'} \circ \tau_{\bm e}, \quad \Gamma_{-\bm e}  = - \Gamma_{\bm e}\circ \tau_{-\bm e}, \quad \bm e,\bm e' \in \Gg_d.
\end{equation}
In terms of these maps, the dense points inside the $(\Omega,\ZM^d,\tau)_W$-generated patterns can be computed iteratively:
\begin{equation}\label{Eq:OmegaToPatterns}
\bm p_0=0, \quad \bm p_{\bm n + \bm e} = \bm p_{\bm n} + \Gamma_{\bm e}(\tau_{\bm n}\omega), \quad \bm n \in \ZM^d, \quad \bm e \in \Gg_d.
\end{equation}
\end{proposition}

\proof As we have seen in Remark~\ref{Re:Constraint1}, $\pi_1\circ W_g$ are all determined by the evaluation at the origin $  (\pi_1 \circ W_g)(\bm 0,\omega)$. Group relations assures us that it is sufficient to specify these functions for the generators only. As such, we define:
\begin{equation}
\Gamma_{\bm e}(\omega) = (\pi_1 \circ W_{\bm e})(\bm 0,\omega), \quad \bm e \in \Gg_d,
\end{equation}
which in turn give the evaluation of $W$ on the generators:
\begin{equation}
W_{\bm e}(\bm x, \omega) = \big ( \bm x + \Gamma_{\bm e} (\omega), \tau_{\bm e} \omega \big ), \quad \bm e \in \Gg_d.
\end{equation}
We must necessarily have $W_{\bm e}\circ W_{\bm e'}=W_{\bm e'}\circ W_{\bm e}$ for any pair $\bm e,\bm e' \in \Gg_d$, which translates to:
\begin{equation}
\Gamma_{\bm e'} + \Gamma_{\bm e}\circ \tau_{\bm e'} = \Gamma_{\bm e} + \Gamma_{\bm e'} \circ \tau_{\bm e}, \quad \bm e, \bm e' \in \Gg_d.
\end{equation}
These relations coincide with the first set of constraints in \eqref{Eq:Consistency}. We must also have $W_{\bm e} \circ W_{-\bm e} = W_{-\bm e} \circ W_{\bm e} = {\rm id}$, hence $W_{-\bm e} = W_{\bm e}^{-1}$, which translates to:
\begin{equation}
\Gamma_{-\bm e}  = - \Gamma_{\bm e}\circ \tau_{-\bm e}, \quad \bm e \in \Gg_d.
\end{equation}
These relations coincide with the second set of constraints in \eqref{Eq:Consistency}. For a generic element $\bm n \in \ZM^d$, then:
\begin{equation}\label{Eq:X1}
W_{\bm n} = W_{\bm e_1}^{n_1} \circ \ldots \circ W_{\bm e_d}^{n_d},
\end{equation}
where the right hand side in \eqref{Eq:X1} does not depend on the order we wrote the terms. Then, by following the general prescription of the previous chapter, we are indeed lead to the iterative algorithm \eqref{Eq:OmegaToPatterns}.\qed

\begin{proposition}\label{Pro-Label} Any family of $(\Omega,\ZM^d,\tau)_W$-generated pattern satisfies:
\begin{equation}\label{Eq-Label}
\sup \big \{|\bm p_{\bm n+\bm m}-\bm p_{\bm n}| , \ \bm n \in \ZM^d\big \} < D |\bm m|,
\end{equation}
for some finite parameter $D$.
\end{proposition}.

\proof Let $\bm k_0, \bm k_1,\ldots, \bm k_M$ be a sequence in $\ZM^d$ such that $\bm k_0=\bm 0$, $\bm k_M=\bm m$ and $\bm k_i- \bm k_{i-1} \in \Gg_d$ . Then:
\begin{equation}
|\bm p_{\bm n+\bm m}-\bm p_{\bm n}| =\big | \sum_{i=1}^M (\bm p_{\bm m+\bm k_i}-p_{\bm m+\bm k_{i-1}}) \big | \leq \sum_{i=1}^M |\bm p_{\bm n+ \bm k_i}- \bm p_{\bm n+ \bm k_{i-1}}|,
\end{equation}
hence:
\begin{equation}
|\bm p_{\bm n+\bm m}-\bm p_{\bm n}| \leq M \sup\{|\Gamma_{\bm e}(\omega)|, \omega \in \Omega, \bm e \in \Gg_d\},
\end{equation}
and note that $M$ can always be chosen smaller than $d|\bm n- \bm m|$.\qed

\section{Dynamically generated uniform point patterns}
\label{Ch:DGUPP}

In this chapter, a new assumption will be added, namely, that the patterns are Delon\'e sets. In the specialized literature, such sets are also referred to as uniform point patterns. In this more restrictive setting, we show that the map $\Pp$ is always continuous of $\omega$. Four explicit examples of dynamically generated uniform point patterns will be provided, which will subsequently serve as working cases for the $K$-theoretic bulk-boundary machinery.

\subsection{Uniform point patterns}

\begin{definition} Let $\Dd=\RM^d \times [-D,D]^{\alpha-d}$ be a domain in $\RM^\alpha$ for some $D \geq 0$. A subset $\Lambda \subset \Dd$ is a Delon\'e set if there exist $r_{min}, \, r_{max} >0$ such that:
\begin{equation}\label{Eq:Delone1}
B(\bm p,r_{min}) \cap \Lambda = \{\bm p\}, \quad \forall \ \bm p\in \Lambda,
\end{equation}
and:
\begin{equation}\label{Eq:Delone2}
B(\bm x,r_{max}) \cap \Lambda \neq \emptyset, \quad \forall \ \bm x\in \Dd.
\end{equation}
\end{definition}

\begin{remark}{\rm Delon\'e subsets consist of isolated points, hence they are closed and belong to the metric space $\Cc(\Dd)$.}
$\Diamond$
\end{remark}

\begin{remark}{\rm The reason we restricted to the domain $\Dd$ and not simply considered $\RM^d$ is because we don't want the patterns to be flat. Let us also remark that Delon\'e sets are called uniform for a good reason, since the two conditions above prevent the points to conglomerate and also exclude the occurrence of large holes.}
$\Diamond$
\end{remark}

\begin{definition}\label{Def-PointPattern} A $(\Omega,\ZM^d,\tau)_W$-generated pattern will be called quasi d-dimensional uniform point pattern if it is a Delon\'e subset of $\Dd$. 
\end{definition}

\begin{example}{\rm  If $\Gamma(\Omega) \subset B(\bm 0,\frac{1}{2})$ in \ref{Ex:X2}, then the pattern is a Delon\'e set.}
$\Diamond$
\end{example}

\begin{remark}{\rm We have seen in Proposition~\ref{Pro-Label} that any $(\Omega,\ZM^d,\tau)_W$-generated pattern satisfies \eqref{Eq:Delone2}. Unfortunately, we cannot formulate optimal conditions on $\Gamma$'s that assures \eqref{Eq:Delone1}. Nevertheless, for all concrete examples considered in this work, \eqref{Eq:Delone1} can be easily verified.
}$\Diamond$
\end{remark}

\begin{proposition}\label{Pro-Continuity} Assume that all $(\Omega,\ZM^d,\tau)_W$-generated patterns are uniform point patterns. Then the map:
\begin{equation}\label{Eq-MapOmegaPp}
\Omega \ni \omega \mapsto \Pp(\omega) \in \Cc(\Dd)
\end{equation}
is continuous. Since we always assume that $\Pp$ is injective, the map is a homeomorphism between $\Omega$ and its image $\Xi$.
\end{proposition}

\proof We wish to show that $\Pp(\omega') \rightarrow \Pp(\omega)$ in $\Cc(\RM^\alpha)$ whenever $\omega' \rightarrow \omega$ in $\Omega$. Let us consider an arbitrary large but nevertheless finite $r$. Since we are dealing with Delon\'e sets and $\Omega$ is compact, there is always $R\in \RM_+$ such that $\bm p_{\bm n}(\omega) \notin \Pp[r]$ whenever $|\bm n|$ exceeds $R$, for all $\omega \in \Omega$. In other words, for each fixed $r$ appearing in the norm \eqref{Eq-PatternMetric}, we only have to deal with a finite number of points. Now, given that $W_{\bm n}$ are continuous, $|\bm p_{\bm n}(\omega') - \bm p_{\bm n}(\omega)|$ eventually becomes less than $\frac{1}{r}$ as $\omega' \rightarrow \omega$, for all $|\bm n| \leq R$. This implies that, as $\omega' \rightarrow \omega$, ${\rm d}_H(\Pp'[r], \Pp[r])$ eventually becomes smaller than $1/r$ and, consequently, $D(\Pp',\Pp)$ becomes less than $1/(r+1)$. Since $r$ was arbitrarily large, this shows that $D(\Pp',\Pp) \rightarrow 0$ as $\omega' \rightarrow \omega$. The last statement follows from Proposition~\ref{Pro:TopEquiv}.\qed

\subsection{The discrete hull of dynamically generated patterns}

We have already seen in chapter~\ref{Ch:PattRez} the signature of the hull on the dynamics of coupled resonators. The Galilean invariance assumption was essential there and, from its very definition, it will become apparent that the concept of the discrete hull is rooted in this invariance.

\begin{definition} The discrete hull of a uniform point pattern $\Pp \in \Cc(\Dd)$ is defined as the closed, hence compact, subspace of $\Cc(\Dd)$:
\begin{equation}
\Xi(\Pp)=\overline{ \{\Pp - \bm p, \; \bm p \in \Pp \} } \in \Kk(\Cc(\Dd)).
\end{equation}
In other words, the discrete hull is the closure of the orbit obtained by rigidly shifting the pattern such that each of its points get to seat at the origin. A point pattern is called homogeneous if $\Xi(\Pp')$ coincides with $\Xi(\Pp)$ for any $\Pp' \in \Xi(\Pp)$.
\end{definition}

\begin{remark}{\rm Note that, while the discrete hull can be defined for generic uniform point patterns, an action of a topological group exists only for specific patterns, particularly, for the dynamically generated ones.}
$\Diamond$
\end{remark}

\begin{definition}\label{Def-MinimalSystems} A topological dynamical system $(\Omega,\ZM^d,\tau)$ is said to be minimal if any orbit $\{\tau_{\bm n} \omega\}_{{\bm n} \in \ZM^d}$, $\omega \in \Omega$, is dense in $\Omega$. It is said to be uniquely ergodic if it possesses a unique invariant and ergodic probability measure. The dynamical system is called strictly ergodic if it is both minimal and uniquely ergodic. 
\end{definition}

\begin{proposition}\label{Pro:HullVsOmega} Consider a family of $(\Omega,\ZM^d,\tau)_W$-generated uniform point patterns. If $(\Omega,\ZM^d,\tau)$ is minimal, then all patterns are homogeneous and their discrete hulls coincide with the image $\Xi$ of $\Omega$ through $\Pp$. 
\end{proposition}

\proof Recall Propositions~\ref{Pro:TopEquiv} and \ref{Pro:DynamicalConjg}. Under the conjugacy defined there, the hull of $\Pp(\omega)$ is identical with the orbit of $\omega$ in $\Omega$ under the $\ZM^d$-action. The closure of this orbit gives the whole $\Omega$, whose image through $\Pp$ is $\Xi$. Then, necessarily, $\Xi\big (\Pp_\omega)\big ) = \Xi$. \qed

\begin{remark}{\rm We reached an important conclusion in our exposition. Indeed, one now can see explicitly that, using the proposed algorithms, we can generate patterns with pre-defined hull.  
}$\Diamond$
\end{remark}

\begin{figure}[t]
\center
\includegraphics[width=\textwidth]{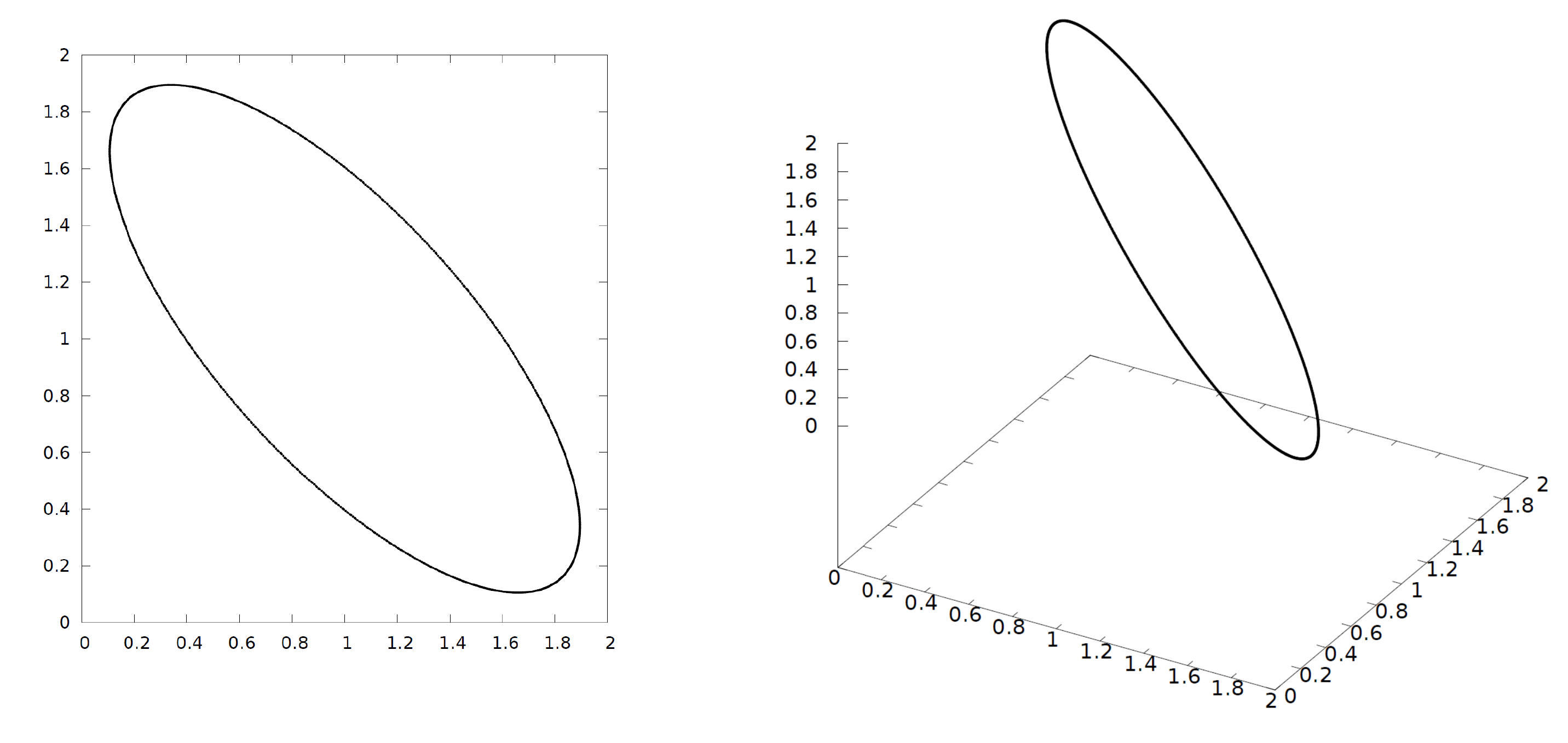}
\caption{{\bf Reconstruction of a hull.} The panels display projections of the orbit $\Pp -p_i$, $i\in \ZM$ on different subspaces of the Hilbert cube, for pattern \ref{Eq:1Dpattern} at irrational value of $\theta$. Particularly, the left panel shows the discrete set of points $\big (p_i-p_{i-1},p_{i+1}-p_i \big)_{i=\overline{0,1000}}$, while the right panel shows $\big (p_i-p_{i-1},p_{i+1}-p_i,p_{i+2}-p_{i+1} \big)_{i=\overline{0,1000}}$.}
\label{Fig-HullExample}
\end{figure}

\begin{remark}{\rm The reverse problem is equally interesting. If given a dynamically generated pattern and nothing is disclosed about the algorithm, is it possible to reproduce $\Omega$? In many cases, the answer is yes, as illustrated in the example below. The particular technique used in this exercise is called the stroboscopic method.}
$\Diamond$
\end{remark}

\begin{example}{\rm Consider the 1-dimensional pattern:
\begin{equation}\label{Eq:1Dpattern}
p_i = i + \sin(i \theta) \in \RM.
\end{equation}
To reproduce numerically the discrete hull of this pattern, without using the algorithm that produced the pattern in the first place, we think of the pattern and its translates as points in the infinite dimensional space $\RM^\ZM$. These points trace a shape in $\RM^\ZM$, which we want to determine. To do so, we render the translates $\Pp - p_i$, $i\in \ZM$, relative to each other such that the coordinates of the points of the $i$-th translated pattern are:
\begin{equation}
 \{p_{n+i}-p_{n+i-1}\}_{i \in \ZM} \in [0,2]^\ZM.
\end{equation}
As one can see, the shape we are looking for is contained in the Hilbert cube, which is a compact subset of $\RM^\ZM$. Now, we can visualize this shape by projecting it onto various subspaces of the Hilbert cube. In Fig.~\ref{Fig-HullExample}, we show the projection onto 2-dimensional and 3-dimensional subspaces. In each case, the projection is a 1-dimensional manifold from where we would conclude that the hull is $\TM^1$. This is indeed the case, as it follows from Proposition~\ref{Pro:HullVsOmega} and the geometric realization of the same pattern given in \ref{ExampleI1}.
}$\Diamond$
\end{example}

\begin{figure}[t]
\center
  \includegraphics[width=\textwidth]{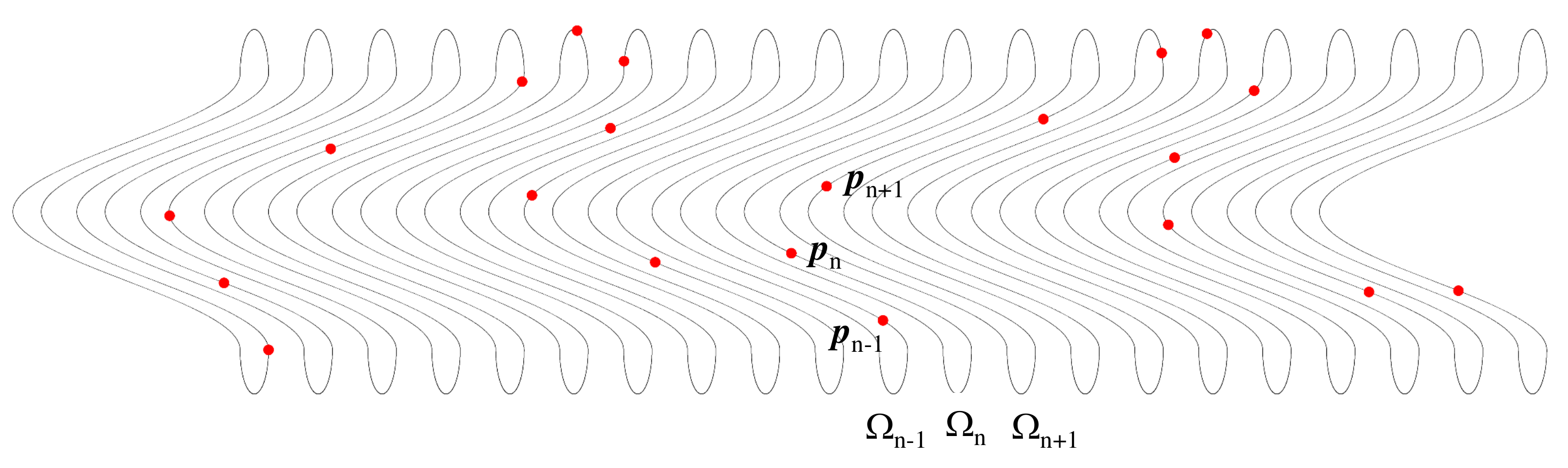}\\
  \caption{\small {\bf Quasi 1-dimensional pattern with non-trivial labeling.} It is generated by the algorithm $\bm p_n=n \bm e_1 + \tau_n \omega - \omega$, $n\in \ZM$, where the dynamical system $(\Omega,\ZM,\tau)$ consists of the closed loop shown in the diagram and the action of $\ZM$ is provided by a displacement along the loop by $\theta L$, with $L$ the total length of the loop. The shape and size of $\Omega$ has been carefully chosen such that the translates $\Omega_n=\Omega + n \bm e_1$ are inter-locked without touching, as in the diagram.}
 \label{Fig-Incomm0}
\end{figure}

\begin{remark}{\rm While the discrete hull can be reproduced from one single pattern, the action of $\ZM^d$ is not always easy to spot. Indeed, although the group action is given by just translations, note that a meaningful and consistent labeling of the points by $\ZM^d$ is required first. In Fig.~\ref{Fig-Incomm0} we show an example of a dynamically generated point pattern that would be virtually impossible to label without the knowledge of the algorithm that produced the pattern in the first place. Sometime, the quasi 1-dimensional patterns can be labeled by just ordering the points relative to the longitudinal coordinates. But for the pattern shown in Fig.~\ref{Fig-Incomm0}, this simple method fails since, for example, the horizontal coordinate of $\bm p_{n-1}$ is larger than the that of $\bm p_{n+1}$. For this reason, it will be difficult, if not impossible to correctly label the pattern by $\ZM$, once we erase the loops $\Omega_n$ from Fig.~\ref{Fig-Incomm0}.
}$\Diamond$
\end{remark}

\begin{remark}{\rm The reader can, perhaps, see that, by increasing the complexity of the loops and of their inter-lockings in Fig.~\ref{Fig-Incomm0}, we can produce, dynamically, extremely complex point patterns. It is, perhaps, this type of examples that highlight the practical value of the main concept introduced in this chapter. We believe that many patterns found in the natural world can be fitted by such dynamically generated patterns and, due to their algorithmic nature, the fitting process can be, perhaps, automated. 
}$\Diamond$
\end{remark}

\begin{remark}{\rm Our last and important remark is that we will not always work with minimal dynamical systems. The reason for it is that, in many cases, extremely useful information is gained when one parameter is varied, such as the angle $\theta$ in \ref{ExampleI1}. During such variations, enforcing minimality is sometimes impossible.  When the minimality is lost, $\Xi$ is strictly larger than the hull of the individual patterns and, instead of working with the family of translates, we choose to work with the whole family of $(\Omega,\ZM^d,\tau)_W$-generated patterns.}
$\Diamond$
\end{remark}

\subsection{Concrete examples}
\label{Sec:Examples}

This section introduces four examples which will serve as working platforms for our applications of $K$-theory. As we shall see, despite their simplicity, they all have non-trivial bulk boundary correspondence.

\subsubsection{}\label{ExampleI1}

\begin{figure}[t]
\center
  \includegraphics[width=0.9\textwidth]{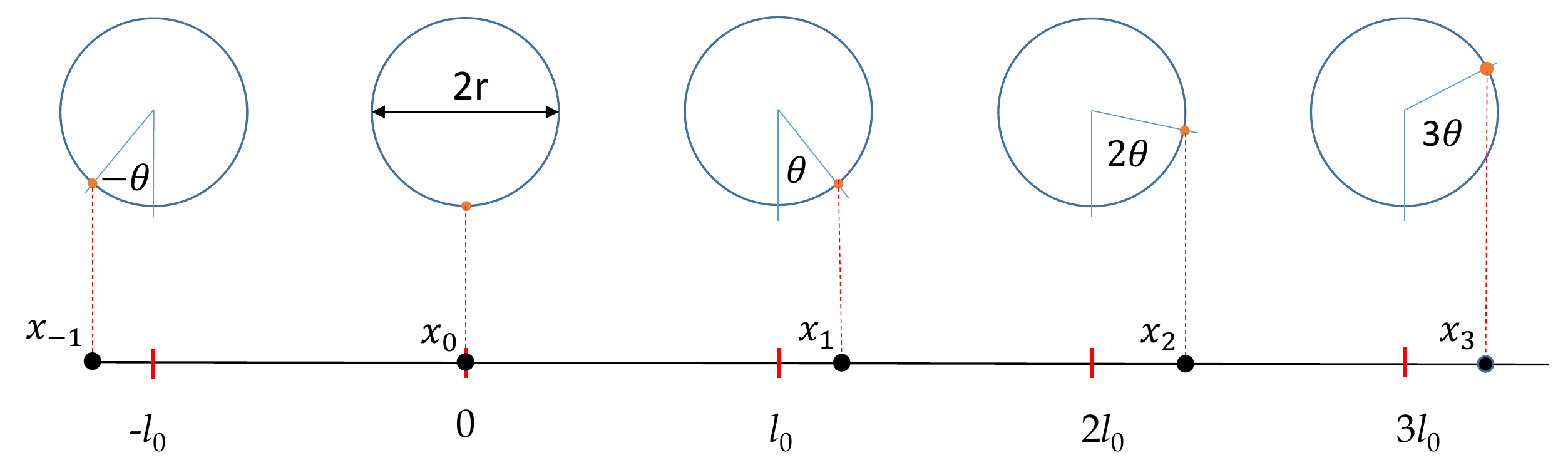}\\
  \caption{\small {\bf Example of a 1-dimensional dynamically generated pattern.} The diagram provides a geometric representation of the point-pattern defined in Example~\ref{ExampleI1}.}
 \label{Fig-Incomm1}
\end{figure}

This is a pattern that leaves in $\RM$ and is indexed by $\ZM$. It consists of the sequence of points:
\begin{equation}\label{Eq-IncommensurateI}
p_n=n + r \big ( \sin(n \, \theta +\omega)-\sin(\omega) \big ), \quad r <\tfrac{1}{2}, \quad n \in \ZM.
\end{equation}
The geometric representation of the pattern is illustrated in Fig.~\ref{Fig-Incomm1} and, from there, it follows that the pattern is generated by the dynamical system $(\Omega,\ZM,\tau)$, where $\Omega$ is the circle of radius $r$ centered at the origin of $\RM^2$ and $\tau$ is the rotation by $\theta$. Furthermore, \eqref{Eq-IncommensurateI} can be equivalently written as:
\begin{equation}
p_n=n+\Pi(\tau_n\omega)-\Pi(\omega),
\end{equation}
where $\Pi$ is the projection of $\Omega$ on the horizontal axis. This establishes a direct relation with Example~\ref{Ex:X2}. We can also see that $\Gamma_1 : \Omega \rightarrow \RM$ is given by:
\begin{equation}
\Gamma_1=1+\Pi \circ \tau-\Pi.
\end{equation}
This is clearly a continuous function of $\omega$.

\vspace{0.2cm}

We have several important observations. First, it is clear that this pattern is a perturbation of the periodic pattern $x_n = n$. Indeed, in the vicinity of each integer $n$ there is one and only one point of the pattern, hence the periodic and the aperiodic patterns can be deformed into one another without changing the labels or the ordering of the coordinates. Secondly, the pattern is part of a much larger class of patterns, obtained by replacing the circle with any smooth closed loop of length $L$, and the rotations by steps along the loop of equal length $\Delta l=\frac{\theta}{2\pi}L$. We will have to make sure that the projection of this loop on the horizontal axis is inside an interval $[-r,r]$ with $r <\frac{1}{2}$. Note that the new dynamical system is conjugate with the original one, so there is virtually no change at the level of dynamical systems, yet the location of the points can be quite different.

\subsubsection{}\label{ExampleII1} 

\begin{figure}[t]
\center
  \includegraphics[width=0.9\textwidth]{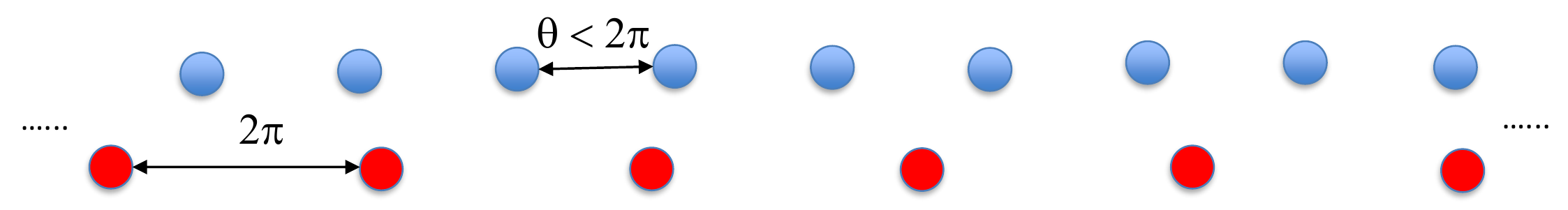}\\
  \caption{\small {\bf Incommensurate pair of periodic chains.} This ideal point pattern is introduced in Example~\ref{ExampleII1}.}
 \label{Fig-Incomm2}
\end{figure}

This pattern leaves in $\RM^2$ but is indexed by $\ZM$ and it is an approximation of the ideal pattern shown in Fig.~\ref{Fig-Incomm2}. The latter consists of two parallel periodic 1-dimensional lattices, of which one has a lattice constant $2\pi$ and the other $\theta<2 \pi$. 

\begin{figure}[b]
\center
  \includegraphics[width=0.7\textwidth]{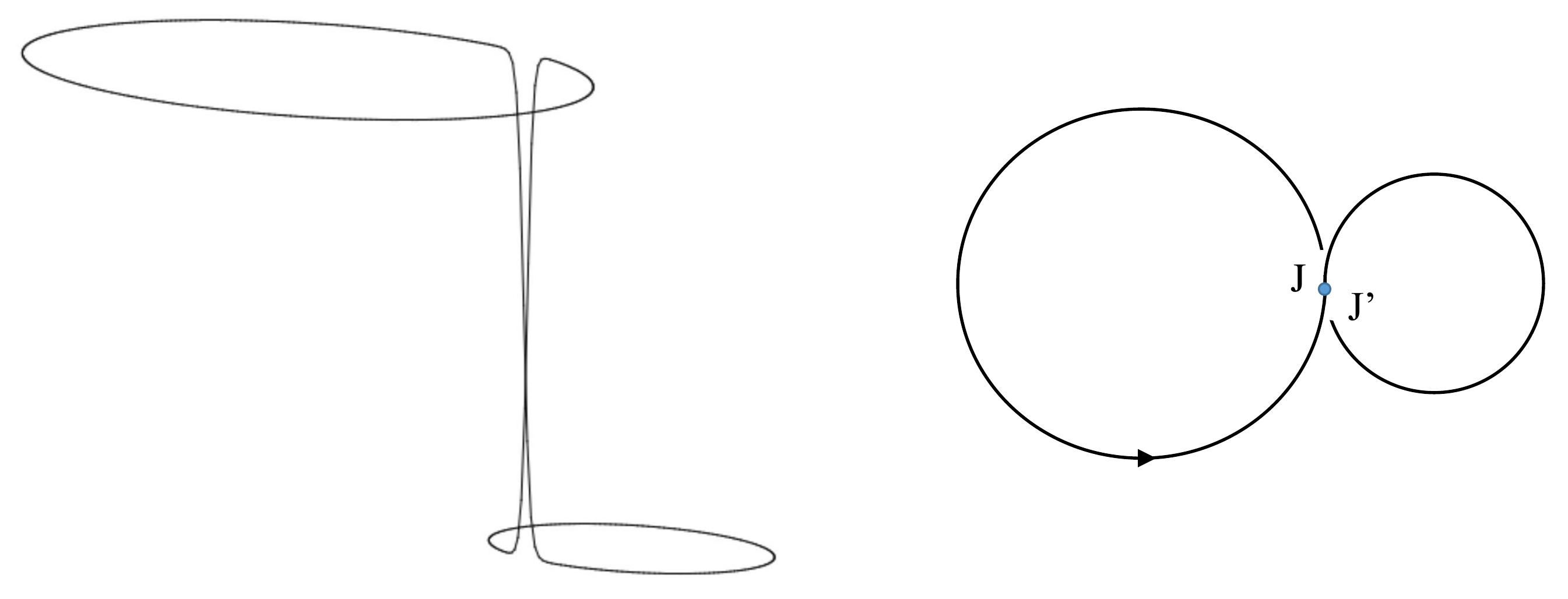}\\
  \caption{\small {\bf The dynamical system for the pattern defined in Example~\ref{ExampleII1}}. It consists of steps of length $\theta$ on the smooth and non self-intersecting loop shown on the left. The right diagram shows a view from the top.}
 \label{Fig-Pretzel}
\end{figure}

We do not know at this moment how to generate the ideal point pattern out of a dynamical system but we can describe an approximate $(\Omega,\ZM,\tau)$-generated pattern that converges to the ideal one in a certain limit. The space $\Omega$ is the closed and smooth loop illustrated in Fig.~\ref{Fig-Pretzel}. Viewed from above, this loop appears as the figure eight, made out of two perfect circles. One of the circles has length $2\pi$ and the other has length $\theta < 2 \pi$. We call this the flat $\Omega$. While walking on the flat $\Omega$ in the positive direction indicated by the arrow in Fig.~\ref{Fig-Pretzel}, one crosses from the large circle to the small circle at point $J$, and then from the small circle back to the large circle at point $J'$. The action of $\ZM$ is implemented on $\Omega$ by stepping by $\theta$ on the flat $\Omega$ in the positive direction. We will adopt the convention that $J$ belongs to the large circle while $J'$ belong to small circle.

\vspace{0.2cm}

We describe now the map $\Gamma_1:\Omega \rightarrow \RM^2$. For $\omega \in \Omega$, let $d_+(\omega)$ be the distance from $\omega$ to $J'$ as moving on the flat $\Omega$ in the positive direction, and $d_-(\omega)$ be the distance from $\omega$ to $J'$ as moving in the negative direction. The vertical coordinate of a point $\omega \in \Omega$ will be denoted by $Y(\omega)$. Then the algorithm for producing the pattern can be described as follows. We start with a point $p$ at the origin of $\RM^2$ and, every time we step by $\theta$ on the flat $\Omega$, arriving at $\tau_1 \omega$, we increase the horizontal and vertical coordinates of $p$ by:
\begin{equation}
\Delta_x(\omega) = \min\{ \theta,d_\pm (\tau_1 \omega)\}, \quad \Delta y(\omega) = Y(\tau_1 \omega) - Y(\omega).
\end{equation}
The crucial observation here is that both quantities are continuous functions of $\omega$. This then leads to:
\begin{equation}
\bm p_{n+1} = \bm p_n + \Big( (\Delta_x \circ \tau_n)(\omega), (\Delta_y \circ \tau_n)(\omega) \Big ),
\end{equation}  
and to the continuous function:
\begin{equation}
\Gamma_1: \Omega \rightarrow \RM^2, \quad \Gamma_1(\omega) = \big (\Delta_x(\omega), \Delta_y(\omega) \big ).
\end{equation}

\vspace{0.2cm}

Let us now describe the algorithm using words. Assume $\omega$ is located on the large circle and that we take a step by $\theta$ in the positive direction on the flat $\Omega$. If $\omega$ remains on the big circle, we increase the horizontal coordinate of $p$ by $\theta$. We repeat this for as long as $\omega$ continues to remain on the large circle. At some point, $\omega$ will step from the large circle to the small circle, in which case we increase the horizontal coordinate of $p$ by only $d_+(\omega)$. With the next step, $\omega$ steps back on the large circle and in this case we increase the horizontal coordinate of $p$ by $d_-(\omega)$. The cycle then repeats from here on. A particular situation is when $\omega$ coincides with $J$, hence it is on the large circle. Then the next step takes $\omega$ into $J'$ which belongs to the small circle, and if we follow the algorithm these two consecutive points have the same horizontal coordinate but note that the vertical coordinates differ. 

\vspace{0.2cm}

The connection with the ideal pattern from Fig.~\ref{Fig-Incomm2} is as follows. If $\omega$ is located on the small circle, it is immediate to check that:
\begin{equation}
d_+(\omega) + d_- \big (\tau_1(\omega) \big ) =\theta.
\end{equation}
Now take the view that point $\bm p_n$ originates from $\tau_n \omega$. Then, all the points of the pattern originating from $\omega$'s located on the large circle are all spaced by $\theta$ on the horizontal axis. Furthermore, one can also see that the points of the pattern originating from $\omega$'s located on the small circle are all spaced by $2 \pi$. Lastly, because of the way the $y$-coordinate of $\omega$ behaves when jumping from one circle to another, it is clear that the pattern generated by $(\Omega,\ZM,\tau)$ is similar to the ideal one. The only difference is that the jump in the vertical direction happens abruptly for the latter and smooth for the former.

\begin{figure}[t]
\center
  \includegraphics[width=\textwidth]{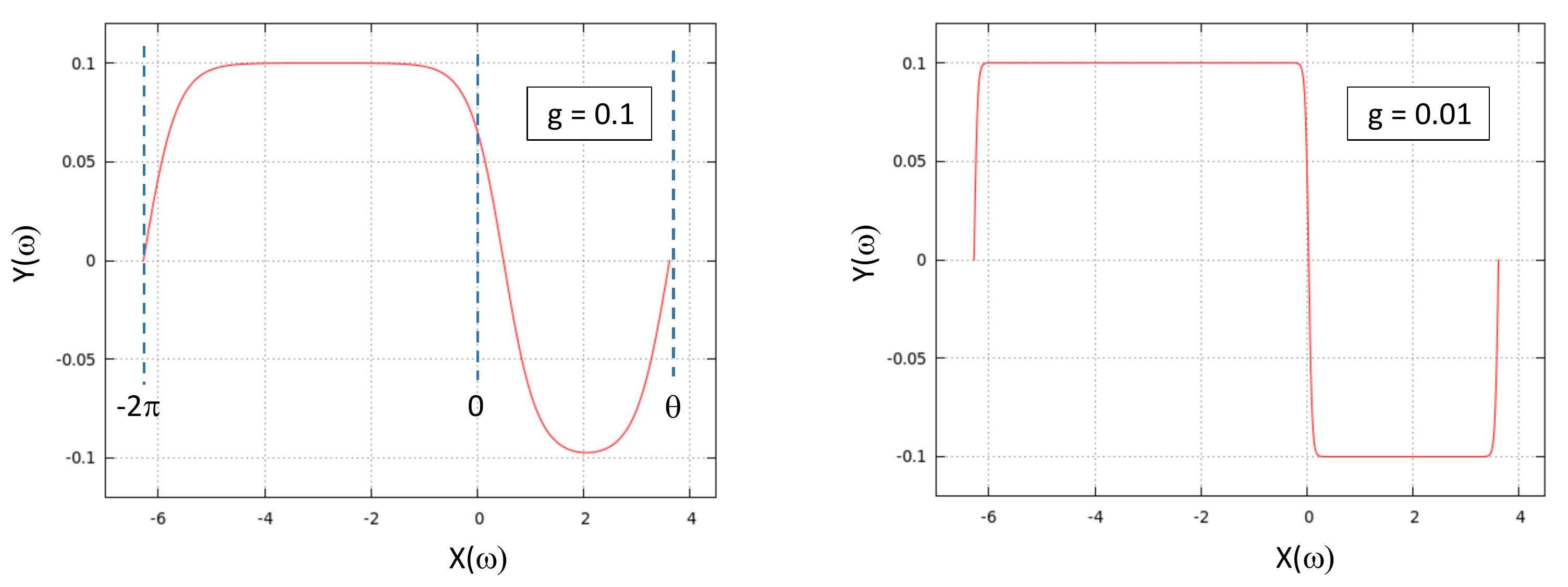}\\
  \caption{\small {\bf Geometric profile.} Plots of Eq.~\ref{Eq-Curve} for different values of the parameter $g$. In both cases, $\delta=0.2$.}
 \label{Fig-Curve}
\end{figure}

\begin{figure}[t]
\center
  \includegraphics[width=1\textwidth]{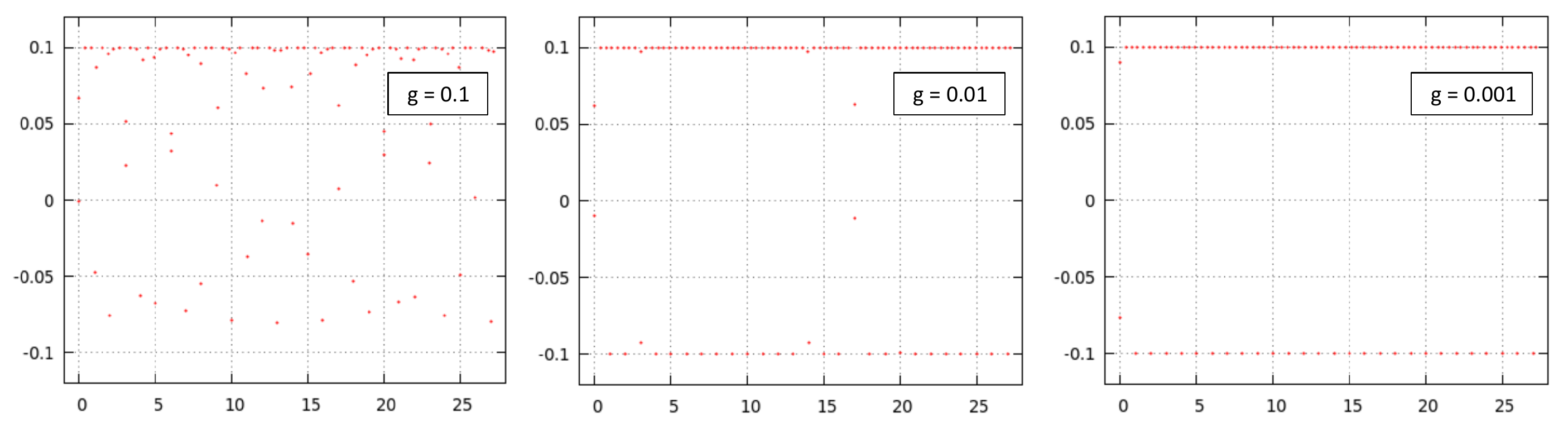}\\
  \caption{\small {\bf Point patterns approximating the ideal point pattern Fig~\ref{Fig-Incomm2}.} Reported are the first 100 points of the patterns generated with Eq.~\ref{Eq-Curve}, for different values of the parameter $g$. The other parameters were fixed as $\delta =0.2$ and $\theta=2\pi /\sqrt{7}$, and the patterns were seeded at $X(\omega)=0$. The unit for the horizontal axis is $2 \pi$.}
 \label{Fig-PatternII}
\end{figure}

\vspace{0.2cm}

To quantify the above affirmation, we now generate the patterns numerically. For this, it is useful to interpret $\Omega$ as a curve in the space $\big (\RM \, {\rm mod}\, (2\pi +\theta) \big ) \times \RM$, as illustrated in Fig.~\ref{Fig-Curve}. The specific curve used in the numerical simulations was generated by the equation:
\begin{equation}\label{Eq-Curve}
Y(\omega)=\tfrac{\delta}{2}\big ( \tanh \big(\tfrac{t+1}{g}\big )-\tanh \big(\tfrac{t-g'}{g}\big )+\tanh\big (\tfrac{t-\theta}{g}\big)\big ), \quad t = X(\omega),
\end{equation}
where $g'=g^{1.1}$ and $g$ is a positive parameter that can be adjusted. Note that, because of the asymmetry introduced by $g'$, the loop in Fig.~\ref{Fig-Pretzel} does not self-intersect. Also, in the limit $g \searrow 0$, the curve becomes a stepped one. The patterns generated with \eqref{Eq-Curve} are shown in Fig.~\ref{Fig-PatternII}. As one can see, as long as we only examine a finite number of points, the $(\Omega,\ZM,\tau)$-generated pattern does converge to the ideal pattern in the limit described above. 

\subsubsection{}\label{ExampleIII1} 
\begin{figure}[t]
\center
  \includegraphics[width=0.4\textwidth]{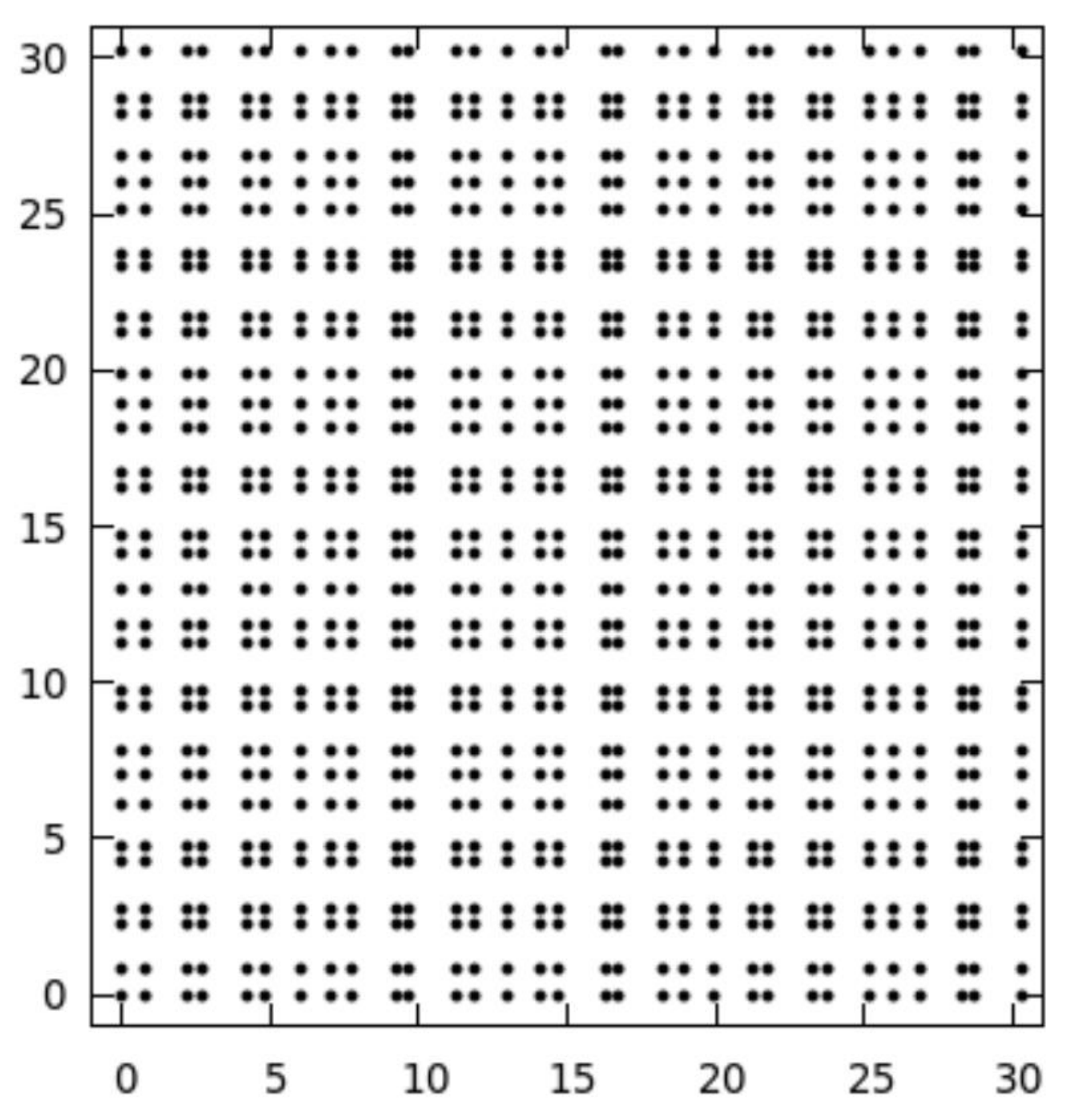}\\
  \caption{\small {\bf Illustration of the pattern defined in Example~\ref{ExampleIII1}}. The parameters were fixed as $r=0.3$ and $\theta_1=\theta_2=\frac{1}{\sqrt{3}}$.}
 \label{Fig-Incomm3}
\end{figure}

In this example, the pattern leaves in $\RM^2$ and is indexed by $\ZM^2$. It is defined by the rule:
\begin{equation}\label{Eq-IncommensurateIII}
\bm p_n= \bm n  + r\,\big ( \sin(n_1 \theta_1+\omega_1)-\sin(\omega_1) \big ) \, \bm e_1 +r\, \big (\sin(n_2 \theta_2 + \omega_2) -\sin(\omega_2) \big )\, \bm e_2,
\end{equation}
where $r <\tfrac{1}{2}$ and $\bm n=(n_1,n_2) \in \ZM^2$. This is a pattern of the same type as in Example~\ref{Ex:X2}, generated by the dynamical system $(\Omega,\ZM^2,\tau)$ with $\Omega= \TM^2$ and the action of $\ZM^2$ implemented by the rotations by $\theta_1$ and $\theta_2$. The function $\Gamma:\Omega \rightarrow \RM^2$ is supplied by:
\begin{equation}
\Gamma(\omega) = r \, \big ( \sin(\omega_1),\sin(\omega_2) \big ).
\end{equation}
A sample of such pattern is reported in Fig.~\ref{Fig-Incomm3}.

\subsubsection{}\label{ExampleIV1} 
\begin{figure}[b]
\center
  \includegraphics[width=0.6\textwidth]{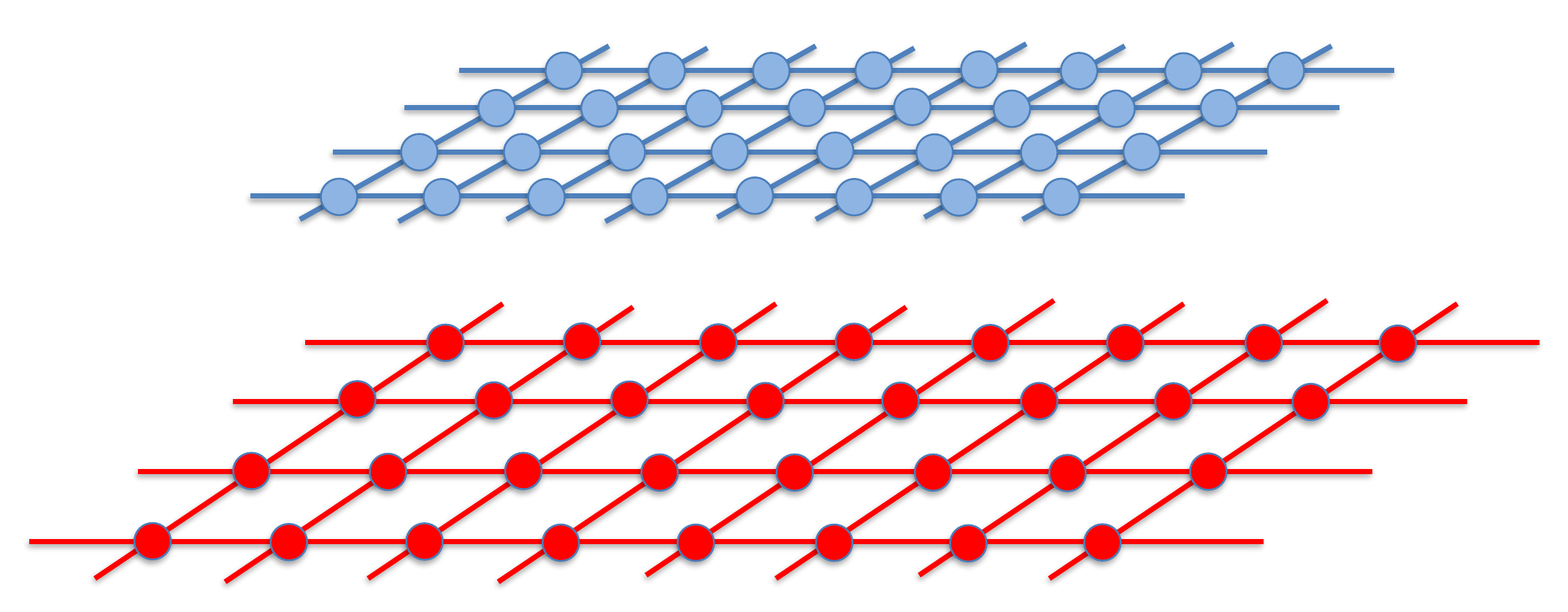}\\
  \caption{\small {\bf Illustration of the ideal pattern defined in Example~\ref{ExampleIV1}.} It consists of two vertically separated incommensurate square lattices.}
 \label{Fig-Incomm4}
\end{figure}

We use this example to showcase the diversity of point patterns that can be constructed with a dynamically-generated algorithm. As in Example~\ref{ExampleII1}, it approximates an ideal point pattern, which this time consists of two periodic square lattices, of which one has a lattice constant $2 \pi$ and the other $\theta< 2 \pi$. The periodic patterns are located on two parallel planes which are spaced by $\delta>0$, as illustrated in Fig.~\ref{Fig-Incomm4}. 

\vspace{0.2cm} 

\begin{figure}[t]
\center
  \includegraphics[width=0.9\textwidth]{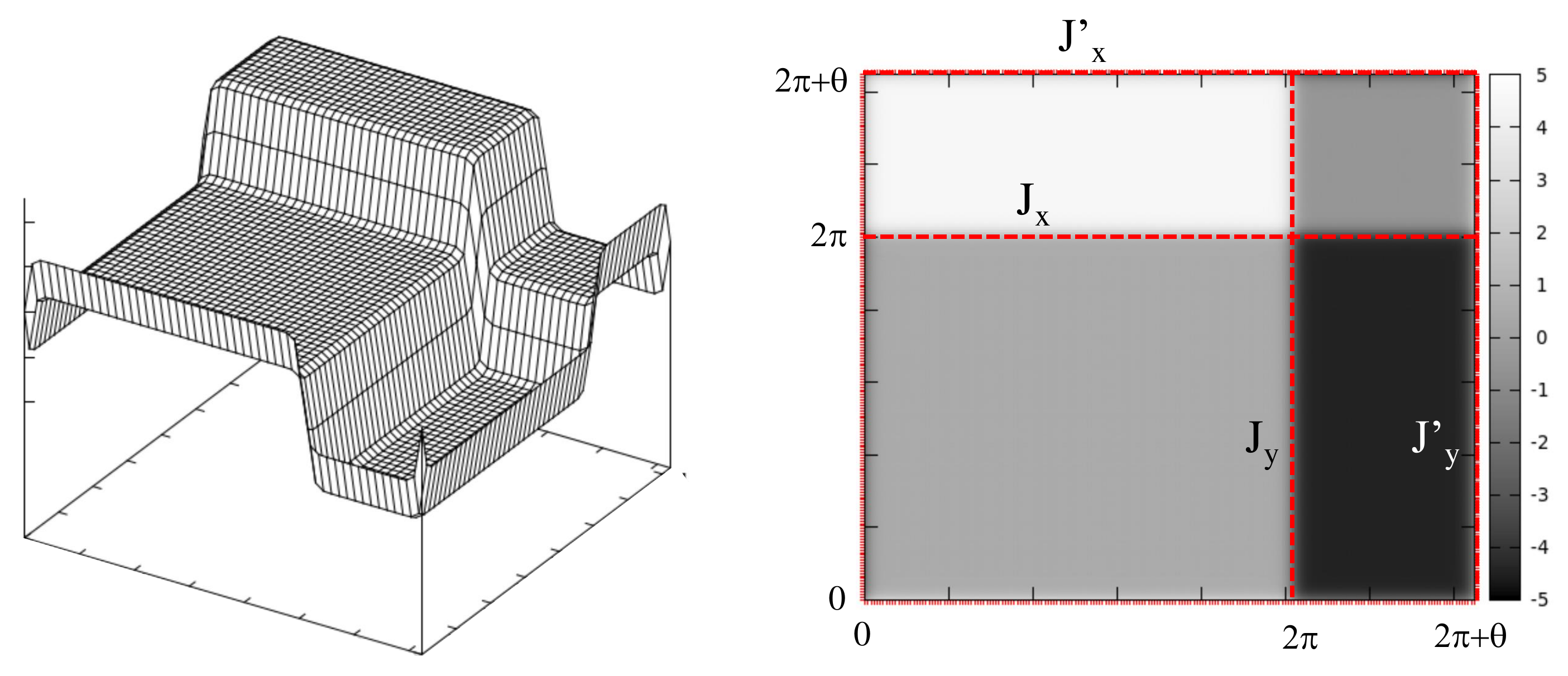}\\
  \caption{\small {\bf The dynamical system for Example~\ref{ExampleIV1}}. It is defined on the surface shown on the left. Viewed from above, this surface appears as a flat torus of sides $2\pi + \theta$.}
 \label{Fig-Surface}
\end{figure}

It is perhaps impossible to label the above ideal pattern by $\ZM^2$ and fulfill \eqref{Eq-Label} in the same time. Yet, the approximate pattern introduced next offers a reasonable solution to this impass. It is dynamically generated by $(\Omega,\ZM^2,\tau)$, where $\Omega$ is the closed surface shown in Fig.~\ref{Fig-Surface}. As before, $X(\omega)$, $Y(\omega)$ and $Z(\omega)$ will represent the Euclidean coordinates of the point $\omega \in \Omega$. Viewed from above, $\Omega$ is a flat torus of side lengths $2\pi +\theta$. The action of $\ZM^2$ is implemented by stepping by $\theta$ in either $x$ or $y$ direction on this flat $\Omega$. In Fig.~\ref{Fig-Surface}, we parametrize this torus and identify the vertical axes $J_y$ and $J'_y$, as well as the horizontal axes $J_x$ and $J'_x$, which will help us define the functions $\Gamma_{\bm e}$. Following somewhat the procedure from Example~\ref{ExampleII1}, we introduce the notation $D_+^x(\omega)$ for the distance from $\omega$ to $J'_y$ as going in the positive direction, and $D_-^x(\omega)$ for the distance from $\omega$ to $J'_y$ as going in the negative direction. Similarly, $D_+^y(\omega)$ denotes the distance from $\omega$ to $J'_x$ as going in the positive direction, and $D_-^y(\omega)$ the distance from $\omega$ to $J'x$ as going in the negative direction. Note that these distances can be expressed analytically as:
\begin{align}
 & D_+^x(\omega)=2\pi + \theta - X(\omega), \quad D_-^x(\omega) = X(\omega), \\ 
& D_+^y(\omega)=2\pi + \theta - Y(\omega), \quad D_-^y(\omega) = Y(\omega).
\end{align}
To generate the pattern, we start from some point of $\Omega$ and we place a point $\bm p$ at the origin of $\RM^3$. Then, every time we jump by $\theta$ on the flat $\Omega$ in the $x$-direction, landing at $\tau_{\bm e_1}\omega$, we increase the coordinates of $\bm p$ by:
\begin{equation}
\Delta^{\bm e_1}_x(\omega)= \min\{\theta,D_\pm^x(\tau_{\bm e_1}\omega)\}, \quad \Delta^{\bm e_1}_y(\omega)=0, \quad \Delta^{\bm e_1}_z(\omega) = Z(\tau_{\bm e_1}\omega) - Z(\omega),
\end{equation}
and, similarly, if the step is in the $y$-direction, landing at  $\tau_{\bm e_2}\omega$:
\begin{equation}
\Delta^{\bm e_2}_x(\omega)= 0, \quad \Delta^{\bm e_2}_y(\omega)=\min\{\theta,D_\pm^y(\tau_{\bm e_2}\omega)\}, \quad \Delta^{\bm e_2}_z(\omega) = Z(\tau_{\bm e_2}\omega) -Z(\omega).
\end{equation}
It is immediate to see that all these functions are continuous of $\omega$. They lead to the algorithm:
\begin{equation}\label{Eq-AlgorithmIV}
\bm p_{\bm n + \bm e} = \bm p_{\bm n} + \Big ( (\Delta^{\bm e}_x \circ \tau_{\bm n})(\omega),(\Delta^{\bm e}_y \circ \tau_{\bm n})(\omega),(\Delta^{\bm e}_z \circ \tau_{\bm n})(\omega) \Big ),
\end{equation}
from where we can read the generating functions:
\begin{equation}
\Gamma_{\bm e} : \Omega \rightarrow \RM^3, \quad \Gamma_{\bm e}(\omega) = \big (\Delta_x^{\bm e}(\omega),\Delta_y^{\bm e}(\omega),\Delta_z^{\bm e}(\omega) \big ), \quad \bm e \in \Gg_2.
\end{equation}
It is straightforward to verify that $\Gamma_{\bm e}$'s satisfy the conditions of Proposition~\ref{Pro-Algorithm}.

\vspace{0.2cm}

\begin{figure}[t]
\center
  \includegraphics[width=0.9\textwidth]{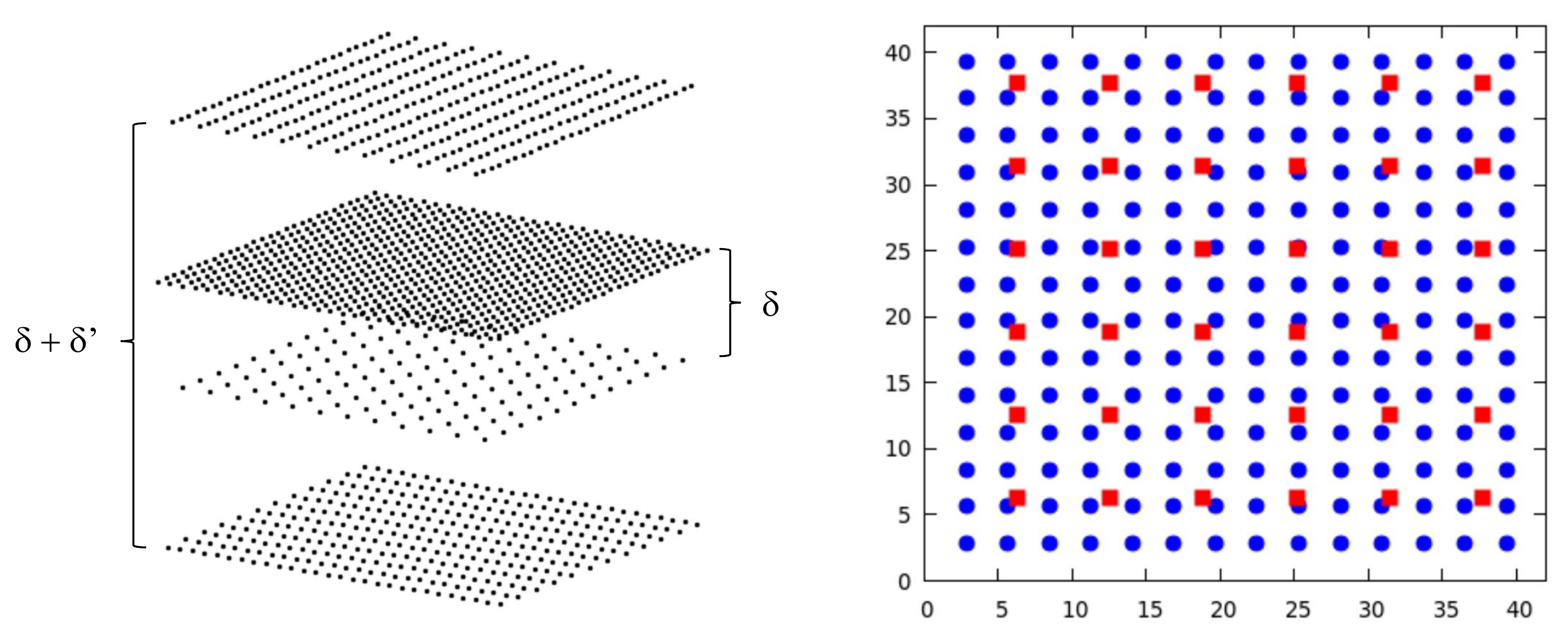}\\
  \caption{\small {\bf Point pattern approximating the ideal pattern from Fig.~\ref{Fig-Incomm4}.} (Left) A point pattern generated with the algorithm~\eqref{Eq-AlgorithmIV}. (Right) A view from the top of the two inner layers.}
 \label{Fig-PatternIV}
\end{figure}

To quantify the above affirmations, we now generate the patterns numerically. The equation for $\Omega$ used in the numerical simulations and in Fig.~\ref{Fig-Surface} is:
\begin{align}\label{Eq-Surface}
Z(\omega)= \tfrac{2\delta + \delta'}{4} & \big ( \tanh \big(\tfrac{u}{g}\big )-\tanh \big(\tfrac{u-2\pi-g'}{g}\big )+\tanh\big (\tfrac{u-2\pi-\theta}{g}\big)\big ) \\
\nonumber  - \tfrac{\delta'}{4} & \big ( \tanh \big(\tfrac{v}{g}\big )-\tanh \big(\tfrac{v-2\pi-g'}{g}\big )+\tanh\big (\tfrac{v-2\pi-\theta}{g}\big)\big )
\end{align}
where $u = X(\omega)$ and $v = Y(\omega)$ and all the other parameters are as in \eqref{Eq-Curve}. The pattern obtained with this $\Omega$ is shown in Fig.~\ref{Fig-PatternIV}. The limit $g \searrow 0$ has virtually the same effect as in Example~\ref{ExampleII1}. The parameter $\delta'$ is used to space out the two outer layers in Fig.~\ref{Fig-PatternIV}, which do not belong to the ideal pattern we want to mimic.

\subsubsection{} The projections on the flat $\Omega$'s together with appropriate units provide dynamical conjugacies, which are used below to bring the dynamical systems to more standard forms. 

\begin{corollary}\label{Cor-DynSys} The following apply:
\begin{enumerate}[{\rm i)}]
\item For Example~\ref{ExampleI1}: 
\begin{equation}
\Omega \simeq \RM/\ZM, \quad \tau_n(x) = (x+n\theta)\, {\rm mod}\, 1.
\end{equation}
\item For Example~\ref{ExampleII1}: 
\begin{equation}
\Omega \simeq \RM/(1+\theta)\ZM, \quad \tau_n(x) = (x+n\theta) \, {\rm mod}\, (1+\theta).
\end{equation}
\item For example~\ref{ExampleIII1}:
\begin{align}
&\Omega \simeq \big (\RM/\ZM \big ) \times \big ( \RM/\ZM \big ), \\ 
& \tau_n(x,y)= ( x+n_1 \theta_1, x+n_2 \theta_2) \, {\rm mod}\, 1.
\end{align}
\item For example~\ref{ExampleIV1}:
\begin{align}
& \Omega \simeq \Big ( \RM/(1+\theta)\ZM \Big )  \times \Big ( \RM/(1+\theta)\ZM \Big ), \\ 
\nonumber & \tau_n(x,y)= ( x+n_1 \theta, x+n_2 \theta ) \, {\rm mod}\, (1 + \theta).
\end{align}

\end{enumerate}

\end{corollary}

\begin{remark}{\rm When $\theta$ is an irrational number then the dynamical systems i) and ii) are minimal. Same can be said about dynamical systems iii) and iv) when both $\theta_1$ and $\theta_2$ are irrational.}$\Diamond$
\end{remark} 

\section{Algebra of bulk physical observables}\label{Ch-BulkAlgebra}

A crossed product $C^\ast$-algebra can be  canonically associated to any topological dynamical system. In this chapter, we define and characterize this algebra, as well as show that its representations generate all physical models over a dynamically generated point pattern. The relation between the elements of the algebra and their physical representations is investigated to some detail. Then the algebra of physical observables is computed explicitly for the examples introduced in section~\ref{Sec:Examples}, and shown to coincides with the non-commutative torus in various dimensions. This in turn enables us to fully characterize their $K$-theories and make various predictions, which are verified numerically.

\subsection{Definition and physical representations}

Associated to the classical topological system $(\Omega,\ZM^d,\tau)$ of a pattern, there is the dual $C^\ast$-dynamical system consisting of the tuple: 
\begin{equation}
\Big (C(\Omega), \tau: \ZM^d \rightarrow {\rm Aut}\big (C(\Omega) \big ) \Big),
\end{equation} 
where $C(\Omega)$ is the $C^\ast$-algebra of complex-valued continuous functions over $\Omega$ and $\tau$ is the group homomorphism provided by the dual action of $\ZM^d$ on $C(\Omega)$: 
\begin{equation}
\tau_{\bm n} (f) = f \circ \tau_{\bm n}, \quad \bm n \in \ZM^d, \quad f \in C(\Omega).
\end{equation} 
Recall that $C(\Omega)$ is equipped with the sup-norm:
\begin{equation}\label{Eq:SupNorm}
\|f \| = \sup_{\omega \in \Omega} |f(\omega)|, \quad f \in C(\Omega).
\end{equation}

\begin{remark}\label{Re:CNOmega}{\rm Later on, we will be dealing with the algebra $C_N(\Omega)$ of continuous functions from $\Omega$ to $\CM^N$. Note that $C_N(\Omega) \simeq \Mm_N \otimes C(\Omega)$ and that the norm on $C_N(\Omega)$ is given by the same \eqref{Eq:SupNorm}, where $|\cdot|$ is understood as the uniform norm on $\Mm_N$. This will be automatically assumed whenever we pass from $C(\Omega)$ to $C_N(\Omega)$.}
$\Diamond$
\end{remark}

A covariant representation of the dual dynamical system on a Hilbert space $\Hh$ is a pair $(\pi,U)$ of a $C^\ast$-representation of $C(\Omega)$ and a unitary representation of $\ZM^d$ such that:
\begin{equation}
U_{\bm n}^\ast \, \pi(f) \, U_{\bm n} =  \pi \big (\tau_{\bm n}(f) \big ), \quad \bm n \in \ZM^d, \quad f \in C(\Omega).
\end{equation}
These covariant representations generate all linear physical models over a point pattern, introduced and discussed in chapter~\ref{Ch:PattRez}. The crossed product algebra $C(\Omega) \rtimes_\tau \ZM^d$ is, by definition \cite{Wil}, the $C^\ast$-algebra that generates all covariant representations of $(C(\Omega),\ZM^d,\tau)$, hence, all the physical models over the pattern. It can be defined in several ways, and here we adopt a definition which is preferred by physicists.

\begin{definition}[\cite{NS}] The crossed product algebra $C(\Omega) \rtimes_\tau \ZM^d$ is the universal $C^\ast$-algebra:
\begin{equation}
\Aa_d = C^\ast \big ( C(\Omega),u_1,\ldots,u_d \big ),
\end{equation} 
generated by a copy of $C(\Omega)$, $d$ commuting unitary operators:
\begin{equation}
u_i u_i^\ast = u_i^\ast u_i = 1, \quad u_i u_j = u_j u_i, \quad i,j=1,\ldots,d,
\end{equation} 
and by the relations:
\begin{equation}
f u^{\bm n} = u^{\bm n} (f \circ \tau_{\bm n}), \quad u^{\bm n}=u_1^{n_1} \ldots u_d^{n_d}, \quad \bm n \in \ZM^d, \quad f \in C(\Omega).
\end{equation} 
\label{Def-BulkAlgebra}
\end{definition}

\begin{remark}{\rm The definition of the universal algebra generated by a set of relations can be found in many standard textbooks, {\it e.g.} \cite{Bla2}[p.~158]. A generic element of the algebra $\Aa_d$ can be presented uniquely in the form:
\begin{equation}\label{Eq-ElementPresentation}
a = \sum_{\bm q \in \ZM^d
} a_{\bm q} u^{\bm q}, \quad a_{\bm q} \in C(\Omega).
\end{equation} 
The $a_{\bm q}$'s are called the Fourier coefficients of $a$ and their norm in $C(\Omega)$ must display a certain decay as $|\bm q| \rightarrow \infty$, in order for the sum in \eqref{Eq-ElementPresentation} to converge in the norm. This is further explained below.
}$\Diamond$
\end{remark}

\begin{remark}{\rm The extension to $\Mm_N \otimes \Aa_d$ will be needed below. Note that the generic elements of these algebras take the form \eqref{Eq-ElementPresentation} with $a_{\bm q}$ in $C_N(\Omega)$ rather than $C(\Omega)$.}
$\Diamond$
\end{remark}

\begin{proposition}\label{Pro-BulkRep} The crossed product algebra accepts a continuous field of canonical representations $\{\pi_\omega\}_{\omega \in \Omega}$ on $\ell^2(\ZM^d)$:
\begin{equation}
 \pi_\omega(u^{\bm q}) = U_{\bm q}, \quad  U_{\bm q}|\bm n \rangle = |\bm n + \bm q \rangle \quad \bm n,\bm q\in \ZM^d ,
 \end{equation}
 and
 \begin{equation} 
 \pi_\omega(f) = \sum_{\bm n \in \ZM^d} f (\tau_{\bm n}\omega)\, |\bm n \rangle \langle \bm n |, \quad f \in C(\Omega).
\end{equation}
The continuity of the field of representations is w.r.t. the strong topology on $\BM\big (\ell^2(\ZM^d)\big )$.
\end{proposition}

\proof It is straightforward to verify that each $\pi_\omega$ preserves the defining relations of the algebra $\Aa_d$. Hence, we focus on the continuity w.r.t. $\omega$. By a general property of the homomorphisms of $C^\ast$-algebras \cite{Dav}[p.~14], all $\pi_\omega$'s are contractions, hence it is enough to establish their continuity w.r.t. $\omega$ on a dense subset of the crossed product. We take this dense subset to be the algebra of polynomials $\sum_{\bm q} a_{\bm q} u^{\bm q}$, where the sum is over compact subsets of $\ZM^d$. Since only a finite number of terms appear in these sums, it is actually enough to establish the continuity of $\pi_\omega$ over the generators, more precisely over $C(\Omega)$. The unit ball of $\BM\big ( \ell^2(\ZM^2) \big )$ with the strong topology is metrizable, {\it e.g.} by:
\begin{equation}
{\rm d}_S(A,A') = \sum_{\bm n\in \ZM^d}  2^{-|\bm n|} \big \|(A-A')|\bm n\rangle \big \|
\end{equation}
We have:
\begin{equation}
{\rm d}_S\big ( \pi_\omega(f),\pi_{\omega'}(f)\big ) = \sum_{\bm n\in \ZM^d} 2^{-|\bm n|}|f\big (\tau_{\bm n}(\omega)\big ) - f(\tau_{\bm n}(\omega')\big )|.
\end{equation}
Now, for an arbitrary small but fixed $\epsilon > 0$, take $N_\epsilon$ large enough such that $\sum_{|\bm n| > N_\epsilon} 2^{-|\bm n|} \leq \epsilon/4$,  and choose  a small neighborhood of $\omega$ such that, for any $\omega'$ in this neighborhood, $ |f(\tau_{\bm n}(\omega)) - f(\tau_{\bm n}(\omega'))| \leq \epsilon/2\Gamma$, $\Gamma =\sum_{{\bm n}\in \ZM^d} 2^{-|\bm n|}$, for all $n$ with $|\bm n| \leq N_\epsilon$. The latter is possible because all these $f\circ \tau_{\bm n}$ are continuous and there are a finite number of them. Then:
\begin{equation} 
{\rm d}_S\big ( \pi_\omega(f),\pi_{\omega'}(f)\big ) \leq \epsilon/2 \Gamma \sum_{|\bm n|
\leq N_\epsilon} 2^{-|\bm n|}+2 \sum_{|\bm n| > N_\epsilon} 2^{-|\bm n|}|   \leq \epsilon.
\end{equation}
and the affirmation follows.\qed

\begin{remark}{\rm The continuous field of representations $\pi_\omega$ can be extended to the continuous field of representations ${\rm id}\otimes \pi_\omega$ of $\Mm_N \otimes \Aa_d$ on $\CM^N \otimes \ell^2(\ZM^d)$. Since this extension is trivial, we keep the notation $\pi_\omega$ for it.}
$\Diamond$
\end{remark}

\begin{corollary}[\cite{RS}(p.~291)] Let $h \in \Mm_N \otimes \Aa_d$ be self-adjoint, $h^\ast=h$. Then the spectrum ${\rm Spec}\big (\pi_\omega(h) \big )$ of the bounded operator $\pi_\omega(h)$ on $\CM^N \otimes \ell^2(\ZM^d)$ belongs to the real axis and is semi-continuous w.r.t. $\omega$, in the sense that, if $\omega_k \rightarrow \omega$ in $\Omega$ and $\lambda \in {\rm Spec}\big (\pi_\omega(h) \big )$, then there are $\lambda_k$'s in ${\rm Spec}\big (\pi_{\omega_k}(h) \big )$ such that $\lambda_k \rightarrow \lambda$.
\label{Cor-SigmaStrong}
\end{corollary}

\begin{remark}\label{Re-SigmaStrong}{\rm We should emphasize `semi-continuity' in the above statement and make sure that it is well understood that, in general, it is false that the spectrum is continuous of $\omega$, {\it e.g.} w.r.t. the Hausdorff metric on the compact subsets of the real line. Nevertheless,  one important consequence of the above statement is that, under similar conditions:
\begin{equation}
{\rm Spec} \big ( \pi_\omega(h) \big ) \subseteq \, \overline{\bigcup {\rm Spec}\big (\pi_{\omega_k}(h)\big )},
\end{equation}
where the inclusion can be strict. On the other hand, recall that $\Omega$'s in the concrete examples introduced in section~\ref{Sec:Examples}, as presented in Corollary~\ref{Cor-DynSys}, are metric spaces on which $\tau$ acts isometrically. Since any $f \in C_N(\Omega)$ is actually uniformly continuous, it follows that, for each $\epsilon > 0$, there is $\delta_\epsilon>0$ such that:
\begin{equation}
|f(\tau_{\bm n}\omega) - f(\tau_{\bm n} \omega')| \leq \epsilon
\end{equation}
whenever:
\begin{equation}
{\rm d}_\Omega(\tau_{\bm n}\omega,\tau_{\bm n} \omega') = {\rm d}_\Omega(\omega,\omega') \leq \delta_\epsilon.
\end{equation}
Then:
\begin{align}
\|\pi_\omega(f) - \pi_{\omega'}(f) \| & = \sup_{\bm n \in \ZM^d} |f(\tau_{\bm n}\omega) - f(\tau_{\bm n} \omega')| \leq \epsilon
\end{align}
for any $\omega, \omega' \in \Omega$ with ${\rm d}_\Omega(\omega,\omega') \leq \delta_\epsilon$. In these cases the representations $\pi_\omega$ are continuous of $\omega$ even when $\BM\big (\ell^2(\ZM^d) \big )$ is endowed with the norm topology. Such systems were named almost-periodic in \cite{Bel1} and the spectra of $\pi_\omega(a)$ are continuous of $\omega$ in these situations. We, however, do not make this assumption, in general.
}$\Diamond$
\end{remark}

\begin{remark}{\rm For a generic element $a = \sum_{\bm q \in \ZM^d} a_{\bm q} u^{\bm q} \in \Mm_N \otimes \Aa_d$, the representation gives:
\begin{equation}
\pi_\omega(a) = \sum_{\bm q \in \ZM^d} \sum_{\bm n \in \ZM^d} a_{\bm q}\big (\tau_{\bm n}(\omega)\big )\, |\bm n \rangle \langle \bm n-\bm q |.
\end{equation}
It then becomes apparent that all physical Hamiltonians \eqref{Eq-LatticeHamiltonian} can be generated from the algebra $\Mm_N \otimes \Aa_d$. 
}$\Diamond$
\end{remark}

\begin{remark}\label{Re-BulkDecay}{\rm Since $\ZM^d$ is an amenable group, the $C^\ast$-norm of the crossed product can be described in very simple terms:
\begin{equation}\label{Eq-Norm}
\| a \| = \sup_{\omega \in \Omega} \|\pi_\omega(a)\|,
\end{equation}
where on the right we have the operator norm on $\ell^2(\ZM^d)$. For a generic element, this means that the Fourier coefficients $a_{\bm q} \in C(\Omega)$ must be such that $\pi_\omega(a)$ are all bounded operators on $\ell^2(\ZM^d)$, and this implies a certain decay of $\|a_{\bm q}\|$'s as $|\bm q| \rightarrow \infty$. In particular, all non-commutative polynomials, that is, elements with $a_{\bm q} =0$ for $|\bm q|$ larger than some $R<\infty$, belong to the crossed product. The latter can be also thought as the completion in norm \eqref{Eq-Norm} of the algebra generated by these particular elements.
}$\Diamond$
\end{remark}

\begin{proposition} The field of canonical representations obey the following covariant property:
\begin{equation}\label{Eq-BulkCovariance}
U_{\bm n}^\ast \, \pi_\omega(a) \, U_{\bm n} = \pi_{\tau_{\bm n} \omega}(a), \quad a \in \Aa_d, \quad \bm n \in \ZM^d.
\end{equation}
\end{proposition}

\proof It is enough to verify the statement for the generators. The covariance relation is trivial on the $u$'s and for $f \in C(\Omega)$:
\begin{equation}
U_{\bm n}^\ast \pi_\omega(f) U_{\bm n} = \pi_{\omega}(u_{\bm n}^\ast \, f \, u_{\bm n})=\pi_{\omega}(f \circ \tau_{\bm n}) = \pi_{\tau_{\bm n}\omega}(f),
\end{equation}
 which is the desired result.\qed

\subsection{Spectral properties}

The bulk-boundary principle is a statement about the spectral properties of the physical models. In its $K$-theoretic formulation, however, the statement is rather about the spectral properties of algebra elements that generate the models. As such, we must pay special attention to the relation between the spectral properties of an element of the algebra and of its physical representations. An introduction to the spectral theory for $C^\ast$-algebras can be found in \cite{Arv}.

\begin{definition} Let $a$ be an element of the crossed product algebra. Then:
\begin{enumerate}[\rm 1.]
\item The resolvent set of $a$ is defined as:
\begin{equation}
\rho(a) = \big \{ \lambda \in \CM \, | \, a - \lambda \cdot 1 \ \mbox{invertible in} \ \Aa_d \big \}.
\end{equation}
The resolvent set is always an open subset of the complex plane.
\item The spectrum of $a$ is defined as the complement of the resolvent set:
\begin{equation}
{\rm Spec}(a) = \CM \setminus \rho(a).
\end{equation}
The spectrum is always a non-empty compact subset of the complex plane.
\end{enumerate}
\end{definition}

\begin{remark}{\rm Note that, in general, the spectrum is determined by both the element and by the algebra to which it belongs.
}$\Diamond$
\end{remark}

\begin{proposition} Let $a \in C(\Omega) \rtimes_\tau \ZM^d$. Then:
\begin{equation}\label{Eq-OmegaSpectrum}
{\rm Spec}(a) = \bigcup_{\omega \in \Omega} {\rm Spec}\big ( \pi_\omega(a) \big ),
\end{equation}
where ${\rm Spec}\big ( \pi_\omega(a) \big )$ is the spectrum of $\pi_\omega(a)$ in the algebra of bounded operators over $\CM^N \otimes\ell^2(\ZM^d)$.
\end{proposition}

\proof The representation $\pi=\bigoplus_{\omega \in \Omega} \pi_\omega $ is unital and faithful, hence $\Mm_N \otimes \Aa_d$ is isometric isomorphic to its image (see also Remark~\ref{Re-BulkDecay}). By an important property of $C^\ast$-algebras \cite{Dav}[p.~15], the spectrum of $\pi(a)$ in $\pi(\Mm_N \otimes \Aa_d)$ coincides with the spectrum of $\pi(a)$ in $\bigoplus_{\omega \in \Omega} \BM \big (\CM^N \otimes \ell^2(\ZM^d) \big )$, hence: 
\begin{equation}
{\rm Spec}(a) = {\rm Spec} \Big (\bigoplus_{\omega \in \Omega} \pi_\omega(a)\Big ) = \bigcup_{\omega \in \Omega} {\rm Spec} \big ( \pi_\omega(a) \big ),
\end{equation}
which is the desired result. \qed

\begin{corollary}\label{Pro-BulkSpectralRelation} If the classical dynamical system $(\Omega,\ZM^d,\tau)$ is minimal, then ${\rm Spec}\big (\pi_\omega(a) \big )$ is independent of $\omega$ and:
\begin{equation}
{\rm Spec}\big (\pi_\omega(a) \big ) = {\rm Spec}(a).
\end{equation} 
\end{corollary}

\proof In this case each of the representations $\pi_\omega$ is unital and faithful, hence ${\rm Spec}\big (\pi_\omega(a) \big )={\rm Spec}(a)$ and the statement follows. Another way to prove the statement is to start from \ref{Eq-OmegaSpectrum} and use Corollary~\ref{Cor-SigmaStrong} and Remark~\ref{Re-SigmaStrong}, together with the fact that the orbits are dense in $\Omega$, to write:
\begin{equation}
\sigma(a)=\overline{\bigcup_{\bm n \in \ZM^d} \sigma\big (\pi_{\tau_{\bm n}\omega}(a) \big )}, 
\end{equation}
a relation that holds for any $\omega \in \Omega$. The covariance property \eqref{Eq-BulkCovariance} of the representations tells us that $\pi_{\tau_{\bm n}\omega}(a)$ differ by unitary conjugations. The statement then follows because the spectrum is invariant under such operations.\qed

\begin{remark}{\rm The above statement is extremely important for the physical applications because, in the conditions of Corollary~\ref{Pro-BulkSpectralRelation}, we can be sure that the spectra of the physical Hamiltonians $H_\omega = \pi_\omega(h)$ are independent of $\omega$ and coincide with the spectrum of the element $h \in C_N(\Omega) \rtimes_\tau \ZM^d$ which generates them.
}$\Diamond$
\end{remark}

\subsection{The algebras of the concrete examples}\label{SubSec-Concrete}

The algebras of physical observables for the concrete examples introduced in section~\ref{Sec:Examples} can be computed explicitly and they are all connected to the non-commutative torus algebra.

\begin{definition} Let $\Theta=\{\theta_{ij}\}_{i,j=\overline{1,n}}$ be a $n\times n$ antisymmetric matrix with entries from the interval $[0,1]$. The non-commutative $n$-torus associated to $\Theta$ is the universal $C^\ast$-algebra:
\begin{equation}
\Aa_\Theta = C^\ast(u_1,\ldots,u_n),
\end{equation}
generated by $n$-unitary operators satisfying the relations:
\begin{equation} 
u_i u_j=e^{\imath 2 \pi \theta_{ij}}u_ju_i, \quad i,j = 1, \ldots,n.
\end{equation}
\end{definition}

\begin{example}\label{Ex-Calculation1}{\rm For the pattern \ref{ExampleI1}, we found 
\begin{equation}
\Omega \simeq \RM/\ZM,\quad \tau_n(x)=(x+n\theta)\, {\rm mod} \,1,
\end{equation} 
hence $C(\Omega)$ has a single generator: \begin{equation}
u_1:\Omega \rightarrow \CM, \quad u_1(x) = e^{\imath 2 \pi x}.
\end{equation}
 As such:
\begin{equation}
C^\ast \big ( C(\Omega),u_2 \big ) = C^\ast(u_1,u_2),
\end{equation}
and the defining relation becomes:
\begin{equation}
u_1 u_2 = u_2 (u_1\circ\tau_1) = e^{\imath 2 \pi  \theta} u_2 u_	1,
\end{equation}
because: 
\begin{equation}
(u_1\circ\tau_1)(x) = e^{\imath 2 \pi (x+\theta)}=e^{\imath 2\pi \theta} u_1(x).
\end{equation}
As such, the algebra of bulk physical observables coincides with the non-commutative 2-torus $\Aa_\Theta$, $\theta_{12}=-\theta_{21} =\theta$.
}$\Diamond$
\end{example}

\begin{example}\label{Ex-Calculation2}{\rm For the pattern \ref{ExampleII1}, we found: 
\begin{equation}
\Omega \simeq \RM/(1+\theta)\ZM,\quad \tau_n(x)=(x+n\theta)\, {\rm mod} \,(1+\theta),\end{equation} 
hence $C(\Omega)$ has a single generator: \begin{equation}
u_1:\Omega \rightarrow \CM, \quad u_1(x) = e^{i 2 \pi  x/(1+\theta)}.
\end{equation}
 As such:
\begin{equation}
C^\ast \big ( C(\Omega),u_2 \big ) = C^\ast(u_1,u_2)
\end{equation}
and the defining relation becomes:
\begin{equation}
u_1 u_2 = u_2 (u_1\circ\tau_1) = e^{i 2 \pi \tilde \theta} u_2 u_	1, \quad \tilde \theta = \frac{\theta}{1+\theta}, 
\end{equation}
because:
\begin{equation}
(u_1\circ\tau_1)(x) = e^{i 2 \pi (x+\theta)/(1+\theta)}=e^{i 2 \pi \tilde \theta} u_1(x).
\end{equation}
As such, the algebra of bulk physical observables coincides with the non-commutative 2-torus $\Aa_\Theta$, $\theta_{12} =-\theta_{21}=\tilde \theta$. 
}$\Diamond$
\end{example}

\begin{example}\label{Ex-Calculation3}{\rm For the pattern \ref{ExampleIII1}, we found: 
\begin{equation}
\Omega \simeq \big (\RM/\ZM\big ) \times \big (\RM/ \ZM\big ),\quad \tau_n(x)=\big ( x_1+n_1\theta_1, x_2 + n_2\theta_2 \big )\, {\rm mod} \,1,
\end{equation} 
hence $C(\Omega)$ has two generators: \begin{equation}
u_1:\Omega \rightarrow \CM, \quad u_1(x) = e^{i 2 \pi x_1},
\end{equation}
and
\begin{equation}
u_2:\Omega \rightarrow \CM, \quad u_2(x) = e^{i 2 \pi x_2}.
\end{equation}
 As such:
\begin{equation}
C^\ast \big ( C(\Omega),u_3,u_4 \big ) = C^\ast(u_1,u_2,u_3,u_4)
\end{equation}
and the defining relations becomes:
\begin{equation}
u_1 u_3 = u_3 (u_1\circ\tau_{(1,0)}) = e^{i 2 \pi \theta_1} u_3 u_	1, 
\end{equation}
because:
\begin{equation}
(u_1\circ\tau_{(1,0)})(x) = e^{i 2 \pi (x+\theta_1)}=e^{i 2 \pi \theta_1} u_1(x).
\end{equation}
Similarly:
\begin{equation}
u_2 u_4 = u_4 (u_2\circ\tau_{(0,1)}) = e^{i 2 \pi \theta_2} u_4 u_	2, 
\end{equation}
because:
\begin{equation}
(u_2\circ\tau_{(0,1)})(x) = e^{i 2 \pi (x+\theta_2)}=e^{i 2 \pi \theta_2} u_2(x).
\end{equation}
As such, the algebra of bulk physical observables coincides with the non-commutative 4-torus $\Aa_\Theta$, with $\theta_{13} = -\theta_{31}=\theta_1$, $\theta_{24}=-\theta_{42}=\theta_2$ and $0$ in rest. 
}$\Diamond$
\end{example}

\begin{example}\label{Ex-Calculation4}{\rm For the pattern \ref{ExampleIV1}, we found: 
\begin{align}
& \Omega \simeq \big (\RM/(1+\theta_1)\ZM\big ) \times \big (\RM/ (1+\theta_2)\ZM\big ),\\
& \tau_n(x)=\big ( x_1+n_1\theta_1, x_2 + n_2\theta_2 \big )\, {\rm mod} \,(1+\theta_1,1+\theta_2),
\end{align} 
hence $C(\Omega)$ has two generators: \begin{equation}
u_1:\Omega \rightarrow \CM, \quad u_1(x) = e^{i 2 \pi x_1/(1+\theta_1)},
\end{equation}
and
\begin{equation}
u_2:\Omega \rightarrow \CM, \quad u_2(x) = e^{i 2 \pi x_2/(1+\theta_2)}.
\end{equation}
 As such:
\begin{equation}
C^\ast \big ( C(\Omega),u_3,u_4 \big ) = C^\ast(u_1,u_2,u_3,u_4)
\end{equation}
and the defining relations becomes:
\begin{equation}
u_1 u_3 = u_3 (u_1\circ\tau_{(1,0)}) = e^{i 2 \pi \tilde \theta_1} u_3 u_	1, \quad \tilde \theta_1 = \frac{\theta_1}{1+\theta_1},
\end{equation}
because:
\begin{equation}
(u_1\circ\tau_{(1,0)})(x) = e^{i 2 \pi (x+\theta_1)/(1+\theta_1)}=e^{i 2 \pi \tilde \theta_1} u_1(x).
\end{equation}
Similarly:
\begin{equation}
u_2 u_4 = u_4 (u_2\circ\tau_{(0,1)}) = e^{i 2 \pi \tilde \theta_2} u_4 u_	2, \quad \tilde \theta_2 = \frac{\theta_2}{1+\theta_2}, 
\end{equation}
because:
\begin{equation}
(u_2\circ\tau_{(0,1)})(x) = e^{i 2 \pi (x+\theta_2)/(1+\theta_2)}=e^{i 2 \pi \tilde \theta_2} u_2(x).
\end{equation}
As such, the algebra of physical observables coincides with the non-commutative 4-torus $\Aa_\Theta$, with $\theta_{13} = - \theta_{31} = \tilde \theta_1$, $\theta_{24}= - \theta_{42}= \tilde \theta_2$ and $0$ in rest. 
}$\Diamond$
\end{example}

\section{Elements of $K$-theory}
\label{Ch:KTh}

\subsection{The $K$-groups defined}

In the complex $K$-theory of a $C^\ast$-algebra $\Aa$ there are only two $K$-groups, which can be described as follows. The first one is the $K_0(\Aa)$ group, which classifies the projections:
\begin{equation}
p \in \Mm_\infty \otimes \Aa, \quad p^2 = p^\ast=p,
\end{equation}
with respect to the von~Neumann equivalence relation:
\begin{equation}\label{Eq-EquivRelation}
p \sim p' \ \ \mbox{iff}  \ \ p=vv' \ {\rm and} \ p' = v'v, 
\end{equation}
for some partial isometries $v$ and $v'$ from $\Mm_\infty \otimes \Aa$. Recall that $\Mm_N$ is the algebra of $N \times N$ matrices with complex entries, above, $M_\infty$ is the direct limit of these algebras. The class of $p$ relative to \eqref{Eq-EquivRelation} is denoted by $[p]_0$.

\begin{remark}{\rm For any projection $p$ from $\Mm_\infty \otimes \Aa$ there exists $N \in \NM$ such that $p \in \Mm_N \otimes \Aa$. This property can be seen as a generalization of the fact that any compact projection on a Hilbert space is necessarily finite rank.}
$\Diamond$
\end{remark}

\begin{remark}{\rm There are two additional equivalence relations for projections \cite{Par}: $p \sim_u p'$ iff $p'= u p' u^\ast$ for some unitary element $u$ from $\Mm_\infty \otimes \Aa$, and $p \sim_h p'$ if $p$ and $p'$ can be connected by a projection homotopy in $\Mm_\infty \otimes \Aa$. In general, $\sim_h \Rightarrow \sim_u \rightarrow \sim$ but when the projections come from a stable algebra $\Aa'$, {\it i.e.} $\Mm_\infty \otimes \Aa' \simeq \Aa'$, which is the case for $\Aa' = \Mm_\infty \otimes \Aa$, the three equivalence relations coincide.  
}$\Diamond$
\end{remark}

If $p \in \Mm_N \otimes \Aa$ and $p' \in \Mm_M \otimes \Aa$, then $\begin{pmatrix} p & 0 \\ 0 & p' \end{pmatrix}$ is a projection from $\Mm_{N+M} \otimes \Aa$ and one can define the addition:
\begin{equation}
[p]_0 \oplus [p']_0 = \left [ \begin{matrix} p & 0 \\ 0 & p' \end{matrix} \right ]_0,
\end{equation}
which provides a semigroup structure on the set of equivalence classes. Then $K_0(\Aa)$ is its  enveloping group. For more information the reader can consult the standard textbooks \cite{Bla1,Par,RLL}, or \cite{PS} for an exposition within a condensed matter context.

\begin{example}\label{Ex-K0NC}{\rm The $K_0$-group of the non-commutative $n$-torus is simply:
\begin{equation}
K_0(\Aa_\Theta) = \ZM^{2^{n-1}},
\end{equation}
regardless of $\Theta$. Its generators $[p_J]_0$ can be uniquely labeled by the subsets $J \subseteq \{1,\ldots,n\}$ of even cardinality. This assures us that, for any projection $p$ from $\Mm_\infty \otimes \Aa$, one has:
\begin{equation}\label{Eq-GenExpansion}
[p]_0 = \bigoplus_{|J|={\rm even}} [e_J]_0 \oplus \ldots \oplus [e_J]_0 : = \sum_{|J|={\rm even}} c_J [e_J]_0, \quad c_J \in \ZM,
\end{equation} 
and the integer coefficients $c_J$ do not change as long as $p$ is deformed inside its class. In particular, two homotopic projections will display the same coefficients, hence $\{c_J\}_{|J|={\rm even}}$ can be regarded as the complete set of topological invariants associated to $p$ (better said, to the class of $p$).   
}$\Diamond$
\end{example}

The second group in complex $K$-theory is $K_1(\Aa)$, which classifies the unitary elements:
\begin{equation}
u \in \Mm_\infty \otimes \Aa, \quad u u^\ast = u^\ast u=1,
\end{equation}
with respect to the homotopy equivalence relation. The class of $u\in \Mm_\infty \otimes \Aa$ will be denoted by $[u]_1$. Again, complete information can be found in the standard textbooks \cite{Bla1,Par,RLL}, or \cite{PS} for an exposition within a condensed matter context.

\begin{example}\label{Ex-K1NC}{\rm The $K_1$-group of the non-commutative $n$-torus is simply:
\begin{equation}
K_1(\Aa_\Theta) = \ZM^{2^{n-1}},
\end{equation}
regardless of $\Theta$. Its generators $[u_J]_0$ can be uniquely labeled by the subsets $J \subseteq \{1,\ldots,n\}$ of odd cardinality. This assures us that, for any unitary $u$ from $\Mm_\infty \otimes \Aa$, one has:
\begin{equation}
[u]_1 = \bigoplus_{|J|={\rm even}} [u_J]_0 \odot \ldots \odot [e_J]_0 : =\sum_{|J|={\rm odd}} c_J [u_J]_0, \quad c_J \in \ZM,
\end{equation} 
and the integer coefficients $c_J$ do not change as long as $u$ is deformed inside its class. In particular, two homotopic unitaries will display the same coefficients, hence $\{c_J\}_{|J|={\rm odd}}$ can be regarded as the complete set of topological invariants associated to $u$ (better said, the class of $u$).   
}$\Diamond$
\end{example}

\subsection{Application: The gap labeling}\label{Sec-GapLabeling}

\begin{definition} A trace on $C^\ast$-algebra $\Aa$ is a positive linear map $\Tt: \Aa \rightarrow \CM$ such that $\Tt(a a') = \Tt(a' a)$. The trace is called faithful if $\Tt(a^\ast a) =0$ implies $a=0$. If the algebra has a unit, the traces are always normalized such that $\Tt(1)=1$.
\end{definition}

\begin{example}\label{Ex:Trace0}{\rm Let $\Omega$ be a compact topological space and $\mu$ a measure which is normalized $\mu(\Omega)=1$. Then:
\begin{equation}\label{Eq-TraceCOmega}
\Tt_0(f) = \int_\Omega {\rm d}\mu(\omega)\, f(\omega), \quad f \in C(\Omega),
\end{equation}
defines a trace on $C(\Omega)$. This trace can be extended to a trace on $C_N(\Omega)$ by tensoring with the ordinary trace on $\Mm_N$. Note that the resulting trace is not normalized but rather $\Tt_0(1) = N$. This choice is made because then $\Tt_0$ can be extended to a semi-finite trace on $\Mm_\infty \otimes C(\Omega)$, which is a non-unital algebra.
}$\Diamond$
\end{example}

Every classical dynamic system $(\Omega,\ZM^d,\tau)$ admits a normalized measure which is invariant and ergodic w.r.t $\tau$ \cite{EFHN}. In \ref{Def-MinimalSystems}, we mentioned the strictly ergodic dynamical systems which admit one and only one such measure. If $\theta$'s appearing in the concrete examples defined in section~\ref{Sec:Examples} are irrational, then all those dynamical systems are strictly ergodic and the unique measures are the Haar measures of the tori. In these cases, \eqref{Eq-TraceCOmega} provides the unique faithful trace on $C(\Omega)$ that is invariant to the dual action of $\ZM^d$. We continue our exposition assuming this context until specified otherwise.

\begin{proposition}[\cite{Dav}, p.~229] The invariant trace $\Tt_0$ on $C(\Omega)$ can be lifted to a normalized faithful trace on $\Aa_d=C(\Omega) \rtimes_\tau \ZM^d$ by:
\begin{equation}
\Tt\Big (\sum_q a_q u^q \Big ) = \Tt_0(a_0).
\end{equation}
It can be further extended to a trace on $\Mm_N \otimes \Aa_d$ as explained in \ref{Ex:Trace0}. 
\end{proposition}

\begin{proposition} The trace $\Tt$ computes the trace per volume of the physical observables:
\begin{equation}
\Tt(a) = \lim_{L \rightarrow \infty} \tfrac{1}{|V_L|} \sum_{n \in V_L} \langle n | \pi_\omega(a) | n \rangle, \quad V_L = \{-L,\ldots,L\}^d \subset \ZM^d.
\end{equation}
\end{proposition}

\proof It follows from a direct application of Birkhoff's ergodic theorem \cite{Bir} (see {\it e.g.} \cite{Bel1}). \qed

\begin{definition} Let $H$ be a self-adjoint Hamiltonian on $\CM^N \otimes \ell^2(\ZM^d)$ and $H_L$ be its canonical finite volume approximation $H_L = \Pi_L H \Pi_L^\ast$, where $\Pi_L$ is the standard isometry from $\ell^2(\ZM^d)$ to $\ell^2(V_L)$. Let $E_L(n)$ be the finite set of eigenvalues of $H_L$, counted with their multiplicities. Then the integrated density of states (IDS) at energy $E$ is defined as:
\begin{equation}\label{Eq-IDSDefinition}
IDS(E) = \lim_{L \rightarrow \infty} \frac{|\{E_L(n) \leq E\}|}{|V_L|}.
\end{equation}
\end{definition}

\begin{proposition}[\cite{Bel1}] Let $h$ be a self-adjoint element from $\Mm_N \otimes \Aa_d$ and let $E \notin {\rm Spec}(h)$. Then the IDS of $\pi_\omega(h)$ is independent of $\omega$ and given by:
\begin{equation}
IDS(E) = \Tt \big ( \chi(h \leq E) \big ),
\end{equation}
where $\chi(x \leq E)$ denotes the characteristic function of the interval $(-\infty,E]$.
\end{proposition} 

\begin{proposition}\label{Pro-KTrace} The trace $\Tt$ is constant over the $K_0$-classes. As a result, $\Tt$ defines a homomorphism between the $K_0$-group and the additive group of the real numbers.
\end{proposition}

\proof Recall that any projection from $\Mm_\infty \otimes \Aa_d$ actually belongs to $\Mm_N \otimes \Aa_d$, for some finite $N$. As a consequence, the extension of the trace $\Tt$ is finite on all projection from $\Mm_\infty \otimes \Aa_d$. Let now $p,p' \in \Mm_\infty \otimes \Aa_d$ and assume $p \sim p'$. Then:
\begin{equation}
\Tt(p) = \Tt(v v') = \Tt(v' v) = \Tt(p'),
\end{equation}
where we should mention again that both $v$ and $v'$ can be chosen inside the domain of the trace. \qed

\begin{proposition}[Gap Labeling \cite{Bel1}] Let $h \in \Mm_N \otimes \Aa_d$ be self-adjoint. Let $G$ be a spectral gap of $h$, that is, a connected component of $\RM \setminus {\rm Spec}(h)$. Per previous observation and Proposition~\ref{Pro-KTrace}, $IDS(G)$ is well defined for all $G$'s and depends entirely on the $K_0$-class of the spectral projector $p_G = \chi(h \leq G)$. The label:
\begin{equation}
G \rightarrow \Tt\Big ( [p_G]_0 \Big ) \in \RM
\end{equation}
is unique. 
\end{proposition}

\proof  Note that the set of gaps and their associated spectral projectors can be strictly ordered. Then:
\begin{equation}
 G > G' \Rightarrow \Tt\Big ( [p_G]_0 \Big ) > \Tt\Big ( [p_{G'}]_0 \Big ),
\end{equation}
because $\Tt$ is a faithful positive map. \qed

\begin{remark}{\rm If $h$ belongs to $\Mm_n \otimes \Aa_d$, then
\begin{equation}
0 \geq \Tt([p_G]_0) \leq \Tt([I_N \otimes 1]_0) = N.
\end{equation}
As such, the IDS, when evaluated on the gaps, takes values in the set:
\begin{equation}
\Tt \big ( K_0(C(\Omega \rtimes_\tau \ZM^d) \big ) \cap [0,N].
\end{equation}
For separable algebras, which is the case here, the $K_0$-groups are countable, hence the range of IDS is a countable subgroup of the real line. It is precisely this property which makes $C^\ast$-algebras so valuable in condensed matter physics \cite{Bel1}.
}$\Diamond$
\end{remark}

\begin{figure}[t]
\center
\includegraphics[width=0.8\textwidth]{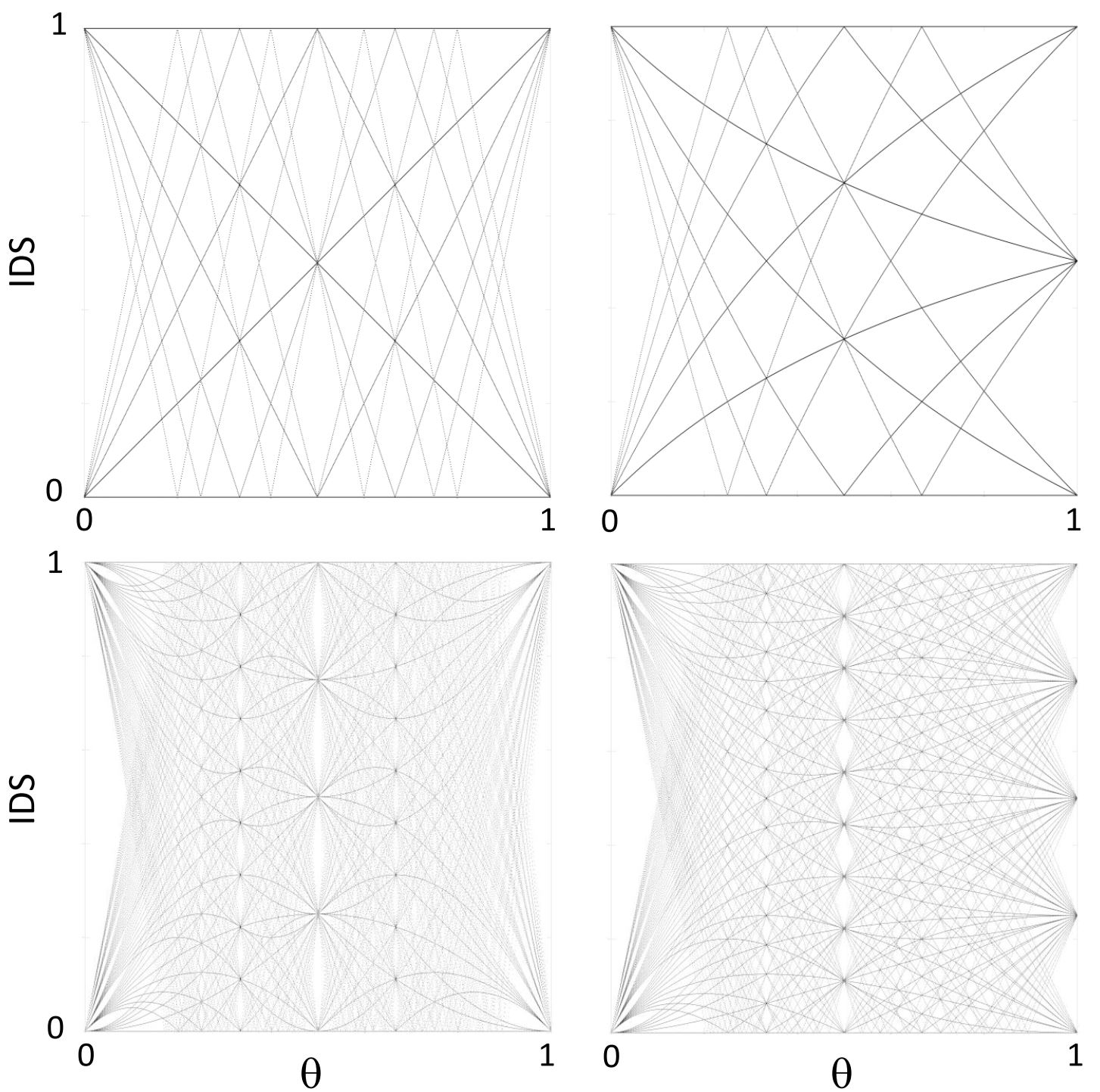}\\
  \caption{\small {\bf Predicted IDS}. IDS values inside the spectral gaps for \ref{ExampleI1} (upper-left), \ref{ExampleII1} (upper-right), \ref{ExampleIII1} (lower-left) and \ref{ExampleIV1} (lower-right). The predictions are based on the statements in Example~\ref{Ex-IDSRange}, where the integer coefficients have been independently varied from $-5$ to $+5$. The plots were rendered so that the curves corresponding to smaller integer coefficients appear darker.
  }
 \label{Fig-IDSPredictions}
\end{figure}

\begin{example}\label{Ex-IDSRange}{\rm For the non-commutative $n$-torus \cite{Ell}:
\begin{equation}
\Tt([p_J]_0) = {\rm Pfaff}(\Theta_J),
\end{equation}
where $\Theta_J$ is the anti-symmetric matrix obtained by restricting $\Theta$ to the indices contained in $J$, and ${\rm Pfaff}$ refers to the Pfaffian of an antisymmetric matrix. As a consequence:
\begin{equation}\label{Eq-IDSGeneric}
\Tt([p]_0) = \sum_{|J|={\rm even}} c_J \, {\rm Pfaff}(\Theta_J).
\end{equation}
This, together with the calculations in Examples~\ref{Ex-Calculation1}-\ref{Ex-Calculation4}, enable us to make the following predictions. When evaluated inside spectral gaps, the IDS of any model with $N$ internal degrees of freedom over the concrete patterns introduced in section~\ref{Sec:Examples} takes values in the following discrete yet dense subsets of $\RM$:
\begin{align*}
& \big \{n+m \theta, \ n,m \in \ZM \big \} \cap [0,N], \ & (\mbox{pattern \ref{ExampleI1}}) \\
& \Big \{n+\frac{m \theta}{1+\theta}, \ n,m \in \ZM \Big \} \cap [0,N], \ & (\mbox {pattern \ref{ExampleII1}})\\
& \big \{n+m \theta_1+k\theta_2 + l \theta_1 \theta_2, \ n,m,k,l \in \ZM \big \} \cap [0,N], \ & (\mbox{pattern \ref{ExampleIII1}}) \\
& \Big \{n+\frac{m \theta_1}{1+\theta_1}+\frac{k \theta_2}{1+\theta_2} +  \frac{l \theta_1 \theta_2}{(1+\theta_1)(1+\theta_2)}, \ n,m,k,l \in \ZM \Big \} \cap [0,N]. \ & (\mbox{pattern \ref{ExampleIV1}})
\end{align*}
The integers appearing in these equations are precisely the coefficients $c_J$ in \eqref{Eq-GenExpansion} or \eqref{Eq-IDSGeneric}, hence they also provide unique labels for the spectral gaps. When, $\Theta$'s are varied and the integer labels are kept constant, the allowed values of the IDS trace specific curves, some of which are displayed in Fig.~\ref{Fig-IDSPredictions}.
}$\Diamond$
\end{example}

\subsection{Numerical examples}\label{Sec-NumericalExamples1}

The numerical calculations are performed for rational $\theta$'s, which enable us to use periodic boundary conditions without introducing defects. Before proceeding with the analysis, we need to address these cases first. The discussion below refers only to the concrete examples introduced in section~\ref{Sec:Examples}. As we shall see later, some of the statements fail if taken from this context.

\vspace{0.2cm}

\begin{figure}[t]
\center
\includegraphics[width=\textwidth]{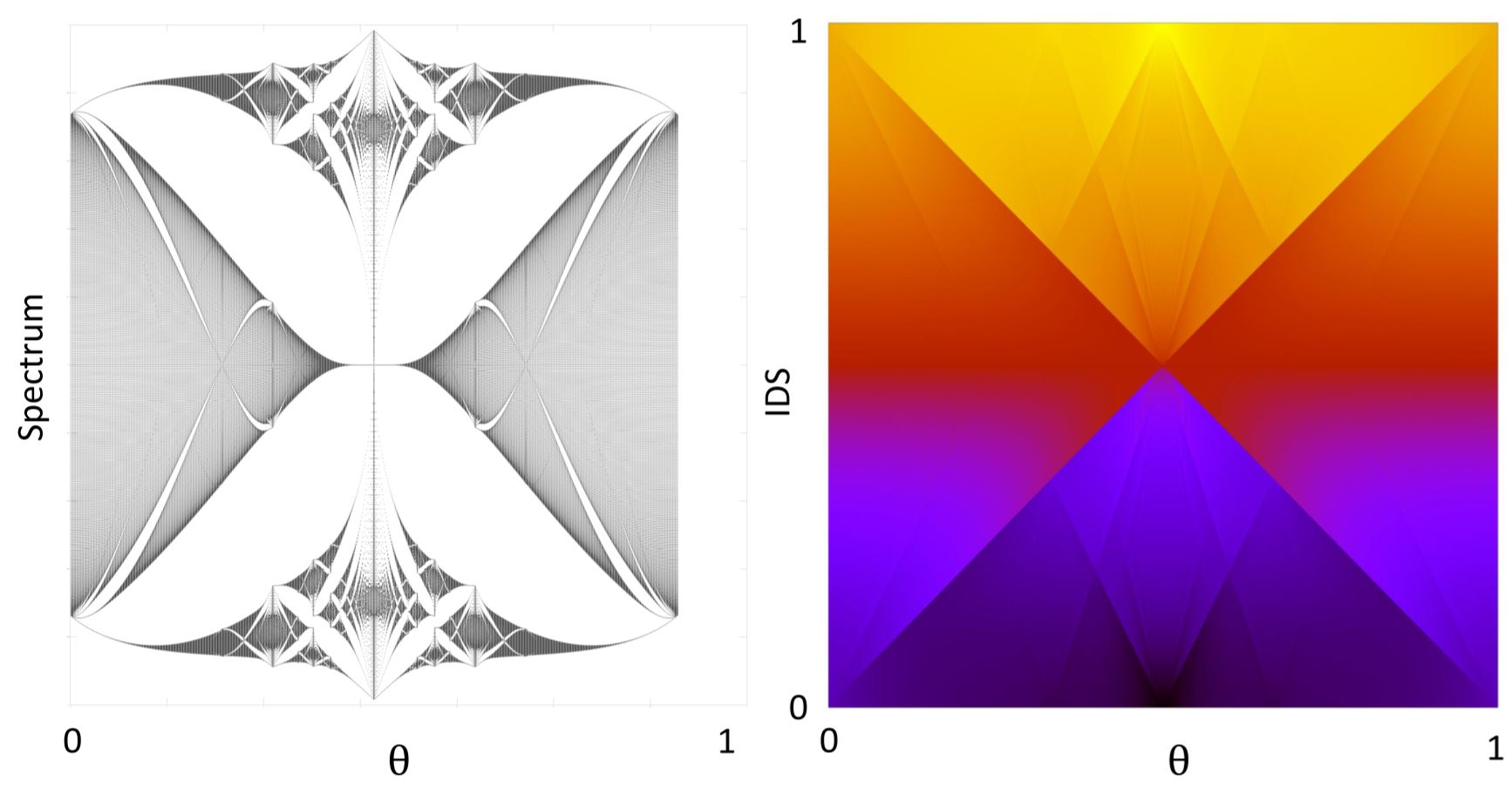}\\
  \caption{\small {\bf Gap labeling for \ref{ExampleI1}.} The spectral (left) and IDS (right) butterflies for the pattern \ref{ExampleI1} ($r=0.4$) and Hamiltonian \eqref{Eq-ModelHam}. The computations were carried on a finite pattern of $840$ sites.
  }
 \label{Fig-ButterflyFigures_Ex1}
\end{figure}

The algebras of physical observables for these models are all isomorphic to the non-commutative torus $\Aa_\Theta$. Let us start by stating that the collection of $\Aa_\Theta$'s, when the entries of $\Theta$ are varied in $[0,1]$, forms a field of continuous $C^\ast$-algebras (for definition, see {\it e.g.} \cite{Dix}[Ch.~10]), where the space $\Gamma$ of continuous fields is spanned by:
\begin{equation}\label{Eq-ThetaElement}
\Aa_\Theta \ni a(\Theta) = \sum_{\bm q} a_{\bm q}(\Theta) u^{\bm q}(\Theta), \quad a_{\bm q}(\Theta) \in \CM,
\end{equation}
with the coefficients $a_{\bm q}(\Theta)$ continuous of $\Theta$. If $\{h(\Theta)\} \in \Gamma$ is a continuous self-adjoint field, then the spectrum $\sigma\big ( h(\Theta) \big )$ is continuous of $\Theta$ in the Housdorff metric on the compact subsets of the real line \cite{Dix}[Proposition~10.3.6] (see also \cite{BB2016}). As a consequence, the spectra of the models at irrational $\Theta$'s can be approximated with arbitrary precision by calculations at rational values. Furthermore, we can continue to use the Haar measure of the tori and generate canonical invariant traces for the crossed products at rational $\theta$'s. Under the isomorphisms to the non-commutative tori, described in \ref{SubSec-Concrete}, this trace becomes:
\begin{equation}
\Tt\big ( a(\Theta) \big ) = a_{\bm 0}(\Theta).
\end{equation}
For a Hamiltonian $h(\Theta)$ as in \eqref{Eq-ThetaElement}, if $E$ is in a gap of $h(\Theta)$ then it will stay in a gap for small variations of $\Theta$. Hence, the spectral projectors $\chi \big ( h(\Theta) \leq E \big )$ are always locally defined and they generate local continuous fields \cite{Dix}[Proposition~10.3.3]. As such, their Fourier coefficients are locally continuous and:
\begin{equation}\label{Eq-IDSTheta}
IDS(E,\Theta)=\Tt \Big ( \chi \big ( h(\Theta) \leq E\big ) \Big )
\end{equation}
is a locally continuous function of $\Theta$ when $E$ is in a gap. The conclusion is that IDS, when evaluate inside the spectral gaps, can be approximated with arbitrary precision using rational $\Theta$'s.

\vspace{0.2cm}

\begin{figure}
\center
\includegraphics[width=\textwidth]{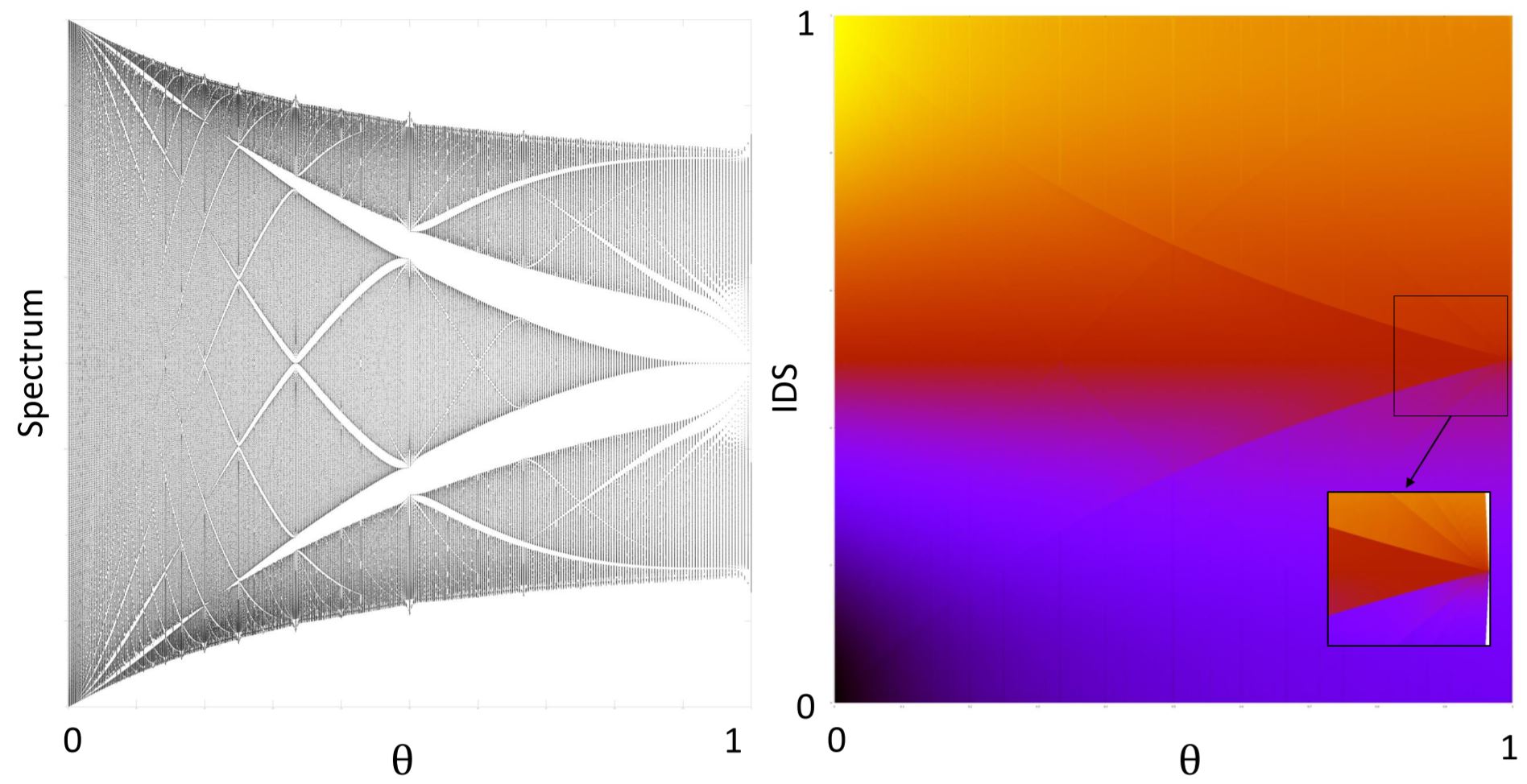}\\
  \caption{\small {\bf Gap labeling for \ref{ExampleII1}.} The spectral (left) and IDS (right) butterflies for the pattern \ref{ExampleII1} ($g=0.01$, $\delta=0.2$) and Hamiltonian \eqref{Eq-ModelHam}. The computations were carried on a finite pattern of $840$ sites.
  }
 \label{Fig-ButterflyFigures_Ex2}
\end{figure}

\begin{figure}
\center
\includegraphics[width=\textwidth]{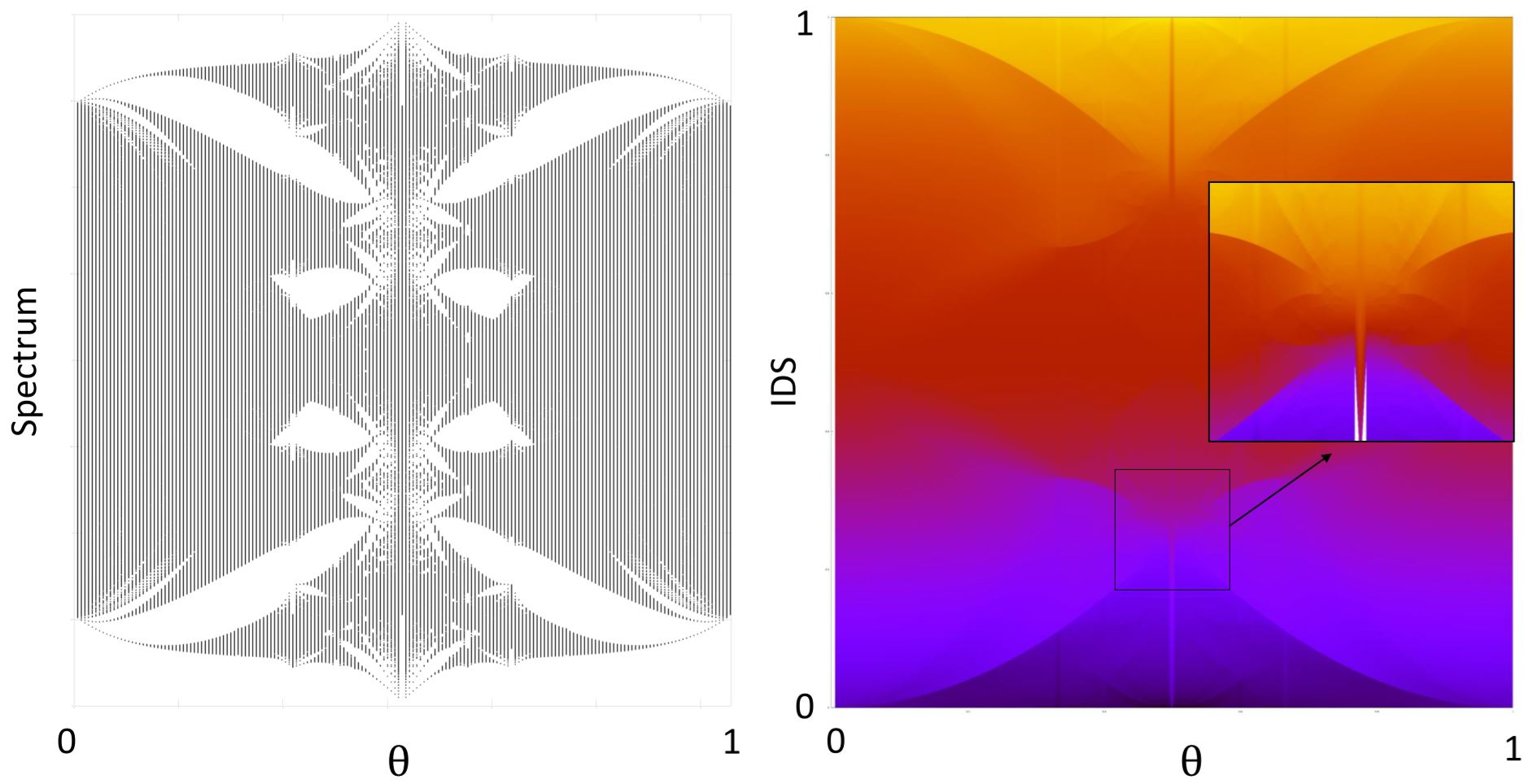}\\
  \caption{\small {\bf Gap labeling for \ref{ExampleIII1}.} The spectral (left) and IDS (right) butterflies for the pattern \ref{ExampleIII1} ($r=0.4$, $\theta_1=\theta_2=\theta$) and Hamiltonian \eqref{Eq-ModelHam}. The computations were carried on a finite pattern of $180 \times 180$ sites.}
 \label{Fig-ButterflyFigures_Ex3}
\end{figure}

Plots of the spectrum of a model against $\Theta$ are best at revealing the spectral complexity. We refer to this kind of plots as spectral butterflies, in analogy with the Hofstadter butterfly \cite{Hof} describing the energy spectrum of electrons on a lattice in a magnetic field. Per the above discussion, the spectral and the IDS butterflies can be represented with arbitrary precision even if we sample only rational values of $\Theta$. For all four patterns \ref{ExampleI1}-\ref{ExampleIV1}, we chose to work with the Hamiltonian:
\begin{equation}\label{Eq-ModelHam}
h=\sum_{\bm q} w_{\bm q} u_{\bm q} \in \Aa_d, \quad w_{\bm q}(\omega)=e^{-|p_{\bm 0}(\omega) - p_{-\bm q}(\omega)|},
\end{equation}
where $|\cdot |$ represents the Euclidean distance. The physical representation of $h$ reads:
\begin{equation}\label{Eq:HamChoice}
\pi_\omega(h) = \sum_{\bm q} \sum_{\bm n \in \ZM^d} e^{-|p_{\bm n}(\omega)-p_{\bm n-\bm q}(\omega)|} \, |\bm n\rangle \langle \bm n- \bm q | \in \BM\big ( \ell^2(\ZM^d) \big ),
\end{equation}  
which is the Hamiltonian for a pattern of single-state resonators coupled via evanescent tails, as discussed in chapter~\ref{Ch:PattRez}. Clearly, $w_q$'s are continuous functions of $\theta$'s. The spectral butterflies are generated by exact diagonalization of $\pi_\omega(h)$ and, since we work at rational $\theta$'s, we must sample $\Omega$ in order to obtain an accurate representation of ${\rm Spec}(h)$ (see \eqref{Eq-OmegaSpectrum}). Also, \eqref{Eq-IDSTheta} can be computed numerically by averaging \eqref{Eq-IDSDefinition} over $\pi_\omega(h)$'s. The results and the details of the computations are reported in Figs.~\ref{Fig-ButterflyFigures_Ex1}-\ref{Fig-ButterflyFigures_Ex3}. They indeed confirm the theoretical predictions summarized in Fig.~\ref{Fig-IDSPredictions}. Unfortunately the point pattern \ref{ExampleIV1} did not produced any spectral gap, hence the IDS could not be computed.

\section{Algebras of the half-space and boundary physical observables}
\label{Ch:BBCorrespondence}

In this chapter, we examine the situation when all the couplings between the resonators with $n_d \geq 0$ and those with $n_d <0$ are turned off. Concentrating on one half of such system, we need to study Hamiltonians of the type:
\begin{equation}\label{Eq-LatticeHamiltonian0}
H(\Pp) = \sum_{\bm n,\bm m\in \ZM^d}^{n_d, m_d \geq 0} h_{\bm n,\bm m}(\Pp) \otimes |\bm n \rangle \langle \bm m |,
\end{equation}  
plus an additional boundary term, which takes into account relaxation or reconstruction processes which inherently occur when a sample is halved. The $C^\ast$-algebraic framework for this problem was developed in \cite{KRS} but here we mainly follow \cite{PS}.

\subsection{Definitions and physical representations}

We collect here the basic and well known constructs and statements, which are the algebras of half-space and boundary physical observables and the exact $C^\ast$-algebra sequence which connects them. Particular care will be devoted on the relation between the spectra of the algebra elements and their physical representations. Another important concept introduced here is that of boundary spectrum.

\begin{definition}[\cite{PS}] The algebra of physical observables for the half-space is defined as the universal algebra:
\begin{equation}
\widehat \Aa_d = C^\ast \big ( C(\Omega),\hat u_1, \ldots, \hat u_d \big )
\end{equation}
generated by a copy of $C(\Omega)$, $d-1$ commuting unitary elements:
\begin{equation}
\hat u_j \hat u_j^\ast = \hat u_j^\ast \hat u_j = 1 , \quad \hat u_i \hat u_j = \hat u_j \hat u_i, \quad i,j = 1,\ldots, d-1,
\end{equation}
and by a partial isometry that commutes with the unitary elements:
\begin{equation}
\hat u_d^\ast \hat u_d = 1, \quad \hat u_d \hat u_d^\ast = 1-\hat p, \quad \hat u_d \hat u_j = \hat u_j \hat u_d, \quad j = 1,\ldots,d-1,
\end{equation}
where $\hat p$ is a proper projection ($\hat p^2=\hat p^\ast=\hat p \neq 1$). The remaining relations of the algebra are:
\begin{equation}
f \hat p = \hat p f, \quad f\, \hat u_j = \hat u_j \, (f \circ \tau_{\bm e_j}), \quad f \hat u_j^\ast = \hat u_j^\ast (f \circ \tau_{-\bm e_j}), \quad j=1,\ldots,d,
\end{equation}
where $\bm e_j$ is the $j$-th generator of $\ZM^d$.
\end{definition}

\begin{remark}{\rm A generic element of the algebra can be presented uniquely in the form:
\begin{equation}\label{Eq-HalfElementPresentation}
\hat a = \sum_{n,m \in \NM} \hat a_{nm} \, \hat u_d^n (\hat u_d^\ast)^m, \quad \hat a_{nm} \in C(\Omega) \rtimes_\tau \ZM^{d-1}.
\end{equation} 
The above sum must converge in norm, hence the norm of the elements $\hat a_{nm}$ in the appropriate algebra must display a certain decay w.r.t. $m$ and $n$. In particular, every element of $\widehat \Aa_d$ can be approximated by a polynomial for which $n$ and $m$ takes finite values. This issue is further examined in Remarks~\ref{Re-TildeDecay} and \ref{Re-HalfSpaceDecay}.
}$\Diamond$
\end{remark}

\begin{definition} The algebra $\widetilde \Aa_d$ of boundary physical observables is defined as the proper two-sided ideal of $\widehat \Aa_d$ generated by $\hat p$: 
\begin{equation}
\widetilde A_d = \widehat \Aa_d \, \hat p \, \widehat \Aa_d = \{\hat a \hat p \hat a' \ | \ \hat a,\hat a' \in \widehat \Aa_d \}.
\end{equation}
\end{definition}

\begin{remark}{\rm A generic element of $\widetilde \Aa_d$ can be presented uniquely in the form:
\begin{equation}\label{Eq-BoundaryElementPresentation}
\tilde a = \sum_{n,m \in \NM} \tilde a_{nm} \hat u_d^n \, \hat p \, (\hat u_d^\ast)^m, \quad \tilde a_{nm} \in C(\Omega) \rtimes_\tau \ZM^{d-1},
\end{equation}
where, again, the sum must converge in norm. As such, the norm of the elements $\tilde a_{nm}$ must display a certain decay w.r.t. $m$ and $n$, in particular, any element of $\widetilde \Aa_d$ can be approximated by polynomials. This is further discussed in Remark~\ref{Re-TildeDecay}.
}$\Diamond$
\end{remark}

\begin{proposition}[\cite{PS}]\label{Pro-BoundaryChar}
Let $C(\Omega) \rtimes_\tau \ZM^{d-1}$ be regarded as a sub-algebra of $C(\Omega) \rtimes_\tau \ZM^d$. By examining the relations in Definition~\ref{Def-BulkAlgebra}, one can see that $C(\Omega) \rtimes_\tau \ZM^{d-1}$ is invariant under the conjugation by $u_d$, hence we can define the automorphism $\alpha_d(\tilde a) = u_d \tilde a u_d^\ast$. Then the map:
\begin{equation}\label{RhoEe}
\widetilde \rho: \widetilde \Aa_d \to \Aa_{d-1}\otimes \KM\big (\ell^2(\NM)\big ), \quad \widetilde \rho(\tilde{a})
\;=\;
\sum_{n,m\geq 0}
\alpha_d^{-n}\big(\tilde{a}_{nm}\big)
\otimes|n\rangle\langle m|
\;,
\end{equation}
is a C$^\ast$-algebra isomorphism. Here, $\KM\big (\ell^2(\NM)\big )$ denotes the algebra of compact operators over $\ell^2(\NM)$.
\end{proposition}

\begin{remark}\label{Re-TildeDecay}{\rm The above statement is important for two reasons. First, since any compact operator can be approximated in norm by a matrix, we can see more explicitly that each element from $\widetilde \Aa_d$ can be approximated in norm by elements for which $n$ and $m$ in the expansion \eqref{Eq-BoundaryElementPresentation} take finite values. Secondly, since $\KM\big (\ell^2(\NM)\big ) \simeq \Mm_\infty$, it follows that the $K$-theory of the boundary algebra $\widetilde \Aa_d$ coincides with the $K$-theory of $C(\Omega) \rtimes_\tau \ZM^{d-1}$. As such, tor the concrete models introduced in section~\ref{Sec:Examples}, the $K$-theory can be described without any additional effort. Indeed, for all cases, the bulk algebra $\Aa_d$ was found in \ref{SubSec-Concrete} to be isomorphic to the non-commutative $n$-torus $\Aa_\Theta$, with $n$ taking appropriate values. If $\widetilde \Theta$ represents the restriction of $\Theta$ to the indices $\{1,\ldots,n-1\}$, then:
\begin{equation}
K_{\alpha}(\widetilde A_d) \simeq K_{\alpha} (\Aa_{\widetilde \Theta}), \quad \alpha=0,1,
\end{equation}
and the latter have been fully characterized in \ref{Ex-K0NC} and \ref{Ex-K1NC}. Specifically, the generators of the $K_{0,1} (\widetilde \Aa_d)\simeq \ZM^{2^{d-1}}$ groups can be chosen to be $[\tilde p_J]_0$ and $[\tilde u_J]_1$, respectively, with $J \subset {1,\ldots,n-1}$ and $|J|$ of appropriate parity. Furthermore, since $\Aa_{\widetilde \Theta} \subset \Aa_\Theta$, these generators can be chosen to coincide with $[p_J]_0$ and $[u_J]_1$ in examples \ref{Ex-K0NC} and \ref{Ex-K1NC}.
}$\Diamond$
\end{remark}

\begin{proposition}
Let $i:\widetilde \Aa_{d} \hookrightarrow \widehat{\Aa}_{d}$ be the embedding homomorphism and $\mbox{\rm ev}:\widehat{\Aa}_{d}\to {\Aa}_{d}$ be the canonical surjective C$^\ast$-algebra homomorphism:
$$
\mbox{\rm ev}(\phi)=\phi\;,
\qquad
\mbox{\rm ev}(\hat{u}_j)=u_j
\;, 
\qquad 
\mbox{\rm ev}(\hat{u}_j^\ast)=u_j^\ast\,,  
$$ 
for $j=1,\ldots d$. Then necessarily $\mbox{\rm ev}(\hat{p})=0$ so that:
\begin{equation}
\label{Eq-ExactSequence}
\begin{diagram}
0 &\rTo &\widetilde \Aa_{d}   &\rTo{i}  &\widehat{\mathcal A}_{d}  &\rTo{ \mathrm{ev}}  &\mathcal A_{d} &\rTo &0
\end{diagram}
\end{equation}
is an exact sequence of C$^\ast$-algebras. 
\end{proposition}

\begin{remark}\label{Re-HalfSpaceDecay}{\rm The above exact sequence of $C^\ast$-algebras is at the heart of bulk-boundary principle. The following identities are direct consequences of the defining relations:
\begin{equation}
\hat{u}_d^n (\hat{u}_d^*)^m
\;=\;
\left\{
\begin{array}{ll} 
\hat{u}_d^{n-m}\big(\one-\sum_{l=0}^{m-1} (\hat u_{d})^l\hat{e}(\hat{u}_{d}^*)^l\big)\;, & \;\;\;\;n\geq m\;,\medskip
\\
\big(\one-\sum_{l=0}^{n-1} (\hat u_{d})^l\hat{e}(\hat{u}_{d}^*)^l\big)(\hat{u}_d^*)^{m-n}\;, & \;\;\;\; n\leq m\;.
\end{array}
\right.
\end{equation}
from where one can see that the sequence \eqref{Eq-ExactSequence} is split as a sequence between linear spaces. The linear splitting map $i': \Aa_d \rightarrow \widehat \Aa_d$ can be defined explicitly by the following action on the monomials:
\begin{equation}\label{Eq-iprime}
i'\big(f \, u^{\bm n} \big )
\;=\;
\left \{
\begin{array}{ll}
f \, \hat u_1^{n_1}\cdots \hat u_{d-1}^{n_{d-1}}\hat u_d^{n_d}, \;\; & \mathrm{if} \ n_d \geq 0,\medskip \\
f \,\hat u_1^{n_1}\cdots \hat u_{d-1}^{n_{d-1}}(\hat u_d^\ast)^{|n_d|},\;\; & \mathrm{if} \ n_d < 0.
\end{array} \right .
\end{equation}
Then each element from $\widehat \Aa_d$ can be written uniquely as:
\begin{equation} 
\hat a = i'(a) + \tilde a, \quad a = {\rm ev}(\hat a) \in \Aa_d, \quad \tilde a = \hat a - (i' \circ {\rm ev})(\hat a) \in \widetilde \Aa_d.
\end{equation} 
Among other things, this splitting together with Remarks~\ref{Re-BulkDecay} and \ref{Re-TildeDecay} completely characterize the behavior of the coefficients $\hat p_{nm}$ w.r.t. $m$ and $n$, in expansion \eqref{Eq-BoundaryElementPresentation}. It will also enable us to define the boundary spectrum in a precise way. 
}$\Diamond$
\end{remark}

\begin{proposition}\label{Pro-HalfSpaceRep} The algebra $\widehat \Aa_d$ accepts a continuous field of canonical $\ast$-representations $\{\hat \pi_\omega\}_{\omega \in \Omega}$ on $\ell^2(\ZM^{d-1} \times \NM)$:
\begin{equation}
 \hat \pi_\omega(\hat u_j) = \widehat U_{\bm e_j},\quad \hat \pi_\omega(\hat u_j^\ast) = \widehat U_{\bm e_j}^\ast, \quad  \widehat U_{\bm e_j}=\Pi^\ast \, U_{\bm e_j} \Pi , \quad j = 1, \ldots , d,
 \end{equation}
 where $\Pi$ is the partial isometry from $\ell^2(\ZM^{d-1} \times \NM)$ to $\ell^2(\ZM^d)$, and:
 \begin{equation} 
 \hat \pi_\omega(f) = \sum_{\bm n \in \ZM^{d-1} \times \NM} f\big (\tau_{\bm n}(\omega)\big)\, |\bm n \rangle \langle \bm n |, \quad f \in C(\Omega).
\end{equation}
Under these representations, the projection $\hat p$ is mapped onto: 
\begin{equation}
\hat \pi_\omega(\hat p) = \sum_{\bm k \in \ZM^{d-1}} |\bm k,0 \rangle \langle \bm k,0 |.
\end{equation}
The continuity of the canonical field of representations is again w.r.t. the strong topology on $\BM\big (\ell^2(\ZM^{d-1} \times \NM)\big )$.
\end{proposition}

\proof Checking that the representations preserve the relations between the generators is a straightforward task (see \cite{PS}[p.~67]), and continuity follows from the same argument as in Proposition~\ref{Pro-BulkRep}.\qed

\begin{remark}{\rm These canonical representations can be trivially extended to the algebra $\Mm_N \otimes \widehat \Aa_d$ as already explained.}
$\Diamond$
\end{remark}

\begin{corollary}[\cite{RS}(p.~291)] Let $\hat h \in \Mm_N \otimes \widehat \Aa_d$ be self-adjoint, $\hat h^\ast=\hat h$. Then the spectrum ${\rm Spec}\big (\hat \pi_\omega(\hat h) \big )$ of the bounded operator $\hat \pi_\omega(\hat h)$ on $\CM^N \otimes \ell^2(\ZM^{d-1} \times \NM)$ belongs to the real axis and is semi-continuous w.r.t. $\omega$.
\end{corollary}

\begin{remark}{\rm The representations introduced in Proposition~\ref{Pro-HalfSpaceRep} are easier to comprehend if one uses the linear splitting map. Indeed:
\begin{equation}
\hat \pi_\omega \big ( i'(a) \big ) = \Pi \pi_\omega(a) \Pi,
\end{equation}
which is just the restriction of $\pi_\omega(a)$ to the half-space with Dirichlet condition at the boundary. Furthermore:
\begin{equation}\label{Eq-BoundaryRep}
\hat \pi_\omega(\tilde a) = \sum_{\bm q,\bm k \in \ZM^{d-1}} \sum_{n,m \in \NM}  \tilde a_{\bm q;nm}(\tau_{\bm k,n}\omega)|\bm k,n\rangle \langle \bm k-\bm q,m|, 
\end{equation}
which is an operator localized near the boundary. Thus, if $\hat a$ is just of the form $i'(a)$, then its physical representation is just that of $\pi_\omega(a)$ with Dirichlet condition at the boundary. But if $\hat a$ contains the additional term $\tilde a$, then the condition at the boundary is modified and can be virtually any allowed boundary condition. As such, a generic element $\hat a$ with ${\rm ev}(\hat a) =a$ generates a physical model which is just the restriction of $\pi_\omega(a)$ on the half-space with a generic boundary conditions.
}$\Diamond$
\end{remark}

\begin{proposition} Let:
\begin{equation}
\widehat U_{\bm n} = \widehat U_{\bm e_1}^{n_1} \ldots \widehat U_{\bm e_d}^{n_d}, \quad \bm n \in \ZM^d, \quad n_d \geq 0.
\end{equation}
Then the field of canonical representations obey the following covariant property:
\begin{equation}\label{Eq-HalfSpaceCovariance}
\hat U_{\bm n}^\ast \, \hat\pi_\omega(\hat a) \, \widehat U_{\bm n} = \hat \pi_{\tau_{\bm n} \omega}(\hat a), \quad \hat a \in \widehat \Aa_d, \quad \bm n \in \ZM^d, \quad n_d \geq 0.
\end{equation}
\end{proposition}

\proof It is, again, enough to verify the statement for the generators. The covariance relation is trivial for $\hat u_j$, $j=1,\ldots,d-1$, because all $\widehat U_{e_j}$, $j=1,\ldots,d-1$, commute with $\widehat U_{\bm n}$. For the isometry $\hat u_d$:
\begin{equation}
\hat \pi_\omega(\hat u_d) = \hat \pi_{\tau_{\bm n}\omega}(\hat u_d) = \widehat U_{\bm e_d}, \quad \hat \pi_\omega(\hat u_d^\ast) = \hat \pi_{\tau_{\bm n}\omega}(\hat u_d^\ast) = \widehat U_{\bm e_d}^\ast,
\end{equation}
and:
\begin{equation}
\hat U_{\bm n}^\ast \, \widehat U_{\bm e_d} \, \widehat U_{\bm n} = \big (\widehat U_{\bm e_d}^\ast\big )^{n_d} \, \widehat U_{\bm e_d} \, \widehat U_{\bm e_d}^{n_d} = \widehat U_{\bm e_d}.
\end{equation}
By conjugation, we also find that:
\begin{equation}
\hat U_{\bm n}^\ast \, \widehat U_{\bm e_d}^\ast \, \widehat U_{\bm n}=\widehat U_{\bm e_d}^\ast.
\end{equation}
Lastly, for $f \in C(\Omega)$:
\begin{equation}
\widehat U_{\bm n}^\ast \, \hat \pi_\omega(f) \widehat U_{\bm n} = \hat \pi_{\omega}(\hat u_{\bm n}^\ast \, f \, \hat u_{\bm n})=\hat \pi_{\omega}(f \circ \tau_{\bm n}) = \hat \pi_{\tau_{\bm n}\omega}(f),
\end{equation}
 which is the desired result.\qed
 
 \begin{remark}{\rm Note that the covariant relation \eqref{Eq-HalfSpaceCovariance} fails if $n_d$ is allowed to take negative values. Indeed, while $ f \, \hat u_d = \hat u_d \, (f \circ \tau_{\bm e_d})$ can be multiplied by $\hat u_d^\ast$ to the left to obtain $ \hat u_d^\ast \, f \, \hat u_d = f \circ \tau_{\bm e_d}$, this is not the case for the commutation relation $ f \, \hat u_d^\ast = \hat u_d^\ast \, (f \circ \tau_{-\bm e_d})$, because $\hat u_d \, \hat u_d^\ast \neq 1$.
 }$\Diamond$
 \end{remark}

\subsection{Spectral properties}

We establish here the connection between the spectral properties of the elements from the half-space algebra and of their physical representations. But first a central definition.

\begin{definition} Given the second homomorphism in \eqref{Eq-ExactSequence}, we can automatically conclude that ${\rm Spec}(h) \subseteq {\rm Spec}(\hat h)$ for any pair $(h,\hat h)$ such that ${\rm ev}(\hat h) = h$. The boundary spectrum is defined as the excess spectrum of $\hat h$ relative to $h$:
\begin{equation}
{\rm Spec}_B(\hat h) = {\rm Spec}(\hat h) \setminus {\rm Spec}(h).
\end{equation}
The emergence of such extra spectrum can be rightfully attributed to the presence of a boundary.
\end{definition}

\begin{proposition} Let $\hat a \in \widehat \Mm_N \otimes \Aa_d$. Then:
\begin{equation}\label{Eq-HalfSpaceOmegaSpectrum}
{\rm Spec}(\hat a) = \bigcup_{\omega \in \Omega} {\rm Spec}\big ( \hat \pi_\omega(\hat a) \big ),
\end{equation}
where ${\rm Spec}\big ( \hat \pi_\omega(\hat a) \big )$ is the spectrum of $\hat \pi_\omega(\hat a)$ in the algebra of bounded operators over $\CM^N \otimes \ell^2(\ZM^{d-1} \times \NM)$.
\end{proposition}

\proof The representation $\bigoplus_{\omega \in \Omega} \hat \pi_\omega$ is unital and faithful, which can be verified explicitly on the non-commutative polynomials.\qed

\begin{proposition} If $(\Omega,\ZM^d,\tau)$ is minimal, then:
\begin{equation}\label{Eq-HalfSpaceOmegaSpectrumP}
{\rm Spec}(\hat h) = \overline{\bigcup_{k \in \NM} {\rm Spec}\big ( \hat \pi_{\tau_{\bm 0,k}\omega}(\hat h) \big )},
\end{equation}
for any self-adjoint element $\hat h \in \Mm_N \otimes \widehat \Aa_d$.
\end{proposition}

\proof First let us observe that, in contradistinction to the bulk representations $\pi_\omega$, the half-space representations $\hat \pi_\omega$ are not faithful even if the dynamical system is minimal (see {\it e.g.} \eqref{Eq-BoundaryRep}). However, since the spectrum of $\hat \pi_\omega(\hat h)$ is semi-continuous w.r.t. $\omega$ and the orbits are dense in $\Omega$, we can write:
\begin{equation}
\bigcup_{\omega \in \Omega} {\rm Spec}\big ( \hat \pi_\omega(\hat h) \big ) = \overline{ \bigcup_{\bm n \in \ZM^{d-1} \times \NM} {\rm Spec}\big ( \hat \pi_{\tau_{\bm n}\omega}(\hat h) \big )}.
\end{equation}
Using the covariance property \eqref{Eq-HalfSpaceCovariance}, we see that $\hat \pi_{\tau_{\bm n}\omega}(\hat h)$ and $\hat \pi_{\tau_{\bm 0,n_d}\omega}(\hat h)$ are connected by a unitary conjugation. The affirmation follows.\qed

\begin{remark}\label{Re-BoundarySpec}{\rm From the practical point of view, it is important to notice that the Hamiltonian $\widehat H_{\tau_{\bm 0,k}\omega} = \hat \pi_{\tau_{\bm 0,k}\omega}(\hat h)$ is unitarilly equivalent to the half-space Hamiltonian obtained from the bulk Hamiltonian $H_\omega = \pi_\omega(h)$ but with the boundary moved from $n_d=0$ to $n_d=k$.
}$\Diamond$
\end{remark}

\section{The $K$-theoretic bulk-boundary principle}

\subsection{The engine of the bulk-boundary principle}

The exact sequence \eqref{Eq-ExactSequence} sets in motion the following 6-term exact sequence at the level of $K$-theory \cite{Bla1,RLL,WO}:
\begin{equation}\label{Eq-SixTermDiagram}
\begin{diagram}
& K_0(\widetilde \Aa_d) & \rTo{ i_\ast \ \ \ } & K_0(\widehat \Aa_d) & \rTo{\ \ {\rm ev}_\ast \ \ } & K_0(\Aa_d) &\\
& \uTo{\rm Ind} & \  &  \ & \ & \dTo{\rm Exp} & \\
& K_1(\Aa_d)  & \lTo{{\rm ev}_\ast} & K_1(\widehat \Aa_d) & \lTo{\ \ \ i_\ast} & K_1(\widetilde \Aa_d) &
\end{diagram}
\end{equation}
For the applications considered in the present work, only the right side of this diagram is important. The definition of the connecting maps can be found in the standard textbooks and here we follow \cite{KRS,PS}, since our task is to formulate the exponential connecting map in terms of our input data, which is the pair $(h,\hat h)$ of bulk and half-space Hamiltonians.

\begin{proposition}[\cite{KRS,PS}] Let $h \in \Mm_N \otimes \Aa_d$ be a self-adjoint element and assume ${\rm Spec}(h)$ displays a spectral gap $G$. Then the exponential connecting map in \eqref{Eq-SixTermDiagram} acts on the class of the spectral projector $p_G = \chi(h \leq G)$ in $K_0(\Aa_d)$ in the following way:
\begin{equation}\label{Eq-ExponentialMap}
{\rm Exp}\big ([p_G]_0 \big ) = \big [ e^{2 \pi i \Phi(\hat h)}]_1 \in K_1(\widetilde \Aa_d), \quad {\rm ev}(\hat h) =h.
\end{equation}
where $\Phi: \RM \rightarrow \RM_+$ is any continuous non-increasing map such that $\Phi =1/0$ below/above the spectral gap $G$. 
\end{proposition} 

\begin{remark}{\rm The reader should appreciate the generality of this construction, more precisely, that $\hat h$ in \eqref{Eq-ExponentialMap} can be generated with any boundary term $\tilde h \in \widetilde \Aa_d$: $\hat h = i'(h) + \tilde h$, and that the domain where $\Phi$ display variations can be any open sub-interval of $G$.
}$\Diamond$
\end{remark}

\begin{corollary}\label{Cor-BBPrinciple} Let $h \in \Aa_d$ as above. Assume that ${\rm Exp}\big ( [p_G]_0\big )$ differs from the class of the (adjoined) identity in $K_1(\widetilde \Aa_d)$. Then:
\begin{equation}
{\rm Spec}_B(\hat h) \cap G = G,
\end{equation}
for any $\hat h \in \widehat \Aa_d$, ${\rm ev}(\hat h) = h$.
\end{corollary}

\proof Suppose there exists one $\tilde h \in \widehat \Aa_d$, ${\rm ev}(\hat h)=h$, such that ${\rm Spec}_B(\hat h) \cap G$ does not contain an open sub-interval of $G$. Then we can choose $\Phi$ in \eqref{Eq-ExponentialMap} such that its domain of variation is entirely contained in this interval. In this case, $\Phi$ takes the values $0$ and $1$ on ${\rm Spec}(\hat h)$ and, consequently $e^{2 \pi i \Phi(\hat h)} =1$. But this contradicts the assumption that $\big [ e^{2 \pi i \Phi(\hat h)} \big ]_1 \neq [1]_1$.\qed

\begin{remark}{\rm As emphasized several time already, the bulk-boundary principle formulated in Corollary~\ref{Cor-BBPrinciple} is a statement about the spectral properties of the elements from the generating algebras. The interest is, however, on the spectral properties of the physical representations, hence it is important to translate the statements to that context. We will do so for the case when $(\Omega,\ZM^d,\tau)$ is minimal. In this case, the bulk-boundary principle, together with \eqref{Eq-HalfSpaceOmegaSpectrumP}, assure us that:
\begin{equation}
\bigcup_{k \in \ZM} {\rm Spec} \big (\widehat H_{\tau_{\bm 0,k}\omega} \big ) \cap G = G.
\end{equation}
This can be translated into the following prediction. Assume we are given a dynamically-generated pattern $\omega$ of resonators, whose collective dynamics is determined by a Hamiltonian satisfying the conditions in \ref{Cor-BBPrinciple}. Out of such pattern, one can generate a bundle of half-space patterned resonators by cutting copies of the same pattern along shifted boundaries. Then, according to Remark~\ref{Re-BoundarySpec}, the collective dynamics of this bundle of resonators is determined by $\bigoplus_{k \in \ZM} \widehat H_{\tau_{\bm 0,k}\omega}$ and, consequently, the bundle displays a boundary spectrum that covers entirely the bulk spectral gap. Moreover, peeling off any number of layers from the open end of the bundle produces a Hamiltonian that is unitarily equivalent to the original one, hence the boundary spectrum remains intact. The conclusion is that, indeed, the bundle display topological edge spectrum.
}$\Diamond$
\end{remark}

\subsection{Examples of topological edge spectra}

As we have seen in Corollary~\ref{Cor-BBPrinciple}, to predict the existence of topological edge spectrum one needs to know:
\begin{itemize}
\item The $K_0$-group of the bulk algebra;
\item The $K_1$-group of the boundary algebra;
\item How the exponential map acts between these two groups.
\end{itemize}
For the concrete examples introduced in section~\ref{Sec:Examples}, only the last point remains to be clarified. As we have seen, for all these models, $\Aa_d \simeq \Aa_\Theta$ and $\widetilde \Aa_d \simeq \Aa_{\widetilde \Theta}$, where $\Theta$ is an $n\times n$ anti-symmetric matrix and $\widetilde \Theta$ is the restriction of $\Theta$ to the indices ${1,\ldots,n-1}$.

\begin{proposition}[\cite{PS}] Let $[p_J]_0$, $J \subseteq \{1,\ldots,n\}$, $|J|={\rm even}$, be the generators of $K_0(\Aa_\Theta)$ as in Example~\ref{Ex-K0NC}, and $[\tilde u_J]_1$, $J \subseteq \{1,\ldots,n-1\}$, $|J|={\rm odd}$, be the generators of $K_1(\Aa_{\widetilde \Theta})$ as in Example~\ref{Ex-K1NC} and Remark~\ref{Re-TildeDecay}. Then:
\begin{equation}
{\rm Exp}\big ( [p_{J \cup \{n\}}]_0 \big ) = [\tilde u_J]_1, \quad J\subset \{1,\ldots,n-1\}.
\end{equation}
\end{proposition}

\begin{corollary} Let $h \in \Aa_d$ and $G$ a gap in ${\rm Spec}(h)$. Let:
\begin{equation}
[p_G]_0 = \sum_{|J|={\rm even}} c_J [p_J]_0, \quad J \subseteq \{1,\ldots,n\},
\end{equation}
be the resolution of the spectral projection in $K_0(\Aa_d)$. Then, if any of $c_J$'s with $J \cap \{d\} \neq \emptyset$ is non-zero, then ${\rm Exp}\big ([p_G]_0\big ) \neq [1]_1$ and topological edge spectrum emerges after a cut.
\end{corollary}

\begin{figure}[H]
\center
\includegraphics[width=0.7\textwidth]{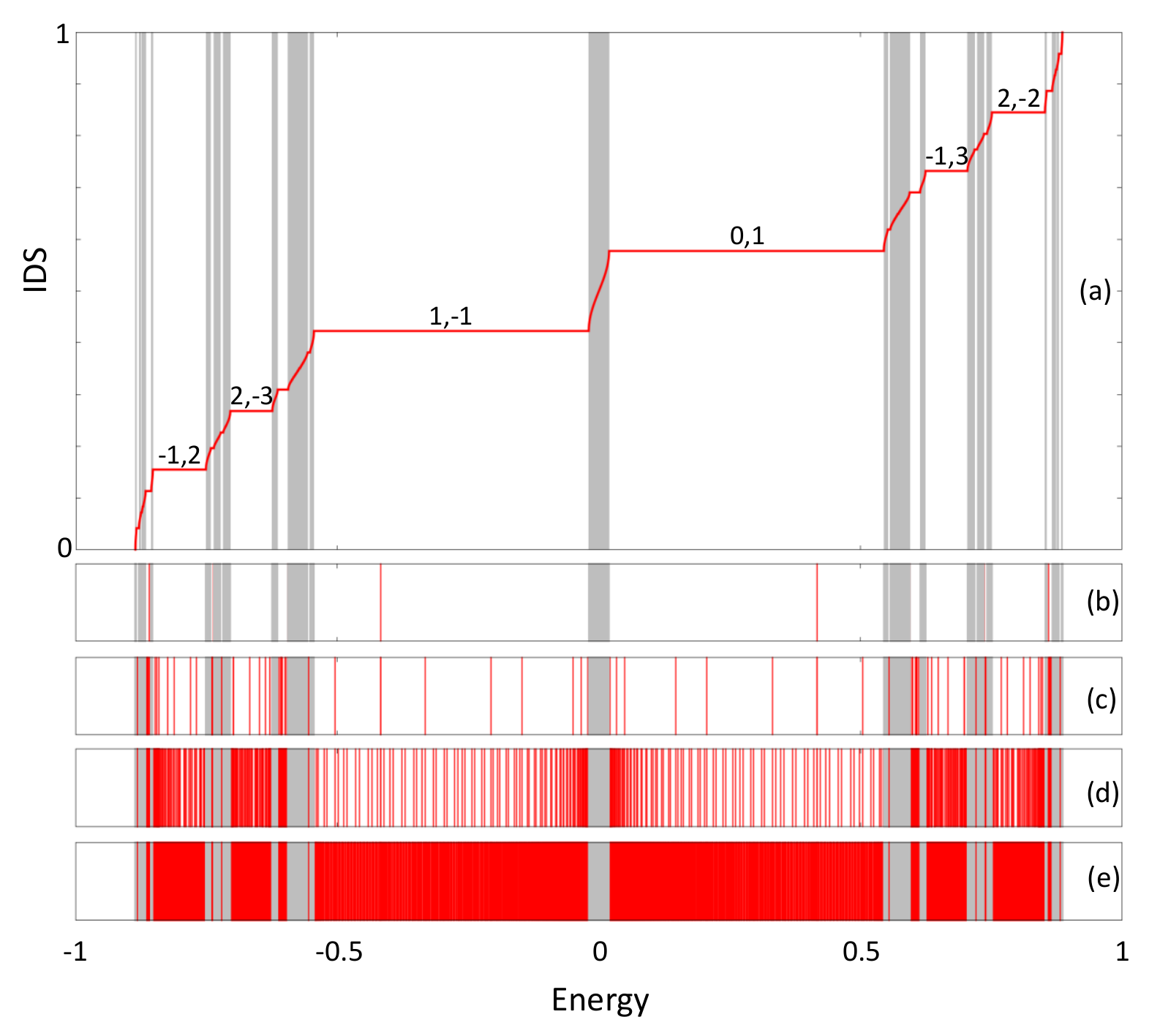}\\
  \caption{\small {\bf Topological edge spectrum for example \ref{ExampleI1}.} (a) Bulk spectrum (shaded regions), integrated density of states (red curve) and gap labels. (b-e) Edge spectrum (red marks) for bundles containing 1 (b), 10 (c), 100 (d) and 1000 (e) patterns. 
  }
 \label{Fig-EdgeSpecEx1Nr5042Rad0p4}
\end{figure}

\begin{example}{\rm The computations in Fig.~\ref{Fig-EdgeSpecEx1Nr5042Rad0p4} relate to Example~\ref{ExampleI1} and they were performed with the Hamiltonian \eqref{Eq-ModelHam}. Only a section of the spectral butterfly reported in Fig.~\ref{Fig-ButterflyFigures_Ex1} was considered. The irrational value $\theta= \frac{1}{\sqrt{3}}$ was chosen, specifically because it accepts the small-denominator rational approximation $\frac{2911}{5042}$, whose error is approximately $-1.13 \times 10^{-8}$. As such, we can generate a good numerical representation of the bulk spectrum by working with a finite pattern of length $L=5042$ and by imposing periodic boundary conditions at the ends. On the same configuration, we can evaluate IDS by the methods described in sections~\ref{Sec-GapLabeling} and \ref{Sec-NumericalExamples1}. The results of this first set of computations are reported in panel (a) of Fig.~\ref{Fig-EdgeSpecEx1Nr5042Rad0p4}, together with the gap labels $(c_\emptyset,c_{\{1,2\}})$. Recall that, for this particular example, \eqref{Eq-IDSGeneric} gives:
\begin{equation}
IDS(G) =\Tt(p_G)= c_{\emptyset} + c_{\{1,2\}} \, \theta. 
\end{equation}
As one can see from Fig.~\ref{Fig-EdgeSpecEx1Nr5042Rad0p4}, all $c_{\{1,2\}}$ labels of the prominent gaps are non-trivial, hence topological edge spectrum is expected. This is confirmed in panels (a)-(e), where the edge spectrum of several bundles is reported. The bundles were obtained by imposing Dirichlet boundary conditions at the fixed sites $1$ and $L$ while shifting the pattern $N$-times, where $N=1$ in panel (b), 10 in panel (c), 100 in panel (d) and 1000 in panel (e). The main observation here is that the bulk gaps with nontrivial labels are sampled finer and finer by the edge spectrum of the bundle, hence supporting the prediction that, in the limit of infinite bundles, the bulk gap is densely sampled by the edge spectrum.  
}
\end{example}

\begin{figure}[H]
\center
\includegraphics[width=0.7\textwidth]{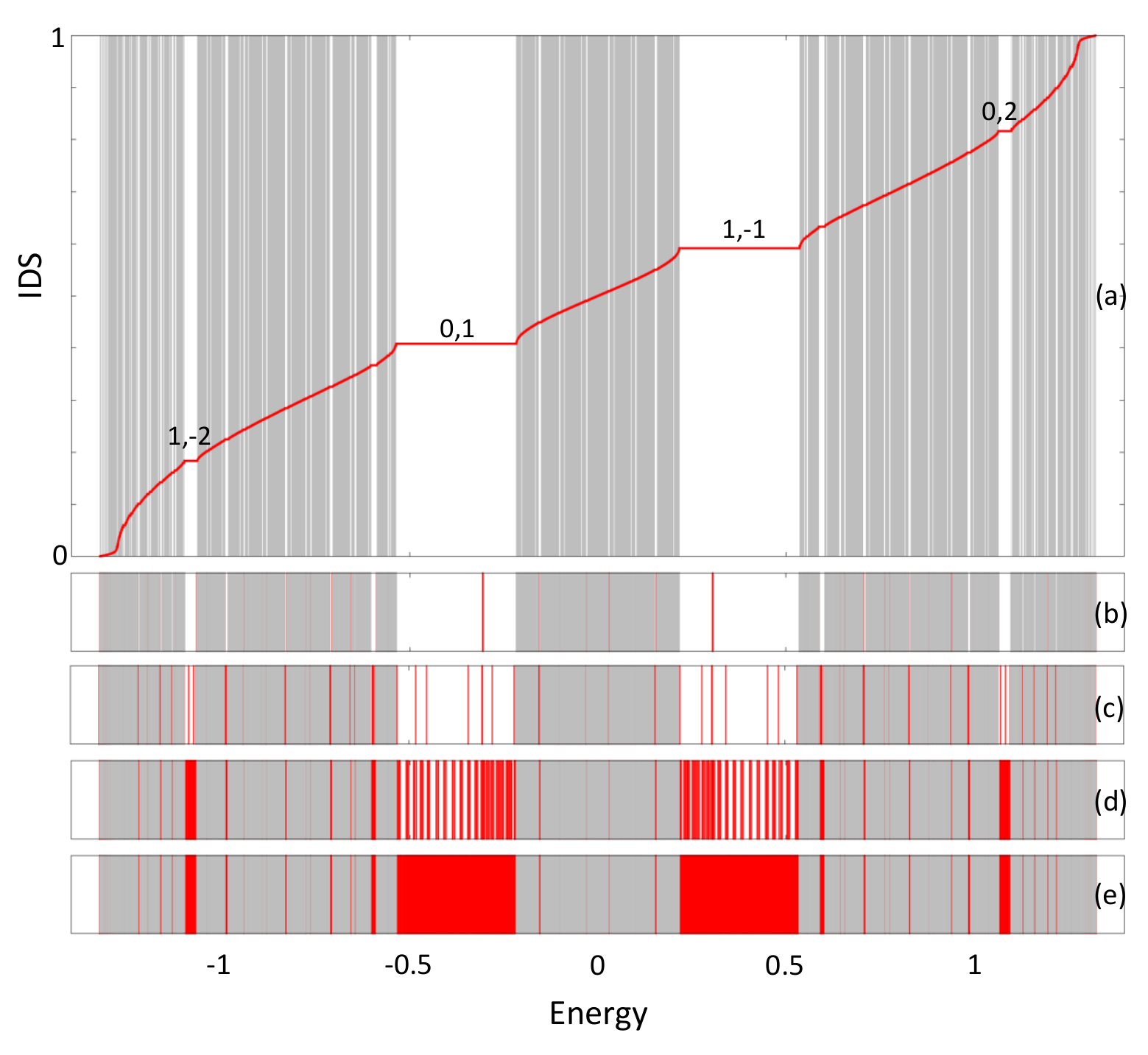}\\
  \caption{\small {\bf Topological edge spectrum for example \ref{ExampleII1}}. (a) Bulk spectrum (shaded regions), integrated density of states (red curve) and gap labels. (b-e) Edge spectrum (red marks) for bundles containing 1 (b), 10 (c), 100 (d) and 1000 (e) patterns.
  }
 \label{Fig-EdgeSpecEx2Nr4801G0p01D0p2}
\end{figure}

\begin{example}{\rm The computations in Fig.~\ref{Fig-EdgeSpecEx2Nr4801G0p01D0p2} relate to Example~\ref{ExampleII1} and they were also performed with the Hamiltonian \eqref{Eq-ModelHam}. Its spectral butterfly was reported in Fig.~\ref{Fig-ButterflyFigures_Ex2} but here we only consider the irrational value $\theta= \frac{1}{\sqrt{6}-1}$, specifically chosen because $\tilde \theta = \frac{\theta}{1+\theta}=\frac{1}{\sqrt{6}}$ accepts the small-denominator rational approximation $\frac{1960}{4801}$, whose error is approximately $-8.85 \times 10^{-9}$. As such, we can generate a good numerical representation of the bulk spectrum and compute IDS by working with a finite pattern of length $L=4801$, as explained in the previous example. The results of this first set of computations are reported in panel (a) of Fig.~\ref{Fig-EdgeSpecEx2Nr4801G0p01D0p2}, together with the gap labels $(c_\emptyset,c_{\{1,2\}})$. Recall that for this model:
\begin{equation}
IDS(G) =\Tt(p_G)= c_{\emptyset} + c_{\{1,2\}} \, \tilde \theta. 
\end{equation}
As one can see, all $c_{\{1,2\}}$ labels of the prominent gaps are non-trivial, hence topological edge spectrum is expected. This is confirmed in panels (b)-(e) of Fig.~\ref{Fig-EdgeSpecEx2Nr4801G0p01D0p2}. The details of these calculations are the same as for the previous example.
}
\end{example}

\begin{figure}[H]
\center
\includegraphics[width=0.7\textwidth]{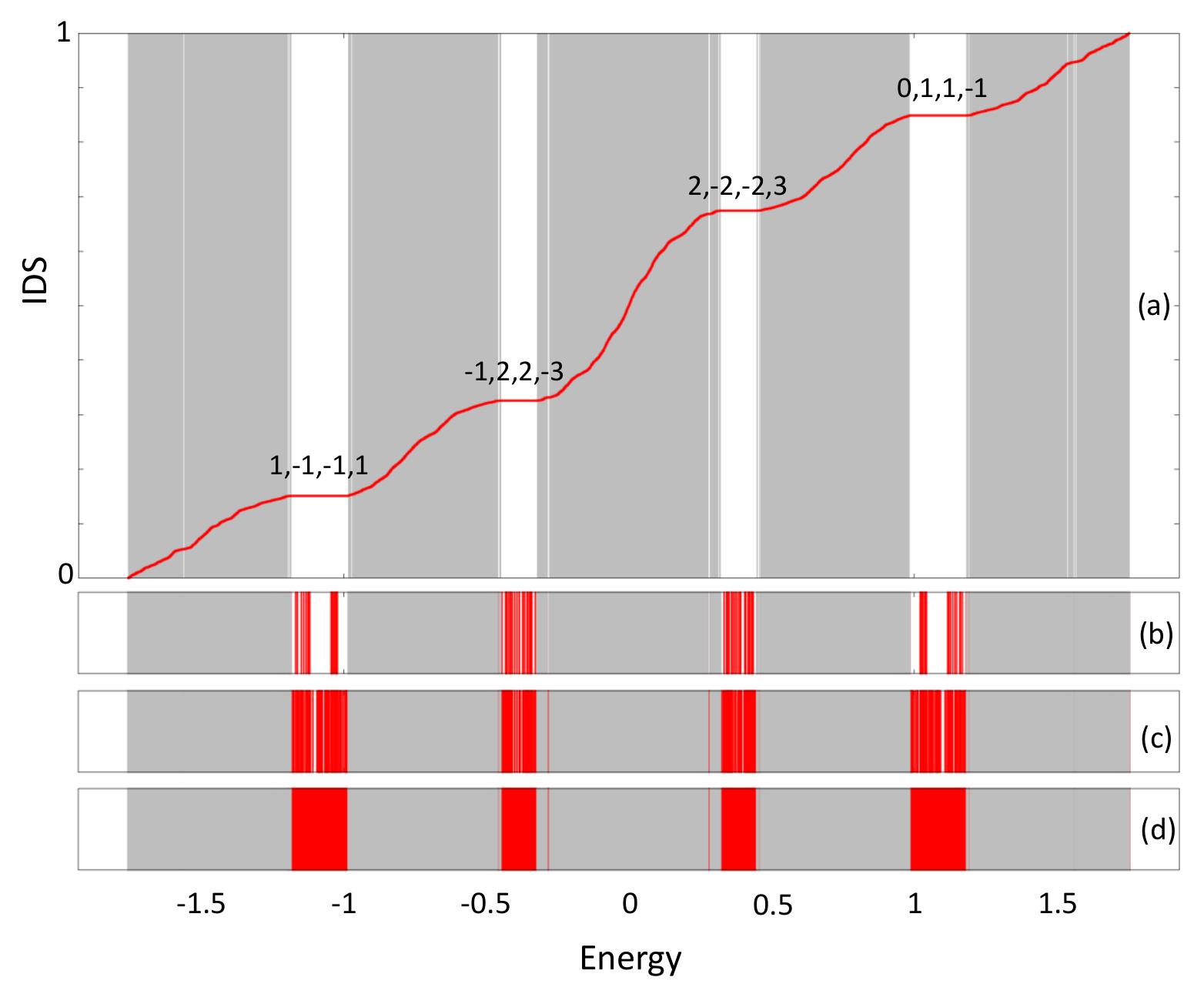}\\
  \caption{\small {\bf Topological edge spectrum for example \ref{ExampleIII1}.} (a) Bulk spectrum (shaded regions), integrated density of states (red curve) and gap labels. (b-e) Edge spectrum (red marks) for bundles containing 1 (b), 10 (c) and 100 (d) patterns.
  }
 \label{Fig-EdgeSpecEx3Nr153Nr148Rad0p4}
\end{figure}

\begin{example}{\rm The computations in Fig.~\ref{Fig-EdgeSpecEx3Nr153Nr148Rad0p4} relate to Example~\ref{ExampleIII1} and they are also performed with the Hamiltonian \eqref{Eq-ModelHam}. Guided by the spectral butterfly \ref{Fig-ButterflyFigures_Ex3}, we fixed $\theta_1= \frac{3-\sqrt{3}}{2} \approx 0.633$ and $\theta_2=\frac{4-\sqrt{5}}{3} \approx 0.587$, which are closed to the values of $\theta$'s in \ref{Fig-ButterflyFigures_Ex3} where prominent spectral gaps can be observed. These $\theta$'s accept the rational approximations $\theta_1 \approx \frac{97}{153}$ and $\theta_2 \approx \frac{87}{148}$, of errors $1.2 \times 10^{-5}$ and $-1.3\times 10^{-4}$, respectively. As such, we can generate reasonable numerical representations of the bulk spectrum and of IDS by working with a finite pattern of size $153 \times 148$ and imposing periodic boundary conditions at the edges. The results of this first set of computations are reported in panel (a) of Fig.~\ref{Fig-EdgeSpecEx3Nr153Nr148Rad0p4}, together with the gap labels $(c_\emptyset,c_{\{1,3\}},c_{\{2,4\}},c_{\{1,2,3,4\}})$. Recall that for this example:
\begin{equation}
IDS(G) =\Tt(p_G)= c_{\emptyset} + c_{\{1,3\}} \, \theta_1+c_{\{2,4\}} \, \theta_2+ c_{\{1,2,3,4\}} \, \theta_1 \, \theta_2. 
\end{equation}
As one can see, all the relevant labels of the prominent gaps are non-trivial, hence topological edge spectrum is expected. This is confirmed in panels (a)-(e) of Fig.~\ref{Fig-EdgeSpecEx3Nr153Nr148Rad0p4}, where the Dirichlet boundary condition was imposed along second spatial direction. Note that, since the top label $c_{\{1,2,3,4\}}$ is non-zero for all the gaps seen in Fig.~\ref{Fig-EdgeSpecEx3Nr153Nr148Rad0p4}, topological spectrum also emerges when the cut is made along the first direction.
}
\end{example}

\section{Singular dynamical systems}
\label{Ch:SingDS}

We have emphasized several times that the range of patterns obtained with the proposed dynamical algorithms is vast. In this chapter we point out that further interesting patterns can be obtained by limiting process. In both cases presented below, the topological dynamical systems become singular and the topological boundary spectrum disappears in one case but it survives in the other. At this moment, we can explain the first case with the tools we already developed but we are forced to leave the second example for future investigations.

\begin{figure}[t]
\center
  \includegraphics[width=0.6\textwidth]{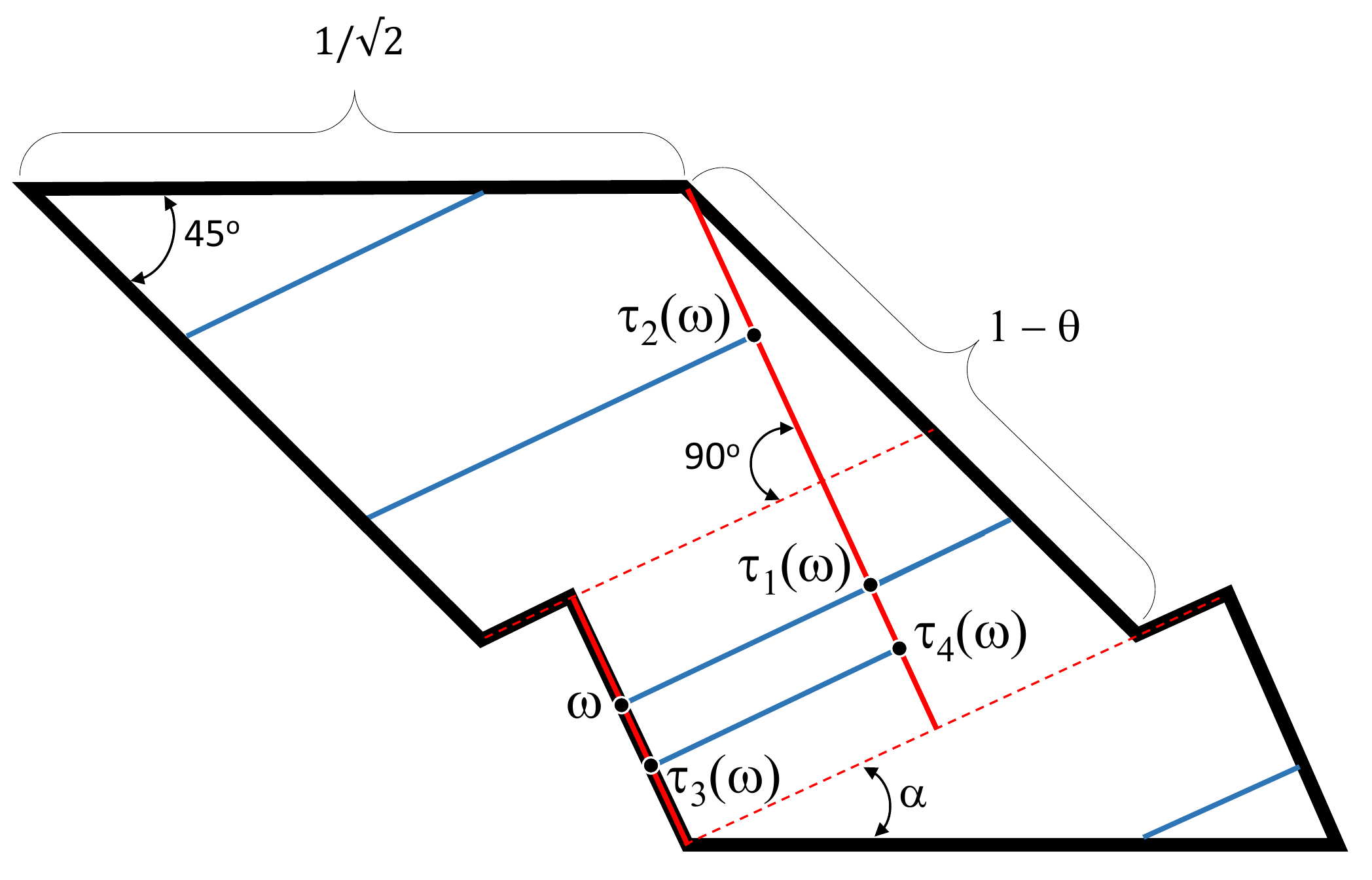}\\
  \caption{\small {\bf The dynamical system for the cut and project sequences.}}
 \label{Fig-IncommensurateF}
\end{figure}
 
\subsection{Cut and project patterns in $d=1$} 

\begin{figure}[b]
\center
\includegraphics[width=\textwidth]{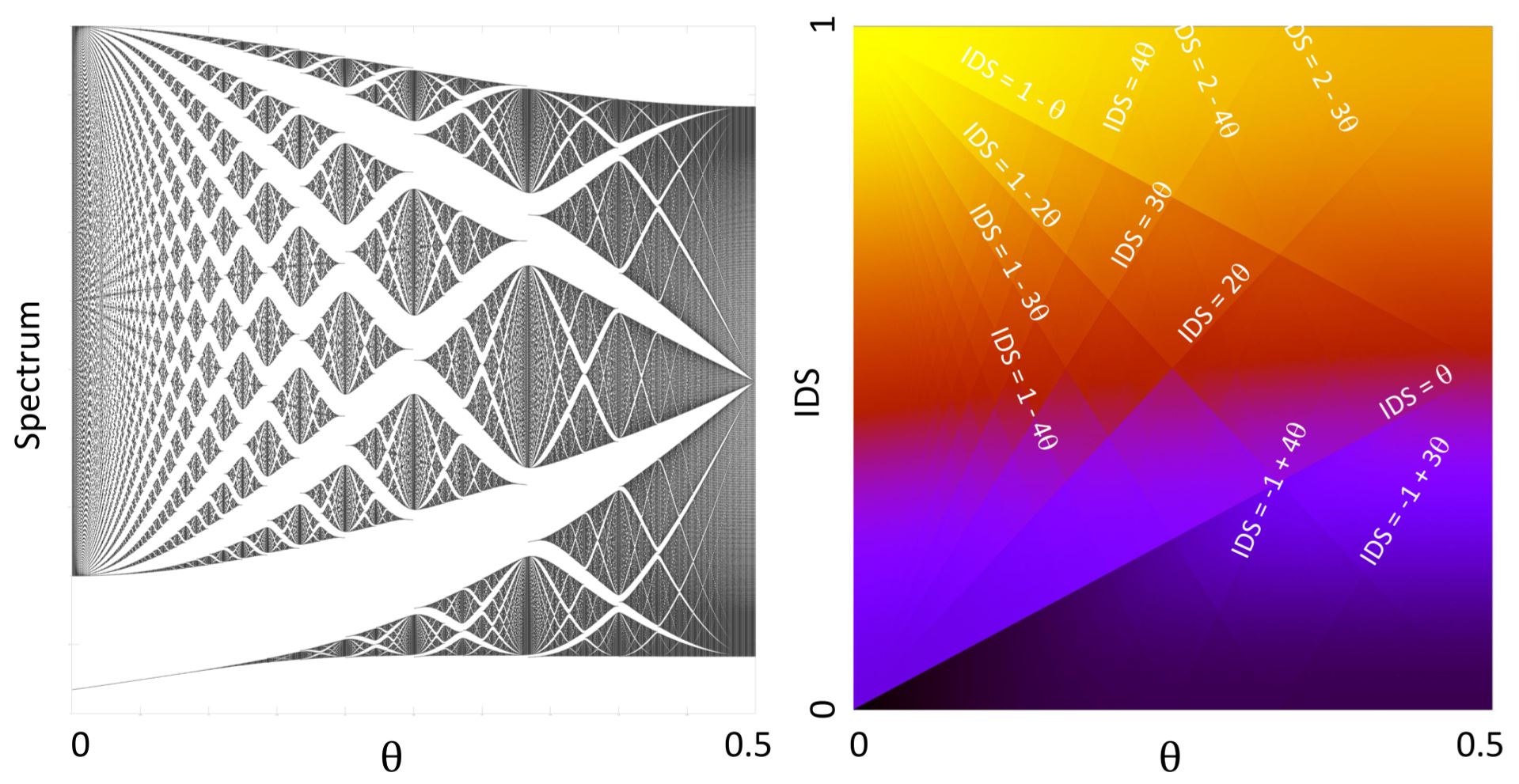}\\
  \caption{\small {\bf Gap labeling for the cut and project patterns}. The spectral (left) and IDS (right) butterflies computed for the Hamiltonian \eqref{Eq-ModelHam}. The computations were carried on a finite pattern of $840$ sites.
  }
 \label{Fig-ButterflyFibonacci}
\end{figure}

\begin{figure}[t]
\center
\includegraphics[width=0.7\textwidth]{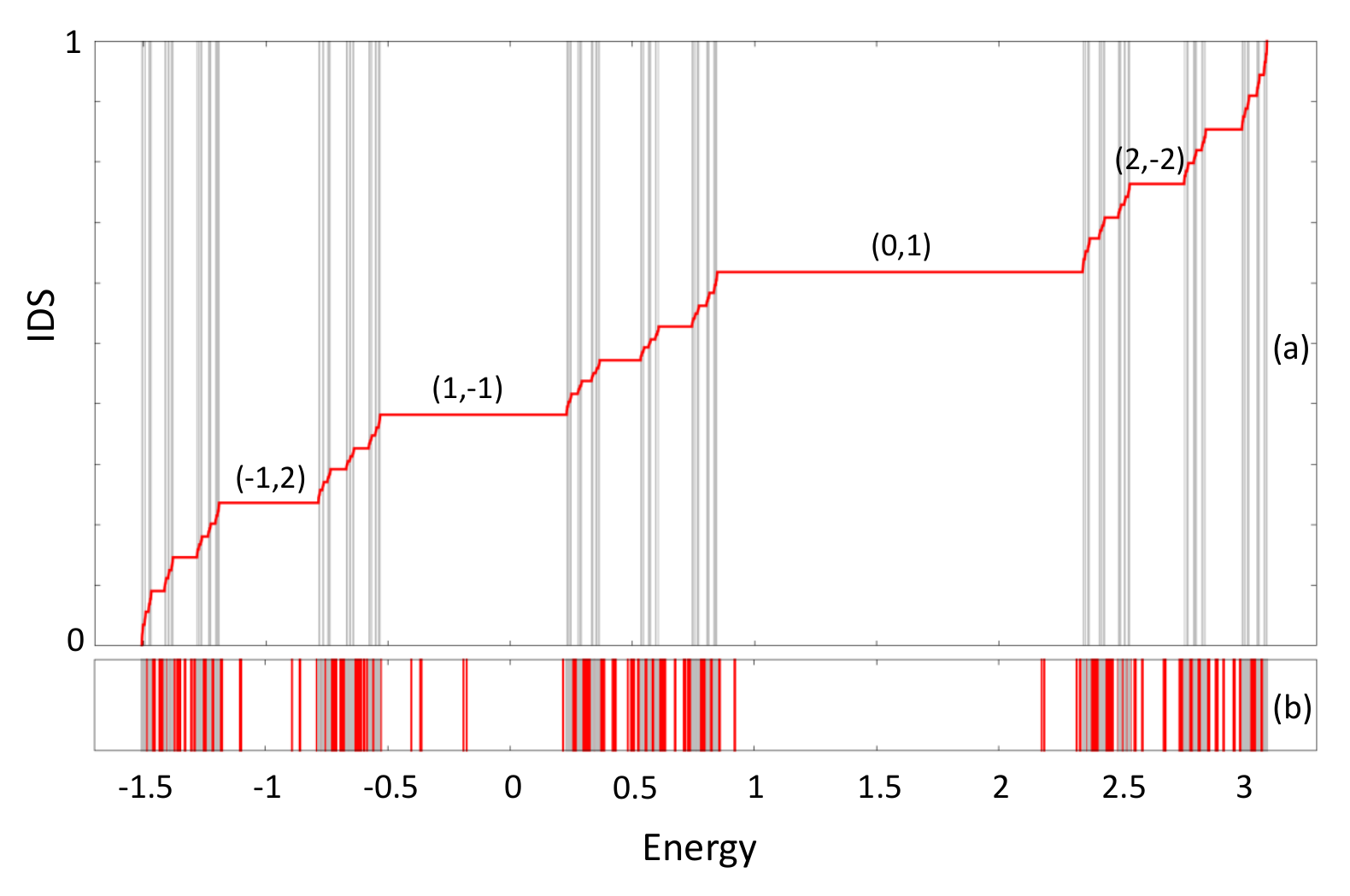}
  \caption{\small {\bf Edge spectrum for a cut and project pattern.} (a) Bulk spectrum (shaded regions), integrated density of states (red curve) and gap labels. (b) Edge spectrum for bundles containing 1000 patterns. The parameter was fixed at $\theta = \frac{3-\sqrt{5}}{2}$, corresponding to the Fibonacci sequence. The computation was performed on a chain of size $6765$ using the rational approximation $\theta = \frac{2584}{6765}-9.77\times 10^{-9}$.}
 \label{Fig-EdgeSpecFibonacciLevel14}
\end{figure}

The bulk boundary correspondence for this type of patterns has been recently studied in \cite{KellendonkArxiv2017}. In Fig.~\ref{Fig-IncommensurateF}, we explain how the cut and project procedure can be formulated as a dynamical algorithm. The shape shown there is in fact the 2-torus $(\RM/\ZM \big ) \times \big ( \RM/\ZM \big )$, when considered with periodic boundary conditions. On this torus, we wrap the blue string and, every time the string meets the red segments, called the transversal, we mark a point on the string. If the parameter $\theta$ is chosen irrational, then the string densely fills the torus without ever closing into itself. At the end of the process, we unwind the string and the marked points provide the point patterns. As illustrated in Fig.~\ref{Fig-IncommensurateF}, this process defines the dynamical system $\tau$ on the transversal. For each seed point $\omega$, we get a different point pattern, and thus we have a collection of point patterns indexed by the transversal. Naively we might think that the discrete hull of these patterns is the whole transversal, and this, in fact, would be the case if the red line were a continuous closed curve on the surface. However, because the red line is singular, we need to remove the $\tau$-orbit of the singular points. When the resulting open set is closed in the topology induced by the space of patterns, the result is that the hull $\Omega$ is a Cantor set \cite{KellendonkArxiv2017}.

\vspace{0.2cm}

\begin{figure}[b]
\center
\includegraphics[width=0.8\textwidth]{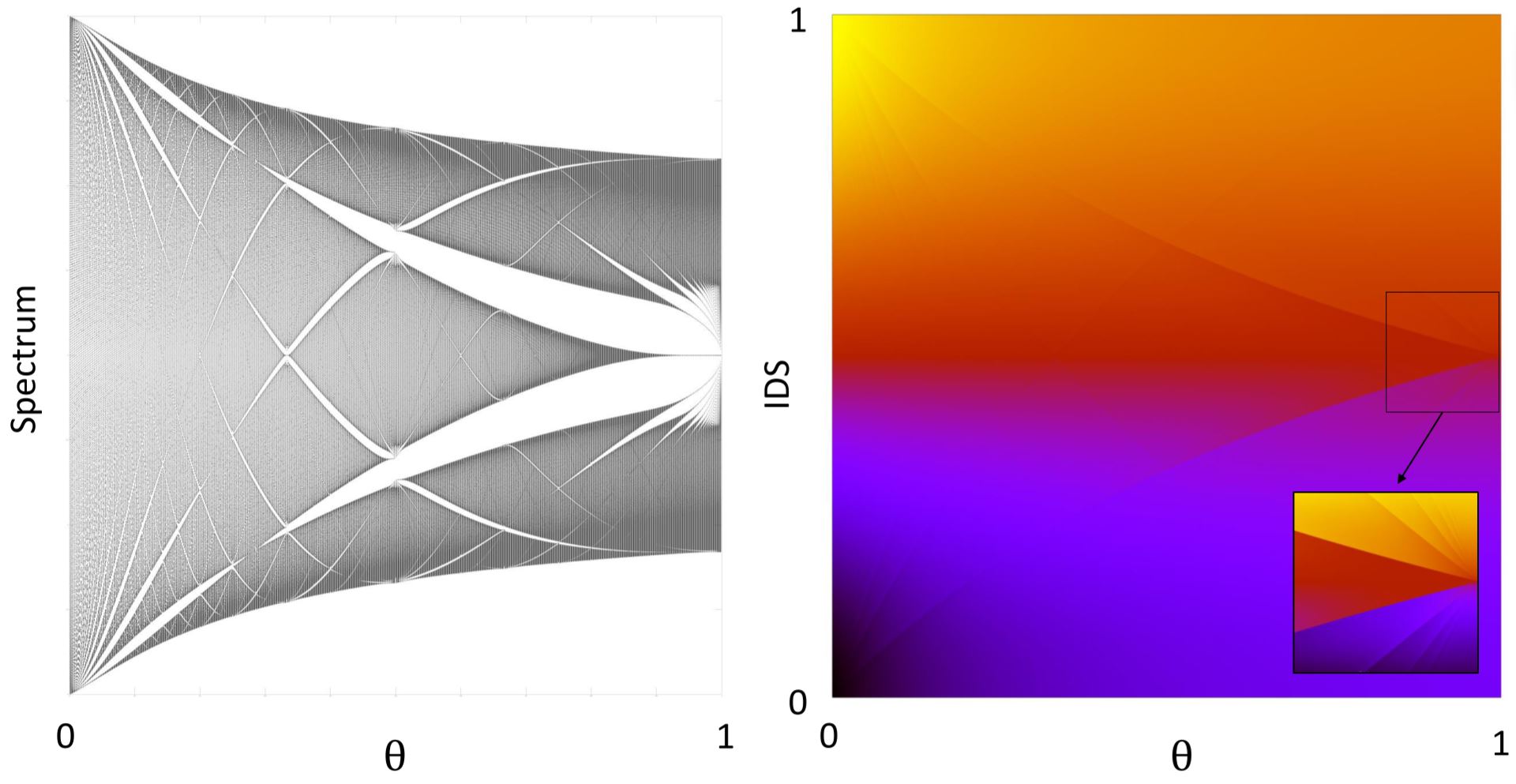}\\
  \caption{\small {\bf Gap labeling for the pattern in Fig.~\ref{Fig-Incomm2}.} The spectral (left) and IDS (right) butterflies, computed exactly as in Fig.~\ref{Fig-ButterflyFigures_Ex2}, except for $g=0$.
  }
 \label{Fig-Butterfly_Ex2Ideal}
\end{figure}

The bulk algebra $\Aa_d$ for these patterns has been analyzed in \cite{Bel95}. In particular, the non-commutative 2-torus is strictly embedded in $\Aa_d$ but the $K$-theories of the two algebras are identical. Furthermore, the generators of both $K$-groups can be represented by elements from the non-commutative 2-torus. As such, the gap labeling is identical with that of Example~\ref{ExampleI1}. This is obvious in Fig.~\ref{Fig-EdgeSpecFibonacciLevel14}, where the spectral and the IDS butterflies have been mapped for the Hamiltonian \eqref{Eq:HamChoice}.

\vspace{0.2cm}

From Proposition \eqref{Pro-BoundaryChar}, we have that the boundary algebra is stably isomorphic to that of continuous functions over the hull $\Omega$. Since the latter is a totally disconnected topological space, the group $K_1(C(\Omega))$ is trivial. As a result, the exponential map is automatically trivial, hence no topological edge spectrum is expected. This is quite evident in the computations reported in Fig.~\ref{Fig-EdgeSpecFibonacciLevel14}. In fact, there is a striking difference when the data is compared with that in Fig.~\ref{Fig-ButterflyFigures_Ex1}.  
          
\begin{figure}[t]
\center
\includegraphics[width=0.7\textwidth]{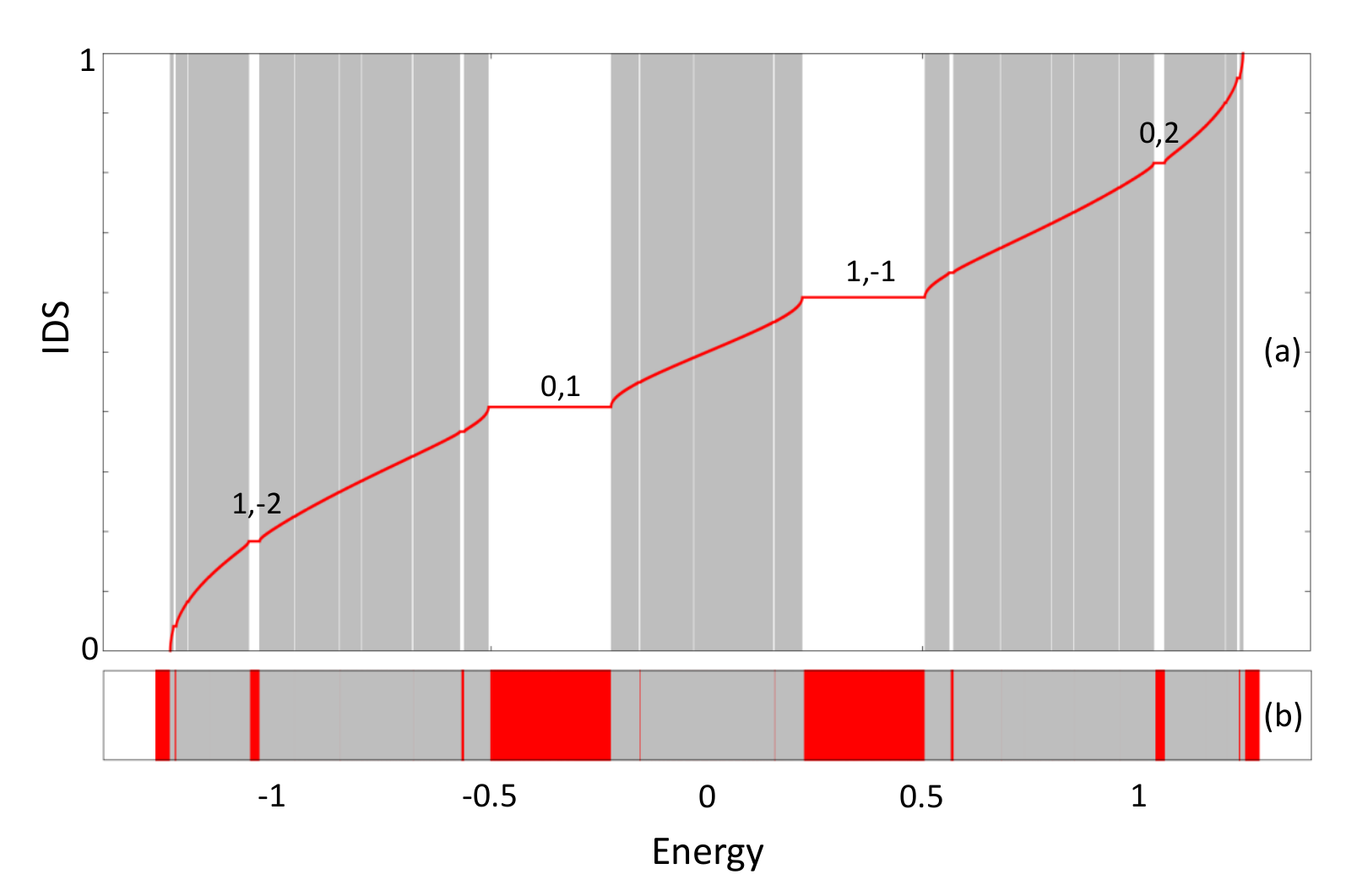}\\
  \caption{\small {\bf Topological edge spectrum for the pattern in Fig.~\ref{Fig-Incomm2}.} (a) Bulk spectrum (shaded regions), integrated density of states (red curve) and gap labels. (b) Edge spectrum (red marks) for bundles containing 1000 patterns.
  }
 \label{Fig-EdgeSpecIdealEx2Nr4801D0p25}
\end{figure}

\subsection{Incommensurate layers of periodic lattices}

We present here the numerical results for the ideal pattern shown in Fig.~\ref{Fig-Incomm2}. This pattern is the $g=0$ limit of the patterns analized in section~\ref{ExampleII1}. Clearly, the hull $\Omega$ becomes singular in this limit yet, quite surprisingly, nothing out of the ordinary happens. Indeed, in Fig~\ref{Fig-Butterfly_Ex2Ideal} we report the spectral and IDS butterflies, which turned out to be very similar to those corresponding to the smooth approximations in Fig.~\ref{Fig-ButterflyFigures_Ex2}. By examining the IDS butterfly, we can conclude with great confidence that the prediction in \ref{Ex-IDSRange} continue to hold, which indicates that the $K$-theory of the bulk algebra remains unchanged in the singular limit. While this was also the case for the cut and project patterns, a closer examination of the spectral butterflies reveal a major difference. While in Fig.~\ref{Fig-ButterflyFibonacci}(a) all the gap regions are connected, in Fig.~\ref{Fig-Butterfly_Ex2Ideal}(a) the gap regions are separated by essential spectrum. Note that the latter was always the case for the patterns analyzed in section~\ref{Sec-NumericalExamples1}. Lastly, the data reported in Fig.~\ref{Fig-EdgeSpecIdealEx2Nr4801D0p25} indicates the topological edge spectrum survives in the singular limit.

\end{document}